\documentclass[twocolumn,tighten]{aastex631}

\usepackage{amsmath}
\usepackage{mathtools}
\usepackage{CJKutf8}
\usepackage{mathptmx,txfonts}
\usepackage{graphicx}
\usepackage{color}
\usepackage{longtable}
\usepackage{array}

\DeclarePairedDelimiter\abs{\lvert}{\rvert}%
\begin{document}

\shorttitle{WD+MS Insights on Rotation}
\shortauthors{Chiti et al.}

\title{Rotation at the Fully Convective Boundary: Insights from Wide WD + MS Binary Systems}

\author[0000-0001-5180-2271]{Federica Chiti}
\affiliation{Institute for Astronomy,
University of Hawai‘i at Mānoa,
2680 Woodlawn Dr., Honolulu, HI 96822, USA}

\author[0000-0002-4284-8638]{Jennifer L. van Saders}
\affiliation{Institute for Astronomy,
University of Hawai‘i at Mānoa,
2680 Woodlawn Dr., Honolulu, HI 96822, USA}

\author[0000-0003-3868-1123]{Tyler M. Heintz}
\affiliation{Department of Astronomy \& Institute for Astrophysical Research, Boston University, 725 Commonwealth Ave, Boston, MA, 02215, USA}

\author[0000-0001-5941-2286]{J. J. Hermes}
\affiliation{Department of Astronomy \& Institute for Astrophysical Research, Boston University, 725 Commonwealth Ave, Boston, MA, 02215, USA}

\author[0000-0001-7664-648X]{J. M. Joel Ong \begin{CJK}{UTF8}{bsmi}
(王加冕)
\end{CJK}}
\affiliation{Institute for Astronomy,
University of Hawai‘i at Mānoa,
2680 Woodlawn Dr., Honolulu, HI 96822, USA}
\affiliation{NASA Hubble Fellow}

\author[0000-0003-3244-5357]{Daniel R. Hey}
\affiliation{Institute for Astronomy,
University of Hawai‘i at Mānoa,
2680 Woodlawn Dr., Honolulu, HI 96822, USA}

\author{Michele M. Ramirez-Weinhouse}
\affiliation{Lāna'i High and Elementary School, 555 Fraser Ave, Lanai City, HI 96763}

\author{Alison Dugas}
\affiliation{Institute for Astronomy,
University of Hawai‘i at Mānoa,
2680 Woodlawn Dr., Honolulu, HI 96822, USA}

\begin{abstract}
Gyrochronology, a valuable tool for determining ages of low-mass stars where other techniques fail, relies on accurate calibration. We present a sample of 185 wide ($>$$100$\,au) white dwarf + main sequence (WD + MS) binaries. Total ages of WDs are computed using all-sky survey photometry, Gaia parallaxes, and WD atmosphere models. Using a magnetic braking law calibrated against open clusters, along with assumptions about initial conditions and angular momentum transport, we construct gyrochrones to predict the rotation periods of MS stars. Both data and models show that, at the fully convective boundary (FCB), MS stars with WD ages up to 7.5\,Gyr and within a $<50\,\mathrm{K}$ effective temperature range experience up to a threefold increase in rotation period relative to stars slightly cooler than the FCB. We suggest that rapid braking at this boundary is driven by a sharp rise in the convective overturn timescale ($\tau_{\mathrm{cz}}$) caused by structural changes between partially and fully convective stars and the $^3 \textrm{He}$ instability occurring at this boundary. While the specific location in mass (or temperature) of this feature varies with model physics, we argue that its existence remains consistent. Stars along this feature exhibit rotation periods that can be mapped, within 1$\sigma$, to a range of gyrochrones spanning  $\approx 6$\, Gyr. Due to current temperature errors ($\simeq$$50\,\mathrm{K}$), this implies that a measured rotation period cannot be uniquely associated to a single gyrochrone, implying that gyrochronology may not be feasible for M dwarfs very close to the FCB. 
\end{abstract}


\section{Introduction} \label{sec:intro}

Ages of stars are critical to our understanding of the evolution of astrophysical systems and yet are one of the most difficult stellar properties to measure. The only star for which we have a precise and accurate age is the Sun; for any other star, age can only be estimated or inferred. There are many techniques to estimate stellar ages, but there is no single method that is applicable to all spectral types \citep{soderblom2010}. 

K and M dwarfs are the most numerous stars in the Galaxy and have lifetimes longer than the age of the Milky Way disk, meaning that they preserve a record of the history of star formation and chemical evolution of the Galaxy. They are, however, resistant to standard age-dating techniques. Isochrone fitting fails to provide constraints on stellar ages when used on low-mass stars, due to their slow nuclear evolution \citep{takeda2017}. Similarly, asteroseismology, which provides precise ages for Sun-like stars, cannot be used to date low-mass stars like K and M dwarfs due to their low oscillation amplitudes \citep{chaplin2011}.

A promising tool in this low mass regime is gyrochronology \citep{barnes2007}, which derives ages for cool MS stars by exploiting the fact that they spin down with time \citep{skumanich1972} due to magnetic braking. Magnetic braking is the mechanism by which a star loses angular momentum to magnetized stellar winds over time as the result of the interaction between mass loss and dynamo-driven stellar magnetic fields. \citet{epstein2014} showed that under Skumanich-type spin down, rotation-based age dating is potentially among the most precise methods available.

Calibration of period-age relations for low mass stars requires a large sample of old, well-dated low mass stars. However, only a handful of stars below 0.8\,$M_{\odot}$ are currently available, and most of them are in young clusters \citep[4\,Gyr at the oldest, ][]{dungee2022}. Furthermore, observations of these clusters have shown that standard braking models fail to reproduce the observed rotational sequences, suggesting that stellar spin-down may not be as simple as it once appeared.

Recent measurements of the rotation period ($P_{\mathrm{rot}}$) of stars in the benchmark open clusters Praesepe \citep[$\approx$700\,Myr,][]{douglas2019} and NGC 6811 \citep[1Gyr,][]{meibom2011,janes2013} show that a simple power law with a braking index $n$ ($n=0.5$ in the Skumanich law) fails at predicting the observed rotational sequences in these clusters \citep{curtis2020}. While solar-type stars in NGC 6811 have longer periods compared to their counterparts in the younger cluster Praesepe, the two sequences merge at $M<0.8$\,$M_{\odot}$ \citep[K0 to M0-type stars,][]{curtis2019, douglas2019}. In other words, the spin-down appears to ``stall" (or reduce) for low-mass stars in NGC 6811. \cite{spada2020} demonstrated that this phenomenon can be explained by relaxing the assumption of solid-body rotation and allowing for angular momentum (AM) transport between the core and the envelope (i.e. radial differential rotation). The spin-down stalling observed in K and early M-type stars in NGC 6811 is putatively an epoch when AM transport from the core to the envelope balances the AM loss at the surface due to winds. The lack of a radiative core in fully convective stars that can support such core-envelope interaction may be responsible for the closure of the intermediate period gap discovered with Kepler by \cite{mcquillan2013} at the fully convective boundary \citep[FCB; ][]{lu2021}.

Near the FCB, another relevant feature is the underdensity of stars observed near $M_G=10.2$ in the main sequence on the Hertzsprung-Russell diagram found by \cite{jao2018} using Gaia DR2 measurements. It has been proposed that this Gaia M-dwarf gap is a manifestation of the location where stars transition from partially to fully convective, which is predicted to occur at a mass of $\sim0.35$\,$M_{\odot}$. Earlier theoretical work by \cite{vansaders2012} demonstrated that stars slightly above this threshold undergo a structural instability due to non-equilibrium $^3$He burning during the first few billion years on the main sequence. This results in the development of a convective core, separated from a deep convective envelope by a thin radiative layer. The continuous accumulation of central $^3$He causes the radiative zone separating them to thin even further, initiating fully convective episodes. Using stellar models and stellar population synthesis, \cite{feiden2021} confirmed that the $^3$He instability is responsible for the appearance of the M-dwarf gap. 

With the distinct structural changes across the FCB, there have been attempts to comprehend whether there are any consequences for the magnetic properties, activity levels, and rotation rates of stars. For instance, \cite{donati2008} and \cite{reiners2009} demonstrated that $0.34-0.36$\,$M_{\odot}$ stars are prone to undergo sudden alterations in their large-scale magnetic topologies, which could lead to observable surface activity signatures. More recently, \cite{jao2023} conducted a high-resolution spectroscopic H$\alpha$ emission survey of M dwarfs spanning the Gaia M-dwarf gap and argued that stars above the top gap edge exhibit H$\alpha$ emission while stars within the gap or below do not display any emission. Thus, stars near the FCB provide a powerful laboratory for testing the physics of M-dwarf stars, including those affected by the $^3$He instability. Moreover, having a reliable spin-down model that can predict the rotational evolution of these stars is crucial for determining a precise period-age relation for old, low-mass stars. 

A primary limitation is the lack of empirical anchors of known age for old K and M-type stars. Open clusters have been the major contributors of calibrators to date; however, due to their short dissipation timescales \citep[$\sim$200\,Myr;][]{wielen1971}, old clusters are rare and tend to be more distant and challenging to observe. The standard gyrochronology calibrators and the recent observations of late K- and early M- dwarfs in M67 \citep{dungee2022} do not extend beyond 4\,Gyr for stars below 0.8\,$M_{\odot}$. Likewise, the asteroseismic calibrator sample that has been important for understanding braking in solar-mass stars \citep{vansaders2016,hall2021} does not extend below $\approx0.8$\,$M_{\odot}$. 

Wide binaries that contain a white dwarf (WD) companion provide a distinctive opportunity to determine the ages of field stars. WDs are the end product of stars with initial masses less than $ 8-10$\,$M_{\odot}$ and, as they no longer undergo nuclear fusion in their core, they gradually cool with time becoming dimmer and colder. Because their effective temperature and mass uniquely correspond to a single cooling age (given a composition), WDs have been utilized as stellar clocks for decades \citep{fontaine2001} to date a variety of stellar populations, such as our Galaxy -- see \citet{garcia2016} and references therein -- and open and globular clusters -- see \cite{winget2009}, \cite{garcia2010}, \cite{jefferey2011}, \cite{hansen2013} for some examples. The advancement of robust cooling models \citep{bergeron1995} allows WDs to serve as precise and dependable age indicators. However, to determine the complete age of a WD, one needs to consider the time from its zero-age main sequence (ZAMS) to its present state as a WD. This involves using a semi-empirical initial-final mass relation (IFMR) in conjunction with stellar evolution model grids to ascertain the progenitor lifetimes from the ZAMS to the WD phase. Using photometry only, \cite{heintz2022} found that WD ages are precise at the 25\% level for WDs with masses $>0.63\, M_{\odot}$.

Wide, coeval binaries are sufficiently distant ($>100$ au) that the two stars can be expected to evolve as single stars without any interaction between them \citep{2001ApJ...556..265W}. Therefore, the WD companion offers an independent age estimate of the entire system, making it possible to extend period-age relationships to the age of the galactic disk for the most common stars in our galaxy. Previous studies have leveraged on wide WD + MS binary systems to investigate the age-metallicity-activity relation \citep{zhao2011}, the age-velocity relation \citep{raddi2022} and the age-metallicity relation \citep{rebassa2016,rebassa2021}. Close WD + MS binaries have also been used to constrain the relations between magnetic activity, rotation, magnetic braking and age in M stars \citep{morgan2012,rebassa2013,skinner2017,rebassa2023}.

The number of wide coeval binary systems has significantly increased since the launch of the Gaia spacecraft. From Gaia DR2, \cite{elbadry2018} constructed a catalog of over $\sim$$53{,}000$ binaries, $\sim$$3000$ of which contained a WD and a main sequence star, which represented a tenfold increase in the number of known coeval binaries \citep{holberg2013}. With the release of Gaia eDR3, \cite{elbadry2021} published an extensive catalog of 1.3 million spatially resolved binary stars within $\approx$$1$\,kpc of the Sun, including more than $16{,}000$ WD + MS binaries, which increased the sample by another order of magnitude. This increase presents an opportunity to infer precision ages for cool, old stars, which are the focus of this work.

In Section \ref{sec:methods} we describe the physics of the rotational evolution models, their calibration and the technique used to determine WD ages. In Section \ref{sec:data}, we present the sample selection of WD + MS systems. The ages of these systems as revealed by our models and WDs are discussed in Section \ref{sec:results}, where we also compare our sample to other datasets. Finally, the conclusions of our results are presented in Section \ref{sec:conclusions}. 

\section{Methods}\label{sec:methods}
\subsection{Gyrochronology Models}
We use the rotation code \texttt{rotevol} \citep{vansaders2013,somers2017} to model the AM evolution of a star. We use as input non-rotating tracks with stellar masses between 0.18 and 1.15 M$_{\odot}$, generated using the Yale Rotating Evolution Code \citep[\texttt{YREC}, see][]{vanSaders2012b, Pinsonneault1989, Bahcall1992}.  The models include helium and heavy element diffusion following \citet{Thoul1994}, but with the diffusion coefficients multiplied by a factor of 0.753 to match the helioseismically determined helium abundance \citep{Basu1995} in the Sun at solar age. We adopt boundary conditions using the \citet{Allard1997} atmospheric tables, OPAL opacities \citep{Iglesias1996} with low-temperature opacities from \citet{Ferguson2005} for a \citet{gs98} solar mixture. We adopt nuclear reaction rates from \citet{Adelberger2011} with weak screening \citep{Salpeter1954} and the equation of state from the OPAL project \citep{Rogers1996,Rogers2002}. We assume no overshooting, and a mixing length theory of convection \citep{Cox1968,Vitense1953}. Our solar-calibrated model at 4.57 Gyr \citep{Bahcall1995} has $Z=0.01709$, $X=0.71642$ and mixing length parameter $\alpha_{ml} = 1.94243$. We run models at a range of surface spot covering fractions in YREC (0\%, 25\% and 50\%) following the prescription of \citet{Somers2015} and \citet{Somers2020} with a spot temperature contrast ratio $x_{spot}=0.8$ but retain otherwise identical physical ingredients in the spotted models.

We run YREC in the non-rotating configuration and compute the rotational evolution post hoc using the tracer code \texttt{rotevol} \citep{vansaders2013,somers2017}. The benefit of this approach is that we can rapidly search parameter space when fitting the magnetic braking law to the observations; the downside is that rotation cannot influence the structure. While this is a reasonable assumption for all but the most rapid rotators, ideally one would actively couple the starspot filling fraction to the rotation rate \citep[e.g.][]{cao2022, cao2023}. We leave this exercise to future work, and instead examine the behavior of a range of fixed spot covering fractions.

To model the rotation period evolution of a star as a function of time, one must choose appropriate initial starting conditions as well as to specify prescriptions for three processes that drive the angular momentum (AM) evolution, early disk interactions, AM loss at the stellar surface through magnetized winds (``braking law"), and internal AM transport. In this section, we describe the ingredients and assumptions of the stellar evolutionary models.

\subsubsection{Initial conditions}
\begin{figure}
    \centering
    \includegraphics[width=0.47\textwidth]{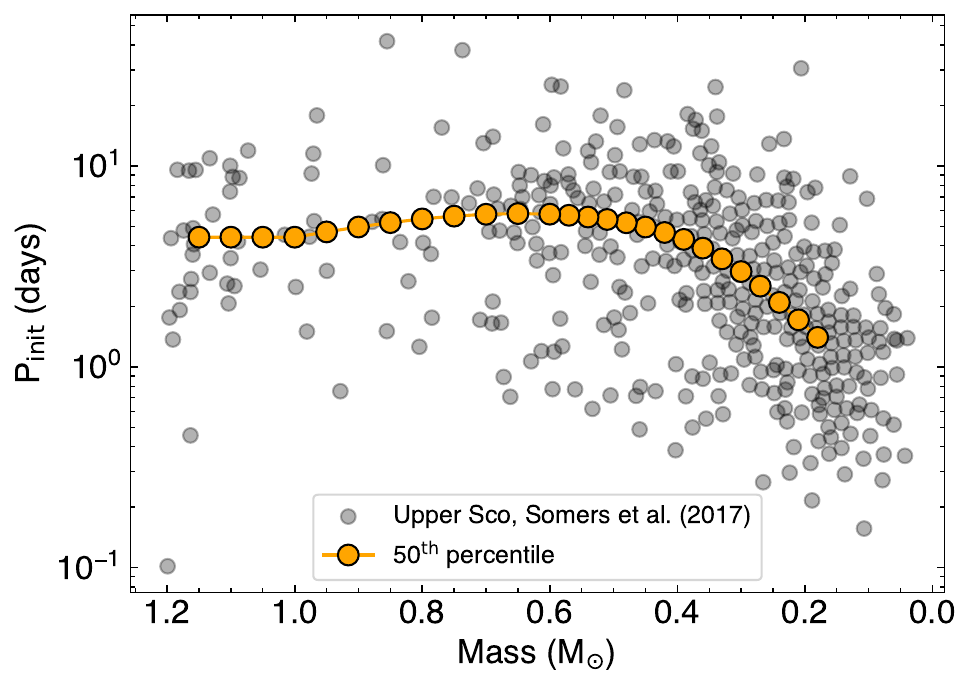}
    \caption{The Upper Sco rotation distribution at 10 Myr \citep{somers2017} from which we adopt the initial rotation periods for the model tracks. The solid yellow line shows a 1-D smoothing spline fit to the median rotation periods. This fit is evaluated at the stellar masses available in our model grid to compute their respective $P_{\mathrm{init}}$.}
    \label{uppersco}
\end{figure}

The initial rotation periods of our models are taken from the observed periods distribution of the young cluster Upper Sco at 10\,Myr. Upper Sco is the most populated cluster sample in the mass range of interest of this work and, by its age, massive accretion disks are nearly absent \citep{williams2011} and significant AM loss has not occurred yet \citep{somers2017, rebull2018}. Therefore, Upper Sco is representative of other young clusters and an ideal dataset from which to infer initial rotational periods. 
We divide the Upper Sco data into mass deciles between the $10^{\mathrm{th}}$ and the $90^{\mathrm{th}}$ percentile masses and compute the median rotation period for each mass bin. The rotation periods are interpolated with respect to the midpoints of the mass bins using a 1-D smoothing spline. We evaluate $P_{\mathrm{init}}$ for each stellar mass available in our model grid, as shown in Fig.~\ref{uppersco}.

\subsubsection{Magnetic braking}
We adopt the \cite{vansaders2013} formulation of the classic wind braking law proposed by \cite{kawaler1988}. We assume that the magnetic field scales as 
\begin{equation}
    \frac{B}{{B_{\odot}}}=\left(\frac{P_{\mathrm{phot}}}{{P_{\mathrm{phot},\odot}}}\right)^{0.5}  \left\{\begin{array}{ll}
     \frac{\omega_{sat}}{\omega_{\odot}} & 
    \omega_{\mathrm{sat}} \leqslant \omega \frac{\tau_{cz}}{\tau_{cz, \odot}} \\
    \frac{\omega \tau_{cz}}{\omega_{\odot} \tau_{cz,\odot}}
     & \omega_{\mathrm{sat}}>\omega \frac{\tau_{cz}}{\tau_{cz,\odot}}
     \end{array},\right.
\label{bfield}
\end{equation}
and tie the mass loss rate to the empirical scaling of $\dot{M}$ with X-ray luminosity from \citet{wood2005}, with X-ray luminosity given by the \citet{pizzolato2003} scaling with Rossby number ($\mathrm{Ro}=P_{\mathrm{rot}}/\tau_{cz} = 2\pi /\omega \tau_{cz}$)
\begin{equation}
    \frac{\dot{M}}{\dot{M_{\odot}}}=\left(\frac{L_{bol}}{L_{bol,\odot}}\right) \left\{\begin{array}{ll}
     \left(\frac{\omega_{sat}}{\omega_{\odot}}\right)^2 & 
    \omega_{\mathrm{sat}} \leqslant \omega \frac{\tau_{cz}}{\tau_{cz, \odot}} \\
    \left(\frac{\omega \tau_{cz}}{\omega_{\odot} \tau_{cz,\odot}}\right)^2
     & \omega_{\mathrm{sat}}>\omega \frac{\tau_{cz}}{\tau_{cz,\odot}}
     \end{array}.\right.
\label{mdot}
\end{equation}
Therefore, we parameterize the AM loss as
\begin{equation}\label{djdt}
\frac{d J}{d t}=\left\{\begin{array}{ll}
f_K K_M \omega\left(\frac{\omega_{\mathrm{ sat}}}{\omega_{\odot}}\right)^2 & \omega_{\mathrm{sat}} \leqslant \omega \frac{\tau_{cz}}{\tau_{cz, \odot}} \\
f_K K_M \omega\left(\frac{\omega \tau_{cz}}{\omega_{\odot} \tau_{cz, \odot}}\right)^2 & \omega_{\mathrm{sat}}>\omega \frac{\tau_{cz}}{\tau_{cz,\odot}}
\end{array},\right.
\end{equation}
where $f_K$ is a normalization constant tuned to reproduce the observed rotation at known age; $\omega$ is the rotation rate; $\omega_{\odot}$ is the rotation rate of the Sun ($2.86\times10^{-6}\,\mathrm{rad\,s}^{-1}$); $\omega_{\mathrm{sat}}$ is the saturation threshold; $\tau_{cz}$ is the convective overturn timescale; $K_M$ is the product
\begin{equation}
\resizebox{\columnwidth}{!}{
    $\frac{K_M}{K_{M, \odot}}=c(\omega)\left(\frac{R}{R_{\odot}}\right)^{3.1}\left(\frac{M}{M_{\odot}}\right)^{-0.22}\left(\frac{L}{L_{\odot}}\right)^{0.56}\left(\frac{P_{\mathrm{phot}}}{P_{\mathrm{phot}, \odot}}\right)^{0.44}$}
\label{km}
\end{equation}
with luminosity $L$, mass $M$, radius $R$, photospheric pressure
$P_{\mathrm{phot}}$; $c(\omega)$ is the centrifugal correction from \cite{matt2012}.

We do not include weakened magnetic braking \citep{vansaders2016} as the low-mass stars in our sample are not expected to reach the relevant Ro within the age of the galactic disk \citep{vansaders2019}. 

\subsubsection{AM Redistribution}
For the internal AM transport, we adopt the prescription for core-envelope coupling as described in \cite{deni2010}. The basic assumption of this model, which was originally proposed by \cite{macgregor1991} as the \textit{double-zone} model, is that the core and the envelope rotate rigidly, but not necessarily at the same rate. This assumption is roughly consistent with the current rotational state of the solar interior \citep{deni2010}. The rate at which the two zones are allowed to exchange angular momentum is defined by the core-envelope coupling timescale $\tau_c$, which is assumed to be constant along the evolution and a function of stellar mass as in \cite{spada2020}
\begin{equation}
    \tau_c = \tau_{c,\odot}\left(\frac{M_*}{M_{\odot}}\right)^{-\alpha}
\end{equation}
where $\tau_{c,\odot}$ is the solar rotational coupling timescale ($\approx 22$ Myr, \citealt{spada2020}) and $\alpha$ is a power-law exponent of this mass-dependent timescale for transport. This scaling was found to remain consistent regardless of the choice of wind braking law and in good agreement with the separate analysis of core-envelope re-coupling by \cite{somers2016}. \cite{lanzafame2015} found $\alpha=7.28$ and \cite{spada2020} refined this estimate to $ \alpha=5.6$ using new data of the clusters Praesepe and NGC 6811 that extended to lower mass stars. More recently, \cite{cao2023} found $ \alpha=11.8$ by using spotted models to fit the rotational sequences in the Pleiades and Praesepe. We leave it as a free parameter in our calibration fits. 

\par 
Thus, our model has five parameters, of which two are set as follows: $P_{\mathrm{init}}$ is given by the Upper Sco rotational distribution at 10 Myr shown in Fig.~\ref{uppersco}; the solar rotational coupling timescale is fixed at 22 Myr. We fit for the remaining three parameters: the normalization constant $f_K$ and the saturation threshold $\omega_{\mathrm{sat}}$ in the braking law, and the exponent $\alpha$ in the core-envelope coupling timescale mass dependence. 
\subsubsection{Model Calibration}
To constrain $f_K$, $\omega_{\mathrm{sat}}$ and $\alpha$, we calibrate the models such that they can reproduce the rotational distributions in the Pleiades \citep[120 Myr;][]{rebull2016}, Praesepe \citep[670 Myr; ][]{douglas2017,douglas2019}, NGC6811 \citep[1 Gyr;][]{curtis2019}, NGC6819 \citep[2.5 Gyr;][]{meibom2015}, Ruprecht 147 \citep[2.7 Gyr;][]{curtis2020} and M67 \cite[4.0 Gyr;][]{dungee2022}. 

For each cluster, we identify the stars suitable for model fitting by selecting those that have converged onto the slowly rotating sequence. To achieve this, we first divide the data into temperature bins and calculate the standard deviation of the rotation periods in each bin, $\sigma_{P_{\mathrm{rot}}}$. Bins with low $\sigma_{P_{\mathrm{rot}}}$ are retained, as they contain stars that have converged onto the cluster's slowly rotating sequence. In contrast, bins with high $\sigma_{P_{\mathrm{rot}}}$ are discarded, as they represent stars that have not yet converged. For the Pleiades, stars with $P_{\mathrm{rot}} < 2$ days are excluded from this analysis, as they belong to the fast rotating sequence. We generally use a temperature bin width of $300\mathrm{K}$, except for Praesepe, which has the largest number of stars and thus requires finer binning, with a width of $100\mathrm{K}$. Each cluster is visually inspected to determine the value of $\sigma_{P_{\mathrm{rot}}}$ that maximizes the number of stars converged onto the slowly rotating sequence. For the Pleiades, the youngest cluster, where $P_{\mathrm{rot}}$ shows the greatest scatter, we set $\sigma_{P_{\mathrm{rot}}} = 1\,\mathrm{day}$. For M67, the oldest cluster, we set $\sigma_{P_{\mathrm{rot}}} = 5\,\mathrm{days}$. For the remaining clusters, $\sigma_{P_{\mathrm{rot}}} = 3\,\mathrm{days}$. This method allows us to avoid imposing a fixed initial rotation period on stars that have not yet converged and continue to exhibit variability in their rotation periods. Note that while stars in Praesepe and NGC 6811 hotter than $6000\mathrm{K}$ and $6200\mathrm{K}$, respectively, have converged onto the clusters' slowly rotating sequences, we exclude such stars from model calibration as we lack model tracks with [Fe/H]=$+0.2$ and solar metallicity with effective temperatures higher than these values.

\begin{figure*}
    \centering
    \includegraphics[width=\textwidth]{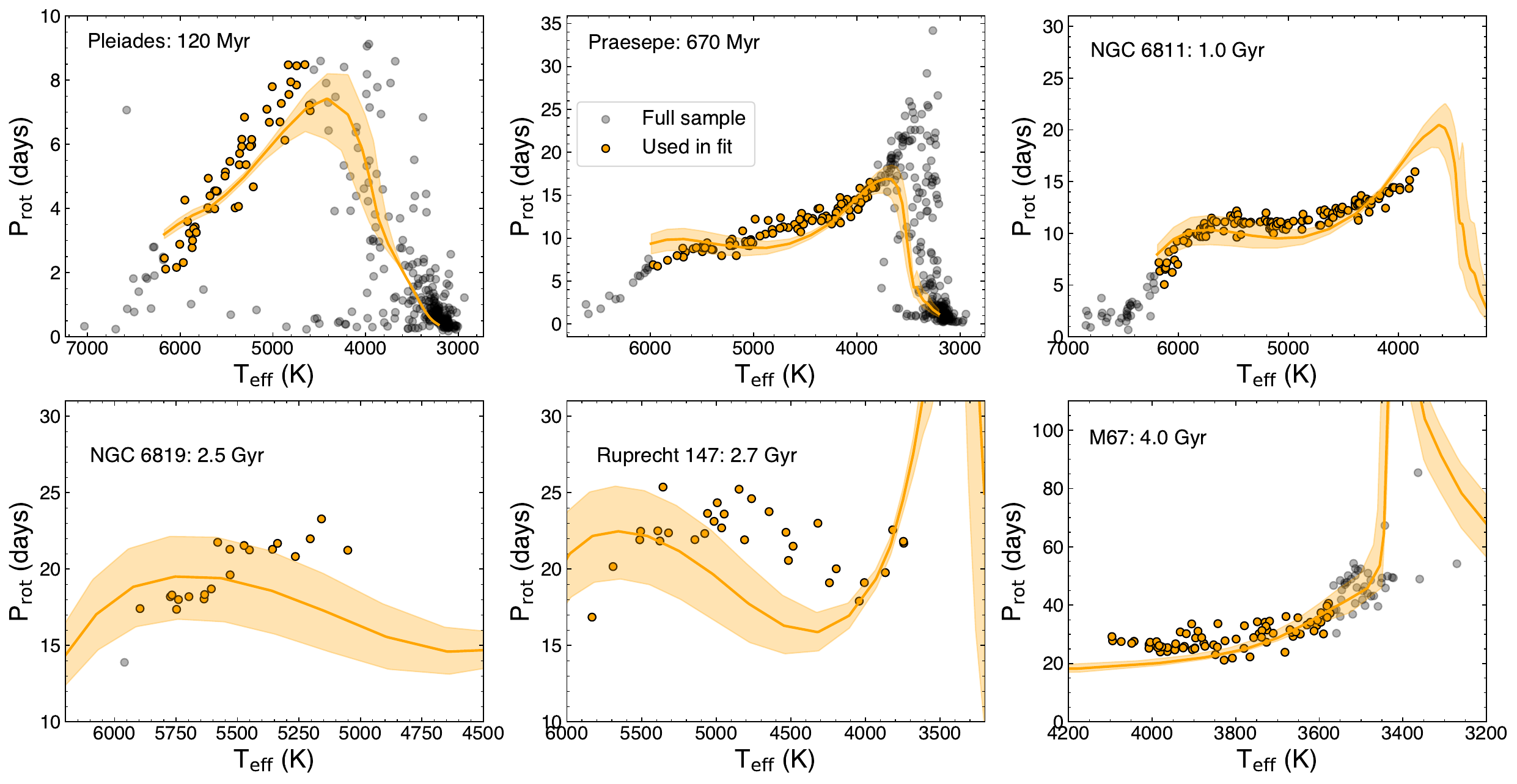}
    \abovecaptionskip=0pt 
    \caption{Gyrochrones launched from the median percentile of the Upper Sco distribution of initial rotation periods are shown as a solid orange line in each plot against observed cluster member rotational sequences. These gyrochrones are fit to the cluster's observed periods of stars that have converged onto the cluster's rotational sequence. Shaded regions around each gyrochrone account for the uncertainty in the cluster's age at which the models are being evaluated. Note that the values of $P_{\mathrm{rot}}$ and $T_{\mathrm{eff}}$ in these clusters do not span the same range.}
    \label{clusters}
\end{figure*}

We initialize non-spotted ($f_{\mathrm{spot}}=0\%$) tracks with masses between 0.18 $M_{\odot}$ and 0.65 $M_{\odot}$ with a 0.03 $M_{\odot}$ step and between 0.65 $M_{\odot}$ and 1.15 $M_{\odot}$ with a 0.05 $M_{\odot}$ step. We match the tracks metallicity to that of the clusters: solar metallicity tracks are used for the Pleiades, NGC6811, NGC6819 and M67; $[\mathrm{Fe/H}]=+0.1$ tracks are used for Ruprecht 147; $[\mathrm{Fe/H}]=+0.2$ tracks are used for Praesepe.

We launch each track using values of $P_{\mathrm{init}}$ obtained from a 1-D smoothing spline fit to the median percentile in Upper Sco (Fig.~\ref{uppersco}) at the track's stellar mass. We allow for core-envelope decoupling for all stars except fully convective ones, which are modeled as rigid rotators due to the lack of a distinct core. We interpolate $T_{\mathrm{eff}}$ and $P_{\mathrm{rot}}$ as a function of age and evaluate them at the age of the clusters. We obtain a gyrochrone by further interpolating $P_{\mathrm{rot}}$ and $T_{\mathrm{eff}}$ across the full stellar mass range.

We quantify agreement between the model and observed $P_{\mathrm{rot}}$ by computing the $\chi^2 = \sum(P_{\mathrm{obs}}-P_{\mathrm{mod}})^2/\sigma^2_{P}$, where $\sigma_{P}$ is the error on $P_{\mathrm{obs}}$ and is assumed to be 10\% of $P_{\mathrm{obs}}$ \citep{epstein2014}. We normalize the $\chi^2$ computed for each cluster by the number of data points used in the cluster's fit such that each cluster's $\chi^2$ has an equal weight in the total $\chi_{\mathrm{tot}}^2 = \chi_{\mathrm{Pleiades}}^2 + \chi_{\mathrm{Praesepe}}^2 + \chi_{\mathrm{NGC6811}}^2 + \chi_{\mathrm{NGC6819}}^2 + \chi_{\mathrm{Rup147}}^2 + \chi_{\mathrm{M67}}^2$. 

Thus, the fit includes 410 data points and three free parameters, $f_k,\, \omega_{\mathrm{sat}},\,\alpha$. The best-fitting values of the parameters ($f_K=10.8$, $\omega_{\mathrm{sat}}=3.83\times10^{-5}$, $\alpha = 9.53$; $\chi^2_{\mathrm{tot}}=25$) are obtained by minimizing $\chi^2_{\mathrm{tot}}$ through the differential evolution (\texttt{DE}) function from the Python library \texttt{yabox} \citep{yabox}. The calibrated gyrochrones evaluated at the clusters ages are shown in Fig.~\ref{clusters}.

\subsection{White Dwarf Cosmochronology}
White dwarfs are the final evolutionary stage of stars with initial masses of less than roughly 8-10 $M_{\odot}$. Because they no longer undergo nuclear fusion in their cores, their evolution consists of a cooling phase dominated by the leaking of residual thermal heat from the non-degenerate ions in the electron-degenerate core. The key idea behind using WDs as cosmochronometers is that their effective temperature and mass map uniquely onto a single cooling age. The effective temperature and surface gravity of the WD, which yield its mass, can be derived from either spectroscopy or photometry coupled with model atmospheres. Once the cooling age has been determined using the WD atmospheric parameters, the next step is to estimate its progenitor MS and post-MS lifetimes. This is done by using IFMRs (e.g., \citealt{cummings2018}) to correlate the final WD mass to initial ZAMS masses, from which the progenitor lifetimes are estimated. The total age of the WD is given by the sum of the cooling age and progenitor MS and post-MS lifetimes. 

For WDs with available spectral classification (99 out of a total of 185 WDs), we adopt the appropriate cooling sequences and atmospheric models. If no spectral information is available, we assume a DA (hydrogen-rich) spectral type. The assumption that a non-DA WD is a DA can introduce a systematic mass error of $10-15$\% \citep{giammichele2012} however, due to the lack of spectral information for these WDs, this is the simplest assumption that we can adopt. Spectral types are listed in Table~\ref{samp}.

Due to the lack of spectroscopic observations for all the WDs in the sample, we use spectral energy distribution (SED) fitting of the mean fluxes in different bands from all-sky surveys including Gaia, the Sloan Digital Sky Survey (SDSS), the Panoramic Survey Telescope and Rapid Response System (PanSTARRS) and the SkyMapper Southern Survey. We compute total ages of the WDs following the methods outlined in \cite{heintz2022}, which we summarize below for completeness. 

\subsubsection{Fitting Routine}\label{sec:fitting}

For DA WDs and WDs with no determined spectral type, we convert model DA white dwarf spectra, spanning effective temperatures of 3000 K to $40{,}000$\,K and surface gravities of 6.25 to 9.5 dex, from \cite{koester2010} to synthetic fluxes by using the sensitivity of each band-pass and the appropriate AB magnitude zeropoints. Following \cite{vincent2024} and \citet{heintz2024}, we use pure He models from \cite{cukanovaite2021} for all DBs, all DQs, and DCs and DZs above $11{,}000$~K and mixed model atmospheres with H/He ratios of $10^{-5}$ from \cite{cukanovaite2021} for all DBAs and DCs and DZs between $5500$~K and $11{,}000$~K. For DCs and DZs below $5500$~K, the same DA models discussed above are used. The observed magnitudes are also converted to absolute fluxes at 10\,pc using AB zeropoints and the weighted mean parallax of the binary from Gaia. For SDSS $u$ and $z$, the magnitudes are shifted $0.04$ and $0.02$ mag, respectively, to account for the shift relative to the AB mag system \citep{abazajian04}. The weighted mean parallax of the binary system is dominated by the brighter MS star and is on average six times more precise than the individual WD parallax which in turn allows for a more precise age determination. The observed fluxes are de-reddened using extinction values from \cite{fusillo2021}, which are obtained using the 3D extinction maps from \cite{lallement2022}. These observed de-reddened fluxes are related to the model synthetic fluxes through the radius of the WD
\citep{bergeron2019} through the following relation
\begin{equation}
F_{X} = \frac{R^2}{(3.08568\times10^{19} \mathrm{cm})^2} F_{X, \mathrm{mod}}(\mathrm{T_{eff}, log\ g})
\end{equation}
where $F_{X}$ is the observed flux at 10\,pc in bandpass $X$, $R$ is the radius of the WD in cm, and $F_{X, mod}$ is the synthetic flux in bandpass $X$ that is a function of effective temperature and surface gravity. The denominator is 10 pc in units of cm.

We use a Markov Chain Monte Carlo approach and make use of the python package \texttt{emcee} \citep{emcee} to get best-fit temperatures and radii, which are represented by the 50$^\mathrm{th}$ percentiles of the MCMC posterior distributions. These are converted to surface gravities, masses, and cooling ages using the ``thick” cooling models from \cite{bedard2020} for the WDs with hydrogen atmospheres, which assume that the WD inherits a thick hydrogen layer from its progenitor and thus retains its DA spectral type throughout its life.  For the mixed and pure He atmosphere fits, the ``thin" cooling models from \cite{bedard2020} are used. We use flat priors for the temperatures and surface gravities that cover the full range of the models. A lower limit on the magnitude uncertainties of 0.03\,mag is set to account for systematics in the conversion of magnitudes to average fluxes. We also impose a lower limit on the uncertainty on the surface gravities of 0.03 dex and a lower uncertainty of 1.2\% on the effective temperatures to account for any unknown systematics in the models (e.g., \citealt{liebert2005}).

\subsubsection{Photometric Cleaning}
There is often a large luminosity contrast between the WD and the MS star in the binary, therefore an added measure of cleaning of the photometry is needed to obtain reliable parameters. We first remove photometry that is flagged for several issues in SDSS, PanSTARRS, and SkyMapper. We remove photometry from SDSS with \texttt{EDGE}, \texttt{PEAKCENTER}, \texttt{SATUR}, and \texttt{NOTCHECKED} flags. We only use photometry from PanSTARRS with rank detections of 0 or 1. We also remove any photometry from SkyMapper that has any raised flags.

Going beyond our method in \citet{heintz2022}, we also systematically remove photometric SED points that are not consistent with the Gaia fluxes in an effort to remove WD photometry that is contaminated by the MS companion. To do this, after running the fitting routine described in Section~\ref{sec:fitting}, we then fit a line to the residuals of the resulting SED fit to search for any incorrect temperature estimates due to contaminated photometry. We take the percent difference between the linear fit and the residuals of the SED fit to find photometric bands that are inconsistent with the Gaia fluxes. Any photometric bands that are more than $3\sigma$ away from the largest percent difference between the Gaia residuals and the linear fit are removed, where $\sigma$ is the uncertainty for the individual photometric band. We then repeat the process with these photometric bands removed and iterate until a consistent list of suspect photometric bands are determined. Then, a final SED fit is performed with these bands removed. This process mainly removes redder photometry that has been contaminated by the MS companion. SDSS $u$-band is not subjected to this stage of photometric cleaning since it can be a strong indicator of whether the WD is a DA or non-DA due to the presence of the Balmer jump in DA WDs. The larger residual of SDSS $u$ can be indicative of a non-DA and not because the photometry is suspect (e.g. \citealt{bergeron2019}). We find that $~30$ WDs show anomalously discrepant SDSS $u$-band photometry that suggests there may be some non-DA in the sample. However, since $>65$\% of WDs in the Gaia magnitude-limited samples are DAs \citep{kleinan2013}, adopting a DA model is a reasonable assumption.

Moreover, \citet{heintz2022} found that, when assuming DA spectral types for all the WD in their sample, the ages are good to 25\% and provided error inflation factors to account for inaccuracies in the WD ages, including the assumption of an incorrect spectral type.
\subsubsection{Progenitor Lifetimes}
To get the progenitor lifetimes of the WDs, we use an IFMR from \citet{heintz2022} which uses a theoretically motivated shape to the IFMR from \citet{fields2016}, fit to WDs in solar metallicity clusters \citep{cummings2015,cummings2016}, in conjunction with the stellar evolutionary tracks from Modules for Experiments in Stellar Astrophysics (MESA, \citealt{paxton2011}, \citealt{paxton2013}, \citealt{paxton2015}). The errors on these values are determined by using the $1\sigma$ uncertainties on the WD mass to determine an upper and lower MS mass. The difference between the central value and the upper and lower MS mass are quoted as upper and lower errors, respectively. The same process is carried out for the progenitor lifetimes as well.

\subsubsection{Precision of Total Ages}\label{precision}
The total age of the WDs in the sample is primarily determined by their mass, and therefore uncertainties in the WD mass have a significant impact on the accuracy of the estimated total ages. \cite{heintz2022} found that the total ages derived from WDs with $M<0.63$\,$M_{\odot}$ become very noisy. Moreover, accurately determining the ages of low-mass WDs poses a challenge due to the lack of well-defined constraints at the lower end of the IFMR. WDs with masses below $0.575$\,$M_{\odot}$ may not provide reliable age estimates since they might not have formed through the evolution of a single star. While their cooling ages can offer a minimum estimate of their total age, it is essential to consider the possibility that their low mass is a result of contamination in the photometry, especially if they are formed through binary interactions. 

Thus, we adopt the following: we ignore total ages obtained from WDs with $M<0.575$\,$M_{\odot}$; for WDs with $M>0.575$\,$M_{\odot}$, we adopt the total ages computed as the sum of the cooling age and progenitor lifetimes. To avoid contamination from the MS companion, we filter sources with a Gaia BP-RP corrected excess factor $<0.1$ \citep{riello2021}. Because formal age uncertainties are often underestimated for higher-mass WDs, we inflate the age uncertainties by a factor computed following the comparison to wide WD+WD described in \cite{heintz2022}.

\begin{figure}
    \centering
    \includegraphics[width=0.47\textwidth]{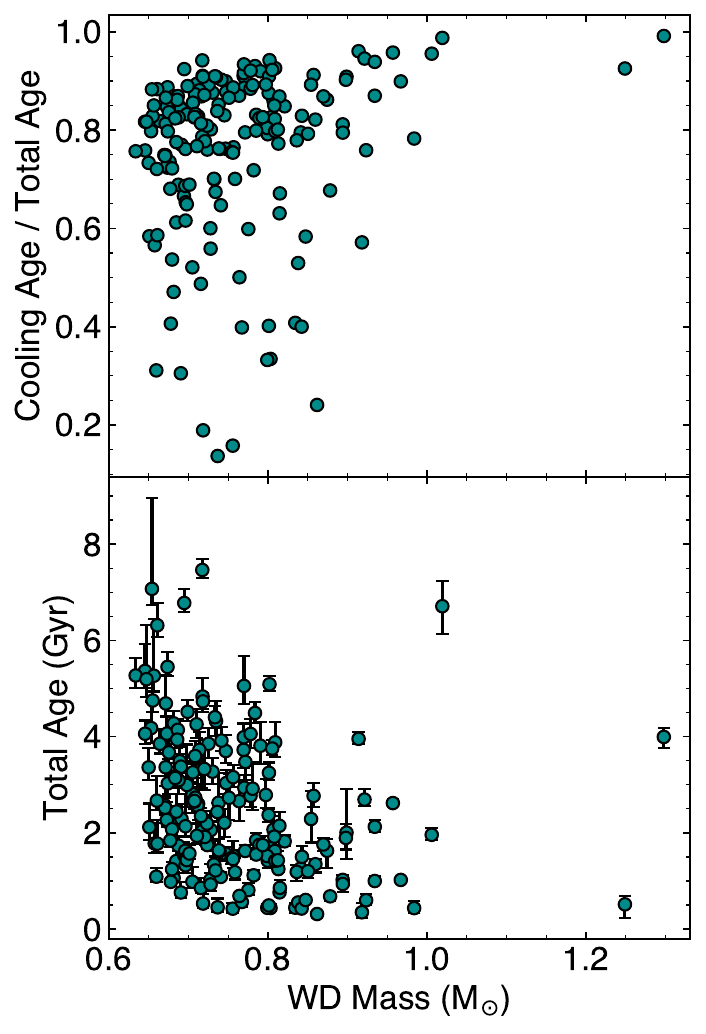}
    \caption{Top: Contribution of the cooling age to the total age of the WD as a function of mass. Bottom: WD total age as a function of mass; the uncertainties on the WD ages have been inflated by empirically determined factors \citep{heintz2022} to compensate for systematics on the total ages. }
    \label{wdages}
\end{figure}

Our final sample predominantly comprises of massive WDs (with masses greater than $0.67$\,$M_{\odot}$), in which the total ages are primarily influenced by cooling rather than the IFMR, as illustrated in Fig.~\ref{wdages}. This results in an average age uncertainty of 10\% prior to inflation and 20\% post-inflation.

\section{Sample Selection}\label{sec:data}
We construct the wide binary sample using the \cite{elbadry2021} catalog, which contains 1.3 (1.1) million binaries with a $>$ 90\% ($>$ 99\%) probability of being bound. Stars are classified as MS or WD based on their location on the Gaia color-absolute magnitude diagram (CMD). The absolute magnitude is defined as $M_{G} = G + 5 \,\mathrm{log}(\omega)-10$, where $G$ is the $G$-band mean magnitude and $\omega$ is the parallax in mas; stars with $M_{G} > 3.25 (G_{\mathrm{BP}}-G_{\mathrm{RP}}) + 9.625$ are classified as WDs; all other stars with measured $G_{\mathrm{BP}}-G_{\mathrm{RP}}$ are classified as MS stars \citep{elbadry2018}. We only select systems containing a WD and an MS star and find $22{,}563$ such systems. 

\begin{figure*}
    \centering\includegraphics[width=\textwidth]{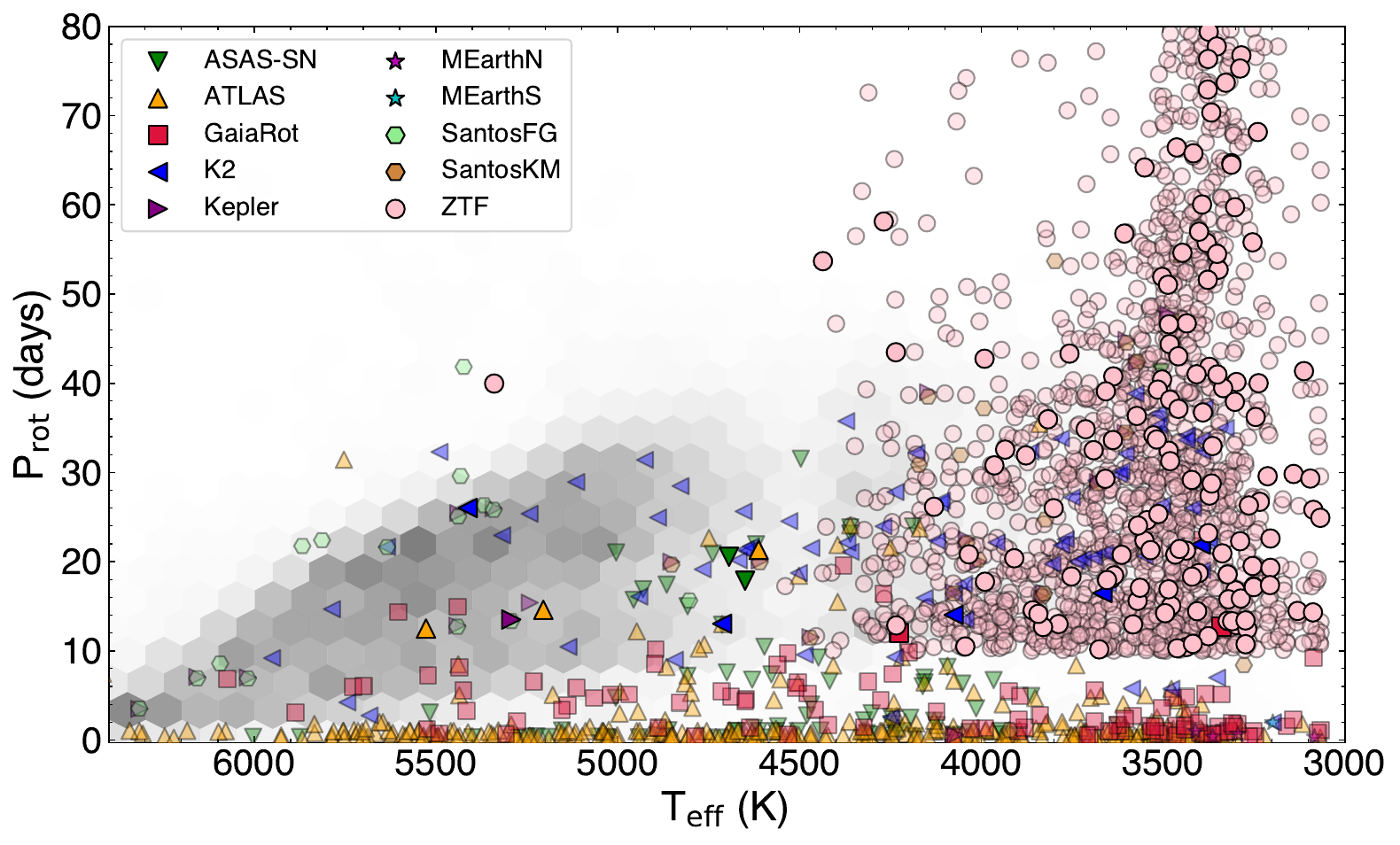}
    \caption{MS stars in the full wide WD+MS sample in a T$_{\mathrm{eff}}-$P$_{\mathrm{rot}}$ space. 1985 rotation periods are retrieved from ZTF; 384 periods are from ATLAS; 102 periods are from Gaia DR3; K2 provides 82 periods; 78 periods are from ASAS-SN; 23 periods are retrieved from Kepler; 15 periods are from \cite{santos2021}; 20 periods are from \cite{santos2020}; 9 periods are from MEarth North and 3 periods are from MEarth South. The most opaque markers represent the MS stars in the WD + MS binaries that made it into the final selection (185 systems) described in Sec.~\ref{sec:data}. The McQuillan Kepler field \citep{mcquillan2014} of 34,030 MS stars below $6500\,\mathrm{K}$ is shown in the background, together with the additional detection of $15{,}640$ M and K-type Kepler stars by \cite{santos2019}.}
    \label{catalogs}
\end{figure*}

\begin{figure*}[ht]
    \centering
    \includegraphics[width=\textwidth]{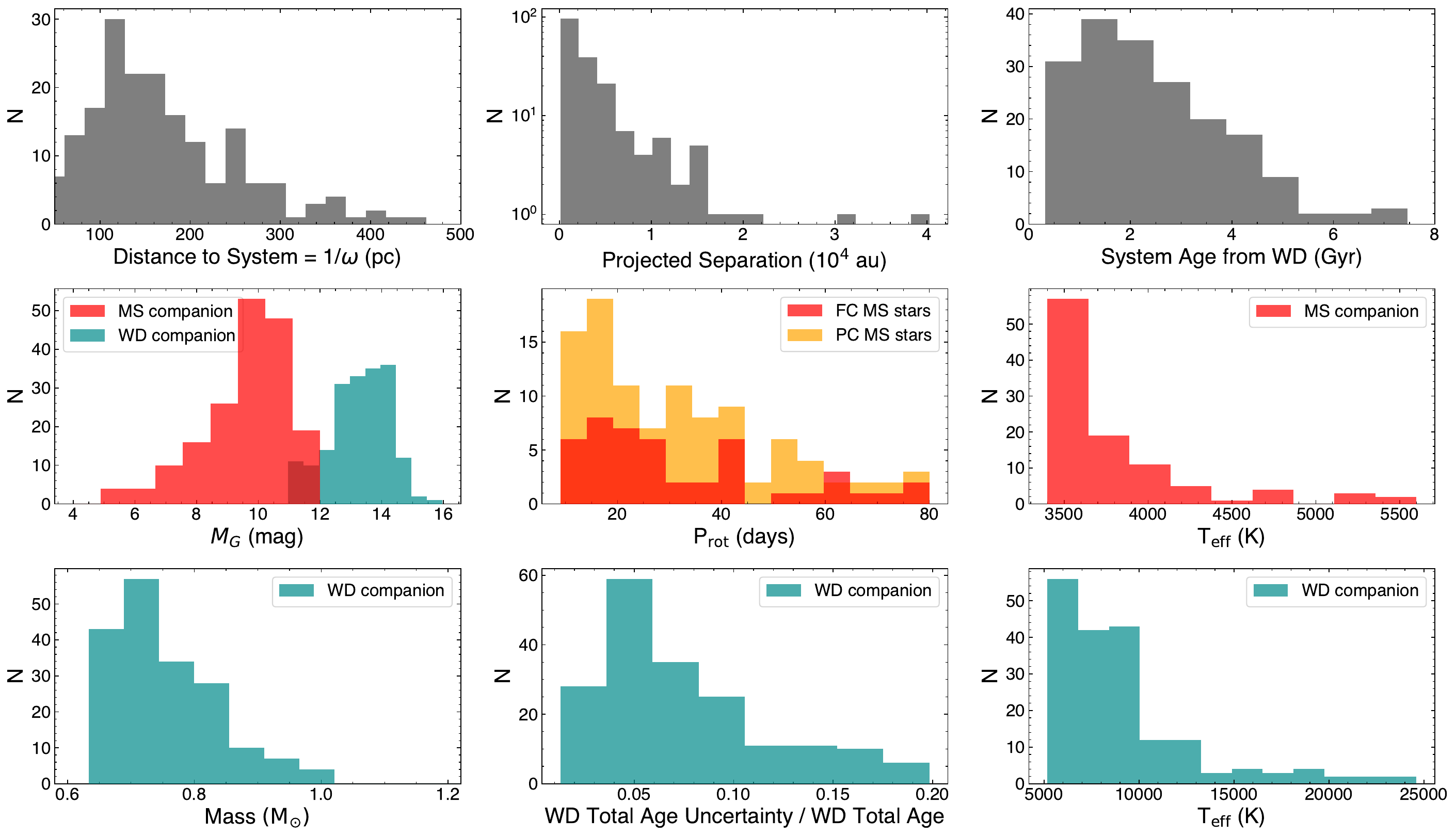}
    \caption{Properties of the 185 binaries with precise WD ages in the sample. We distinguish fully convective (FC) MS stars from partially convective (PC) MS stars based on their absolute Gaia magnitude, $M_G$, and Gaia BP-RP color measurements using Jao’s gap \citep{jao2018}.}
    \label{properties}
\end{figure*}
\begin{figure}
    \centering
    \includegraphics[width=0.47\textwidth]{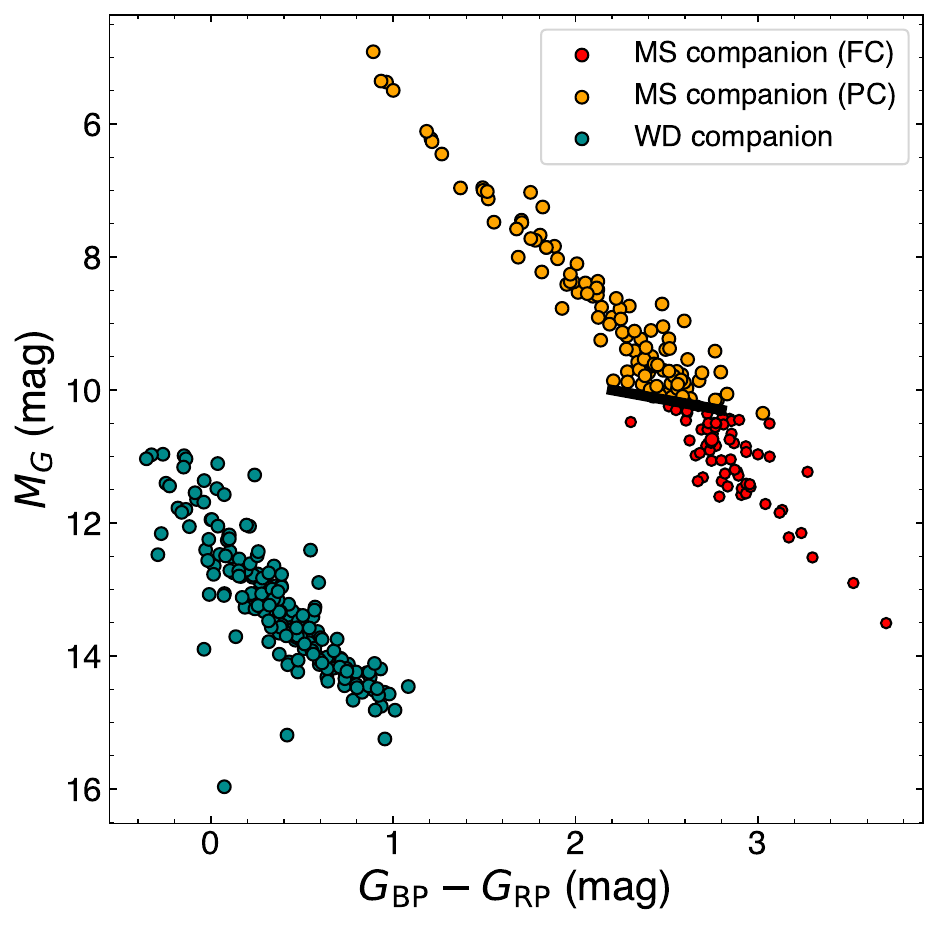}
    \caption{Gaia color-magnitude diagram showing the sample of 185 WD + MS binaries with the most precise WD total ages (average uncertainty $< 10\%$). The \citet{jao2018} gap is shown as a solid black line. }
    \label{cmd}
\end{figure}

We search for rotation periods of these MS stars in several rotation surveys, including the Asteroid Terrestrial-impact Last Alert System (ATLAS) variable stars database \citep{atlas}, the All-Sky Automated Survey for SuperNovae  (ASAS-SN) variable star database \citep{shappee2014, asassn}, the Calar Alto high-Resolution search for M dwarfs with Exoearths with Near-infrared and optical Echelle Spectrographs (CARMENES) catalog \citep{carmenes}, the Gaia third Data Release (DR3, \cite{gaiadr3}), the Hungarian-made Automated Telescope Network (HATNet) Exoplanet survey \citep{hatnet}, the Kilodegree Extremely Little Telescope (KELT) database \citep{kelt}, the Kepler \citep{Kepler, santos2020, santos2021} space mission, the K2 space mission \citep{K2}, the MEarth Observatory \citep{mearthnorth,mearthsouth} and the Zwicky Transient Facility \citep[ZTF,][]{ztf,lu2022}. We find 5005 binaries that feature an MS star with a measured rotation period, with approximately 85\% of the rotation periods sourced from the ZTF catalog. \cite{lu2022} found that nearly 50\% of stars with ZTF periods $<$ 10 days are likely to be incorrect, therefore we exclude all binaries with such ZTF fast rotators. This reduces the number of WD + MS with measured rotation periods to 2701. 

We cross-match these MS stars with various spectroscopic catalogs, including the Galactic Archaeology with HERMES (GALAH) survey \citep{galah}, the Large Sky Area Multi-Object Fibre Spectroscopic Telescope (LAMOST) \citep{lamost}, the Apache Point Observatory Galactic Evolution Experiment (APOGEE) survey \citep{apogee}, and Gaia DR3 \citep{gaiaspec}. We retrieve spectroscopic properties for 430 MS stars. We use spectroscopic data when available, but do not require spectroscopy to be included in the sample.

We obtain WD ages for these 2701 binaries. We require the final sample to meet the following criteria: $P_{\mathrm{rot}}>10$ days to avoid contamination by tidally synchronized binaries \citep{simonian2019}; WD mass $>0.575$\,$M_{\odot}$, since the total ages of the lowest-mass WDs are too uncertain \citep{heintz2022}; Gaia BP-RP excess factor $< 0.1$ (see Sec.~\ref{precision} for more details); chance alignment factor calculated in \cite{elbadry2021} \texttt{R\_chance\_align} $<0.1$ to remove binary pairs with high likelihood of being chance alignments; average of the uninflated low and upper uncertainties less than 20\%. The final sample contains 185 wide binaries with precise WD ages, which we report in Table \ref{samp} in the Appendix. Among these, only 55 binaries have separations less than 1000 au. Furthermore, just 14 of these systems have separations less than 500 au, and only one has a projected separation under 200 au. At these distances, it is unlikely that wind Roche-lobe overflow has influenced the MS companion \citep{willems2004,rebassa2013}, therefore the properties of the MS stars in our sample should resemble those of single MS stars since they have had no influence from their white-dwarf companions.

The distribution of rotation periods across the different catalogs used to create the sample is presented in Fig. \ref{catalogs}. We correct G$_\mathrm{BP}$-G$_{\mathrm{RP}}$ colors for extinction as in \cite{curtis2020} and use them to compute the effective temperatures through the polynomial fit to the empirical color-temperature relation in \cite{curtis2020}. Their color-temperature relation was constructed using nearby benchmark stars, including a sample of low mass stars with 3056\,K~$<$~$T_{\mathrm{eff}}$~$<$~4131\,K and $-0.54$\,dex~$<~\mathrm{[Fe/H]}~<$~+0.53\,dex from \cite{mann2015}. The basic properties and CMD of the sample are shown in Fig.~\ref{properties} and \ref{cmd}, respectively.

\section{Results and Discussion}\label{sec:results}
\subsection{Model Assessment}
In general, our gyrochrones reasonably match the rotation sequences observed in the clusters, with a few exceptions. For instance, for K-dwarfs with effective temperatures between $4250\,\mathrm{K}$ and $5000\,\mathrm{K}$ in Praesepe, our models predict rotation periods that are a few days shorter than those observed. Praesepe is a metal-rich cluster \citep{dorazi2020}. We note that there are no super-solar metallicity atmospheric tables available from \cite{Allard1997}, therefore we have approximated the metal rich case with the solar atmosphere. However, we do not get very different best-fit parameters if we use solar metallicity tracks for all clusters, including Praesepe. Furthermore, the rotational sequences of NGC 6819 and Ruprecht 147, which are roughly coeval, show discrepancies with our fit models.

Between the ages of Praesepe and Ruprecht 147, core-envelope coupling is important. The fit coupling timescale in this work has a strong mass dependence, therefore, lower-mass stars, such as K-dwarfs poorly fit by our models, take an extended period before resuming their spin-down. Here we adopt a constant, rotation-rate independent coupling timescale, although we expect this timescale to change with time. Additionally, our coupling model is not tuned for clusters older than 1 Gyr \citep{spada2020}. \cite{curtis2020} also found that the \cite{spada2020} model underestimated the age for stars with $M < 0.7 M_{\odot}$. We suspect that this coupling timescale prescription is contributing to the morphology mismatch between the rotational sequences of Praesepe, NGC 6819 and Ruprecht 147 and our models. The choice of coupling timescale prescription appears to be a more significant concern than differences in atmosphere or metallicity. Exploring more nuanced prescriptions that depend on evolutionary state and rotation rate are well-motivated, but beyond the scope of this work.  

\begin{figure}
    \centering
    \includegraphics[width=0.45\textwidth]{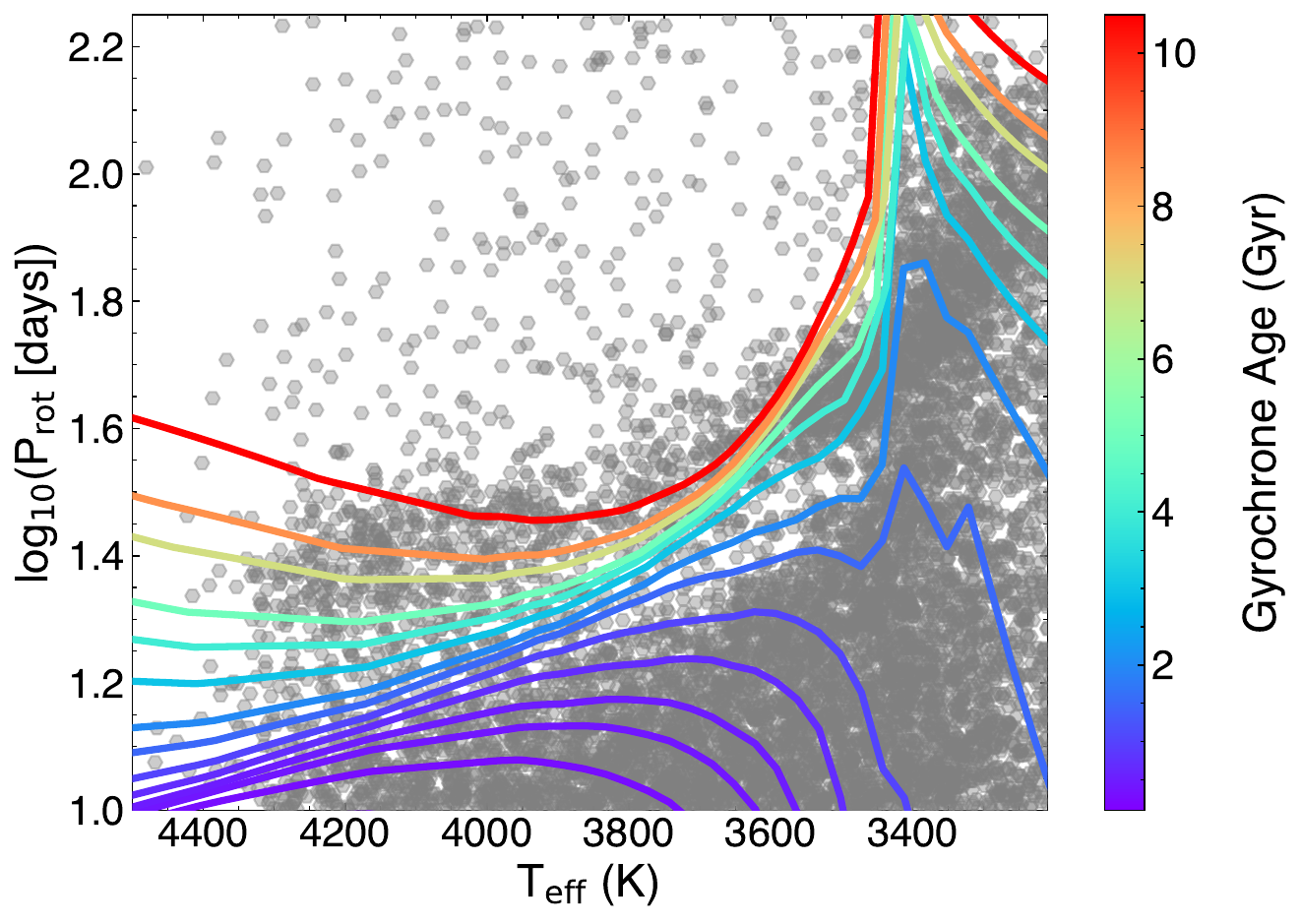}
    \caption{ZTF stars from \cite{lu2022} are plotted as grey circles. Our gyrochrones are shown as solid lines and color-coded by their age. The gyrochrones are constructed using a non-spotted solar-metallicity grid and launched from the median $P_{\mathrm{init}}$ percentile in Upper Sco. For convenience,
selected gyrochrones for a wide range of ages are provided in Appendix \ref{appc}.}
    \label{ztf_models}
\end{figure}

Another assessment of our models is presented in Fig.~\ref{ztf_models}, where we show a set of gyrochrones against the ZTF rotation period catalog from \cite{lu2022}. The ZTF distribution shows an overdensity of fully convective stars rotating slowly ($P_{\mathrm{rot}}>40$ days) past the closing of the intermediate period gap, a period dearth in the T$_{\mathrm{eff}}-\mathrm{P}_{\mathrm{rot}}$ space for low-mass stars that was first detected with Kepler by \cite{mcquillan2013}. This increase in rotation period for such stars is also predicted by gyrochrones older than 2 Gyr. We note that while the apparent agreement with the ZTF is good, the earlier \citet{mcquillan2014} sample contains stars with lower-amplitude modulation and slower rotation at these temperatures that are not fit by our gyrochrones. These stars are presumably older than 4 Gyr, and suggest that our core-envelope coupling prescription may be overly simplistic. Even if this is the case, it does not fundamentally alter the conclusions of this paper.

The best-fit parameters obtained from the clusters fit are in agreement with those from \cite{cao2023} ($f_k = 9.79 \pm 0.37$, \,$\omega_{\mathrm{sat}} = (3.466 \pm 0.085)\times10^{-5} \,\mathrm{rad\,s^{-1}}$ and $\alpha = 11.8 \pm 1.0$), who calibrated spotted models \citep{Somers2020} to the Pleiades and Praesepe. We find a higher value of $\alpha$ compared to the value reported in \cite{spada2020} ($\alpha\approx 5.6$), although their models did not include stellar spots (which alter the mass-temperature relation and therefore apparent mass-dependece of $\alpha$) and assumed a fixed initial rotation period of 8 days for all stellar masses rather than a distribution of $P_{init}$. Moreover, their two-zone model has been calibrated for stellar masses down to $\approx 0.4M_{\odot}$. 

\subsection{A spike in $P_{\mathrm{rot}}$ at the fully convective boundary}\label{rotation_wd}
Using the best-fit parameters, we construct gyrochrones to predict the rotation period of the stars in our sample at the age inferred from their WD companions. To make a direct comparison between model and observed rotation periods, we create a grid of $f_{\mathrm{spot}}=0\%$, solar-metallicity tracks for stellar masses between 0.18 and 1.15 $M_{\odot}$. The grid has a spacing of $0.01\,M_{\odot}$ for masses between $0.18\,M_{\odot}$ and $0.60\,M_{\odot}$ and $0.05\,M_{\odot}$ for masses between $0.60\,M_{\odot}$ and $1.15\,M_{\odot}$. For each stellar mass, we initiate a track with $P_{\mathrm{init}}$ values ranging from the $5^{\mathrm{th}}$ to the $95^{\mathrm{th}}$ percentiles, in steps of one percentile, of the Upper Sco period distribution (Fig.~\ref{uppersco}). For each star in our data sample, the model rotation period $\overline{P_{w}}$ is computed as the likelihood weighted average of the rotation periods in the grid 
\begin{equation}\label{pmodel}
    \overline{P_w}=\frac{\sum_{i=1}^n \Delta t_i \Delta m_i \mathcal{L}_i P_i}{\sum_{i=1}^n \Delta t_i \Delta m_i \mathcal{L}_i},
\end{equation}
where $\Delta t_i$ and $\Delta m_i$ are the time and mass increments between each point on our non-uniformly sampled model grid, respectively. $P_i$ is the $i^{\mathrm{th}}$ model rotation period in the grid; $\mathcal{L}_i$ is its corresponding likelihood, given by
\begin{equation}\label{likelihood}
\mathcal{L}_{i} = \exp\left\{-0.5 \left[ \left(\frac{T_\mathrm{{eff,obs}}-T_\mathrm{{eff,\textit{i}}}}{\sigma_{T_\mathrm{{eff,obs}}}}\right)^2 + \left(\frac{A_\mathrm{{obs}}-A_\mathrm{{\textit{i}}}}{\sigma_{A_\mathrm{{obs}}}}\right)^2 \right]\right\}.
\end{equation}
The likelihood function is computed for every grid point using its associated effective temperature $T_\mathrm{eff,\textit{i}}$ and age $A_\mathrm{\textit{i}}$. It also accounts for the uncertainty $\sigma_{A_\mathrm{{obs}}}$ on the WD age $A_\mathrm{{obs}}$, taken as the average of the inflated (as per \citealt{heintz2022}) lower and upper uncertainties, and the uncertainty $\sigma_{T_\mathrm{{eff,obs}}}$ on the effective temperature $T_\mathrm{{eff,obs}}$ of the MS star. $\sigma_{T_\mathrm{{eff,obs}}}$ is computed as the root sum of the squares of the typical temperature precision ($\pm 50$ K) and the uncertainty obtained from propagation of the Gaia extinction-corrected $G_{\mathrm{BP}}-G_{\mathrm{RP}}$ uncertainties involved in the color-$T_{\mathrm{eff}}$ relation \citep{curtis2020}.

Due to the lack of uncertainties in the observed rotation periods, we compute the likelihood weighted standard deviation $\sigma_{w}$ of the model periods to quantify the constraining power of our models on the rotation periods. $\sigma_{w}$ is defined as
\begin{equation}
\sigma_{w}=\sqrt{\frac{\sum_{\mathrm{i}=1}^{\mathrm{n}} \Delta t_i \Delta m_i \mathcal{L}_{\mathrm{i}}\left(\mathrm{P}_i-\overline{P_w}\right)^2}{\sum_{\mathrm{i}=1}^{\mathrm{n}} \Delta t_i \Delta m_i \mathcal{L}_{\mathrm{i}}}}.
\end{equation}

\begin{figure}
    \centering
    \includegraphics[width=0.42\textwidth]{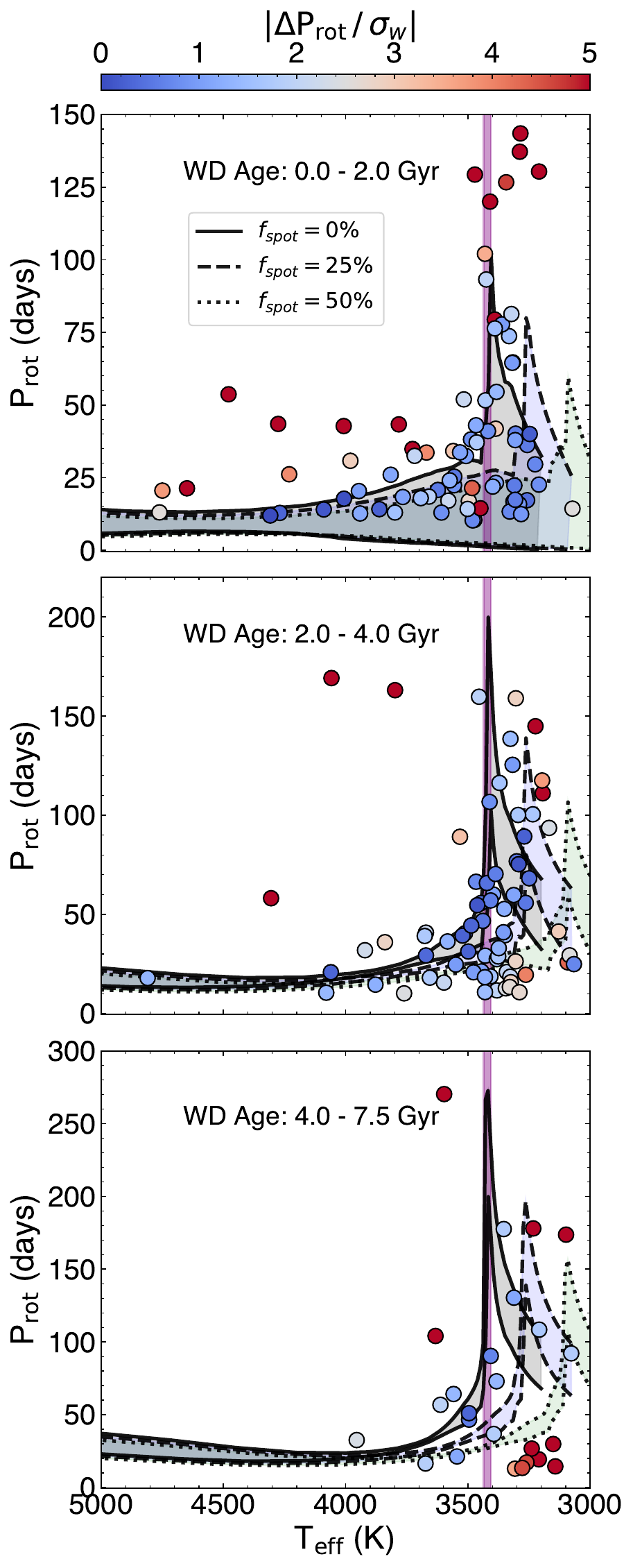}
    \caption{MS stars binned by their companion WDs ages. The grey shaded regions show the range of rotation periods spanned by $f_{\mathrm{spot}}=0$, solar-metallicity gyrochrones launched from the median percentile of Upper Sco at the WD age bounds, which are plotted as black solid line. Gyrochrones with $f_{\mathrm{spot}}=25\%$ and $f_{\mathrm{spot}}=50\%$ launched with the same $P_{\mathrm{init}}$ at the same ages are plotted as black dashed and dotted lines, respectively, and the corresponding region of allowed rotation periods is shown in light pink and green. The vertical purple band represents the FCB. Stars are color coded by $|\Delta P_{\mathrm{rot}}/\sigma_{w}|$ (see Sec.~\ref{rotation_wd}). Blue points show good agreement, while the redder points show worse agreement between the observed and the predicted rotation periods.}
    \label{compare}
\end{figure}

The lower the ratio $\abs{\Delta P_{\mathrm{rot}}/\sigma_w}$, the smaller the discrepancy between the observed and the predicted rotation periods is. The results are presented in Fig.~\ref{compare}. 65\% of the rotation periods predicted by non-spotted, solar-metallicity models are within $3\sigma_w$ from the observed rotation periods and fall in the region bounded by the gyrochrones computed at the lower and upper WD age bounds in each bin. 

At the FCB, our sample shows a rapid increase in the rotation period of MS stars with WD ages up to 7.5 Gyr. The same trend is confirmed by the gyrochrones ([Fe/H]$=+0.0$, $f_{\mathrm{spot}}=0\%$), which span periods between 30 and 100 days across a narrow temperature range ($\sim 50$ K) for ages up to 2.0 Gyr. Similarly, between 2.0 and 4.0 Gyr, the models show a sharp rise in rotation period from 50 to 200 days and up to 270 days between 4.0 and 7.5 Gyr. \textit{Thus, both the models and data suggest that, at the fully convective boundary, stars with relatively long rotation periods are not necessarily old, in contrast to the standard picture of stellar spin-down.} In addition, at this boundary, a measured rotation period cannot be uniquely associated (within reasonable observed errors in $\mathrm{T_{eff}}$ of $\simeq$50 K) to a single gyrochrone -- rather, gyrochrones spanning several billions of years all provide reasonable matches to the observed ($P_{\mathrm{rot}},\,T_{\mathrm{eff}}$) combination. This significantly inflates the age uncertainties on rotation-based ages in this $T_{\mathrm{eff}}$ range as the rotation period of a star along this vertical incline is predicted, within 1$\sigma$, by gyrochrones between 2 Gyr and 8 Gyr.

Beyond the FCB, our models return to a reasonable behavior without distinct features. This suggests the feasibility of applying gyrochronology to the coolest fully convective stars, at least from a model standpoint.

\subsection{Model description of the spike} \label{subsec:spikedesc}
Stellar interior theory predicts that as the stellar mass decreases, the convection zone (CZ) deepens in the interior of the star until the star becomes fully convective at $\sim 0.35 \,\mathrm{M}_{\odot}$. The convective overturn timescale refers to the characteristic timescale of convective motions. In this work, we compute the characteristic convective overturn timescale as the local $\tau_{cz}=H_P/v$, where $H_P$ is the pressure scale height at the base of the convective zone and $v$ is the convective velocity (from a mixing length theory of convection) one pressure scale height above the convective zone boundary. As we approach the FCB, the convective envelope gets deeper, occupying a larger part of the total stellar mass, the pressure scale height increases and the convective velocity decreases, as predicted by the mixing-length theory
\citep{vitense1958}. This leads to an increase in $\tau_{cz}$.

However, it has been shown that the behavior of models near the transition to the fully convective regime is not smooth. The \cite{vansaders2012} instability predicts that low-mass stars at the boundary undergo non-equilibrium $^3$He burning, which gives rise to a small convective core separated from the convective envelope by a thin radiative zone. As the amount of central $^3$He increases, the convective regions grow in mass and the convective envelope deepens, until they merge, leading to a fully convective episode. This process repeats until the total $^3$He concentration is high enough that the star remains fully convective.

\begin{figure}
    \centering
    \includegraphics[width=0.48\textwidth]{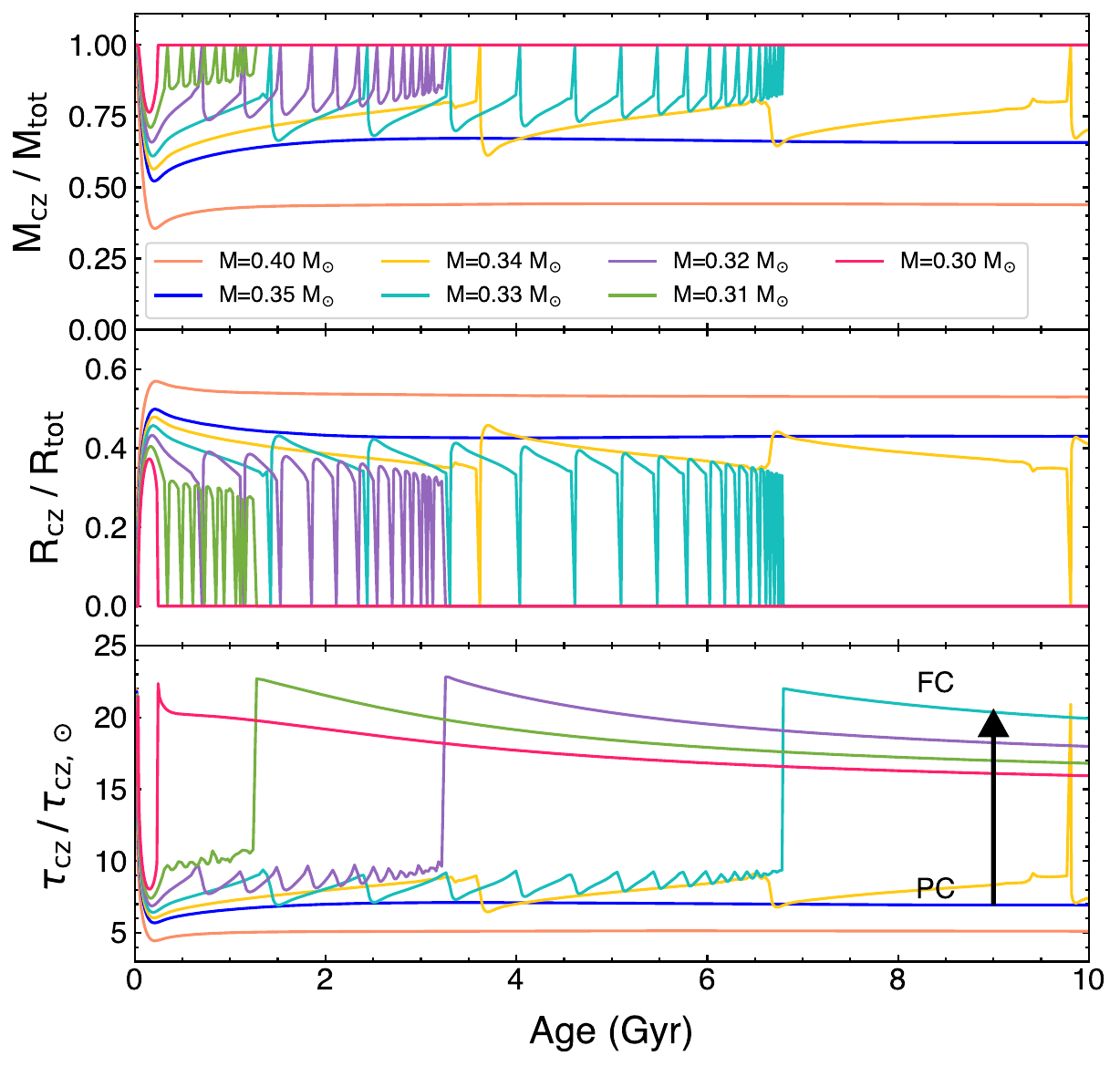}
    \caption{The sizes of the convective zone relative to the total size of the star in mass and radius coordinates as a function of time are plotted in the top and middle panels, respectively. The bottom panel shows the convective overturn timescale normalized by the solar value as a function of time. The black arrow shows the jump between partially convective (PC) and fully convective (FC) stars. In all panels, the saw-toothed curves represent stars undergoing fully convective episodes driven by non-equilibrium $^3$He burning \citep{vansaders2012}. All tracks have solar-metallicity and a 25\% spot covering fraction.}
    \label{tcz_mcz}
\end{figure}

In the models used in this work, the \cite{vansaders2012} instability occurs for masses between 0.30 $M_{\odot}$ and 0.37 $M_{\odot}$, depending on the metallicity and spot covering fraction. For instance, in a solar-metallicity, $f_{\mathrm{spot}}=25\%$ model grid, the \citet{vansaders2012} instability affects models in the $0.31-0.34$ $M_{\odot}$ range and the FCB is at $0.31$ $M_{\odot}$ (i.e. stars with $M<0.31 M_{\odot}$ \textit{never} have a radiative core). This is shown in Fig.~\ref{tcz_mcz}. In the top panel, we see that as we move from a $0.4\,M_{\odot}$ partially convective star to $0.31\,M_{\odot}$, the contribution of the mass of the convective zone to the total stellar mass increases, until the star is fully convective and $M_{cz}/M_{tot}=1$. Similarly, the middle panel shows that the base of the convective zone is eating downward in mass and deepening in the interior as we approach the FCB, which makes $\tau_{\mathrm{cz}}$ longer. Furthermore, due to fully convective episodes initiated by non-equilibrium $^3$He burning, the convective zone base of stars in the range $0.34-0.31$ $M_{\odot}$ suddenly moves from a fractional depth $R_{cz}/R_{tot}=0.4-0.3$ to the center of the star $R_{cz}=0$, which results in discontinuous jumps in $\tau_{\mathrm{cz}}$, as seen in the bottom panel. The convective overturn timescale maxima are in phase with drops in $R_{cz}/R_{tot}$ and peaks in $M_{cz}/M_{tot}$ and correspond to fully convective episodes.

We suggest that it is the rise in $\tau_{\mathrm{cz}}$ due to differences in the structure of partially and fully convective stars that causes the vertical feature in $P_{\mathrm{rot}}$ for low-mass stars older than 1 Gyr, and that its sharpness is caused by the fact that the CZ boundary does not smoothly move towards the core (as a function of mass) until the star is fully convective, but rather jumps from a partially convective configuration in stars undergoing the \cite{vansaders2012} instability. 

\begin{figure}
    \centering
    \includegraphics[width=0.45
\textwidth]{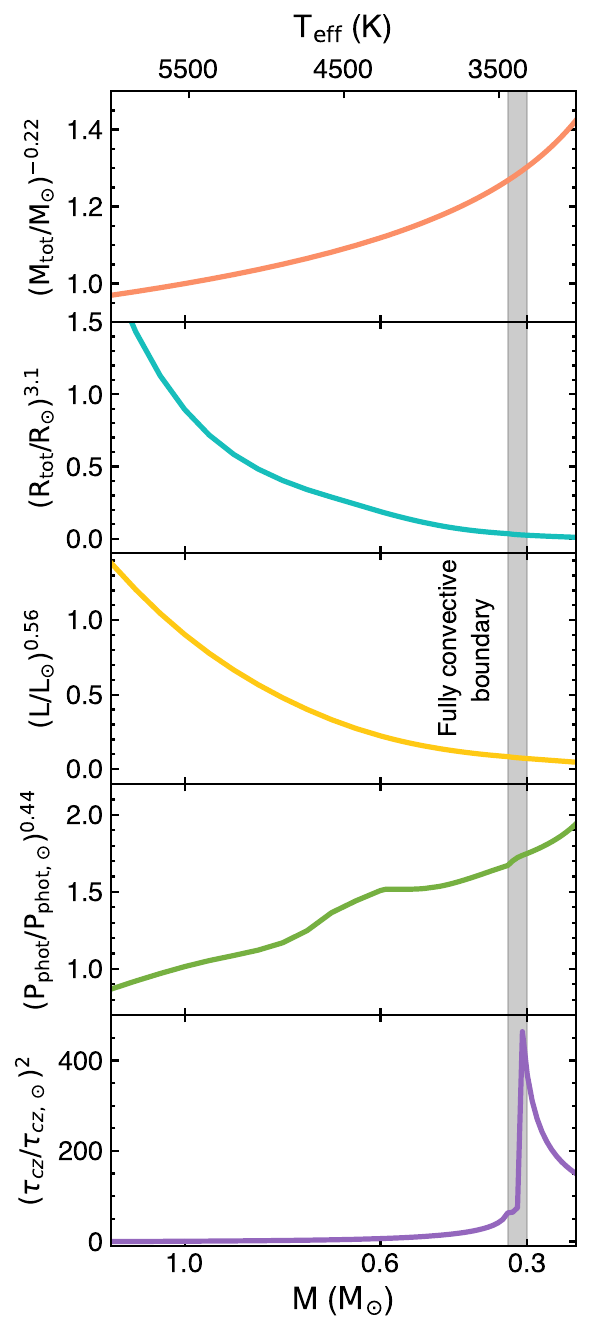}
    \caption{Contributions of the stellar mass $M$, radius $R$, luminosity $L$, photospheric pressure $P_{\mathrm{phot}}$ and convective overturn timescale $\tau_{cz}$ factors to the total $dJ/dt$ using solar-metallicity models with 25\% spot covering fraction. The FCB is at $M = 0.31 M_{\odot}$.}
    \label{braking_factors}
\end{figure}

To show how $\tau_{\mathrm{cz}}$ affects stellar spin down at the FCB, we consider the relative contribution of all terms involved in the braking law. We evaluate the weight of the factors affecting $dJ/dt$, computed as shown in Equations \ref{djdt} and \ref{km}, as a function of stellar mass at the median age of the data sample ($\sim 2$ Gyr). Since the majority of the stars in the sample are M dwarfs that slowly evolve on the main sequence and the structure is very stable after the fully convective episodes, the contribution of the structural terms in the braking law do not significantly depend on the age at which they are evaluated. The results are shown in Fig.~\ref{braking_factors}: while the stellar mass, radius, luminosity and pressure factors vary smoothly for stars with masses between $1.15 M_{\odot}$ and $0.2 M_{\odot}$, the convective overturn timescale factor changes abruptly at the FCB, between $0.33 M_{\odot}$ and $0.31 M_{\odot}$. In this narrow mass range, $\tau_{\mathrm{cz}}$ rapidly increases as we approach $0.33\,M_{\odot}$, reaches a peak at $0.31\,M_{\odot}$, and then drops modestly. No other stellar property exhibits such a distinct feature at this boundary.  

The peak in the $\tau_{\mathrm{cz}}$ curve in the bottom panel of Fig.~\ref{braking_factors} is reached by the model with the largest mass --- and thus largest $H_P$ --- that becomes fully convective, which is the $0.31\,M_{\odot}$ model in the solar-metallicity, $f_{\mathrm{spot}}=25\%$ model grid. Below this point, stars are fully convective since they never have a radiative core. However, they are also smaller in radius and mass of the CZ, therefore  $\tau_{\mathrm{cz}}$ drops. 
Figures \ref{app_fig1} and \ref{app_fig2} showing the rotational evolution of stars affected by the \citet{vansaders2012} instability at the boundary in conjunction with changes in their $\tau_{\mathrm{cz}}$ are included in Appendix \ref{sec:firstapp}.
\subsubsection{Calibrations for $\tau_{cz}$ across the fully convective boundary}
Literature sources do not agree on a single method for computing the convective overturn timescale from a stellar model, but we argue here that the sharp increase in the convective overturn timescale that drives rapid braking is a ubiquitous feature across common prescriptions for inferring the timescale. 
\begin{figure*}
    \centering
    \includegraphics[width=1.0
\textwidth]{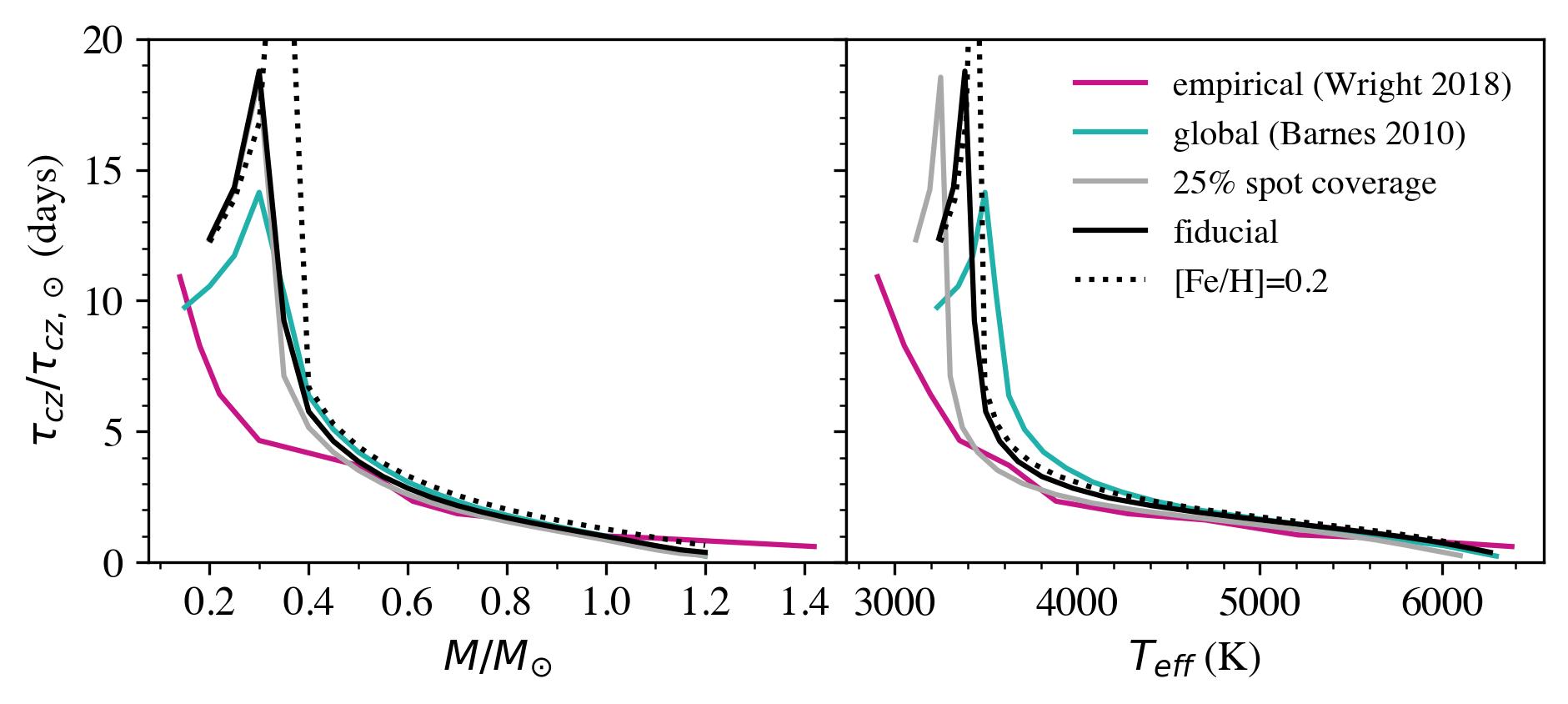}
    \caption{Comparisons of $\tau_{cz}$ values as a function of mass (left panel) and effective temperature (right panel). We show our fiducial, solar metallicity unspotted stellar models as a solid black curve. Models with identical physics but a 25\% spot covering fraction are shown in gray. The \citet{Wright2018} empirical calibration is shown in purple, and the \citet{Barnes2010} ``global'' model-based $\tau_{cz}$ are shown in turquoise. Metal-rich ($\textrm{[Fe/H]}=+0.2$) models are shown as the dotted curve.} 
    \label{lit_tau}
\end{figure*}
When using models to compute convective overturn timescales, there are two primary approaches: the ``local" prescription (used here) and a ``global" prescription, where one instead computes some suitable average $\tau_{cz}$ over the entire convection zone. Both approaches yield fundamentally the same behavior modulo a scale factor, since the deep portions of the CZ probed by the local approach are also the most heavily weighted in the global average \citep{Kim1996}. We show in Fig.~\ref{lit_tau} that our fiducial model, which uses a local approach, displays fundamentally the same behavior as that in \citet{Barnes2010}, which utilizes a global approach. Once normalized by their respective solar convective overturn timescales, both methods show the same behavior as a function of mass and a rapid increase in $\tau_{cz}$ near the FCB as the $\tau_{cz}$ computation begins to probe the structure of the near core. 

\begin{figure*}
    \centering
    \includegraphics[width=\textwidth]{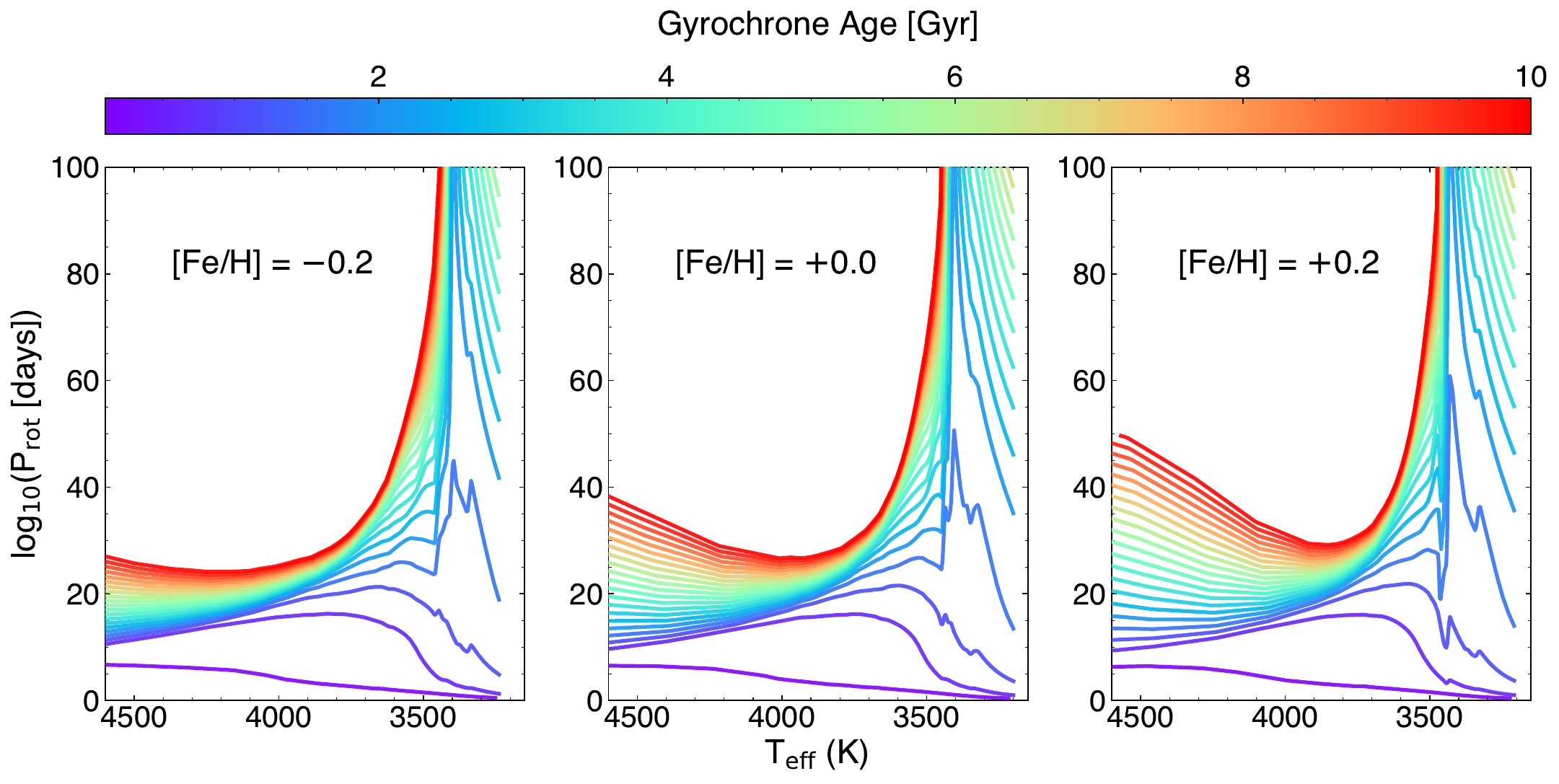}
    \caption{Gyrochrones constructed using a metal-poor (left), a solar metallicity (center) and a metal-rich (right) non-spotted model grid. We show gyrochrones for ages between 0.1 and 10 Gyr, with a step of 0.5 Gyr. While the location of the steep rise in $P_{\mathrm{rot}}$ is shifted to slightly higher masses in metal-rich stars, the feature does not vary significantly in temperature and occurs at $\approx$$3500$\,K across all model grids.}
    \label{models}
\end{figure*}

\begin{figure}
    \centering
    \includegraphics[width=0.45\textwidth]{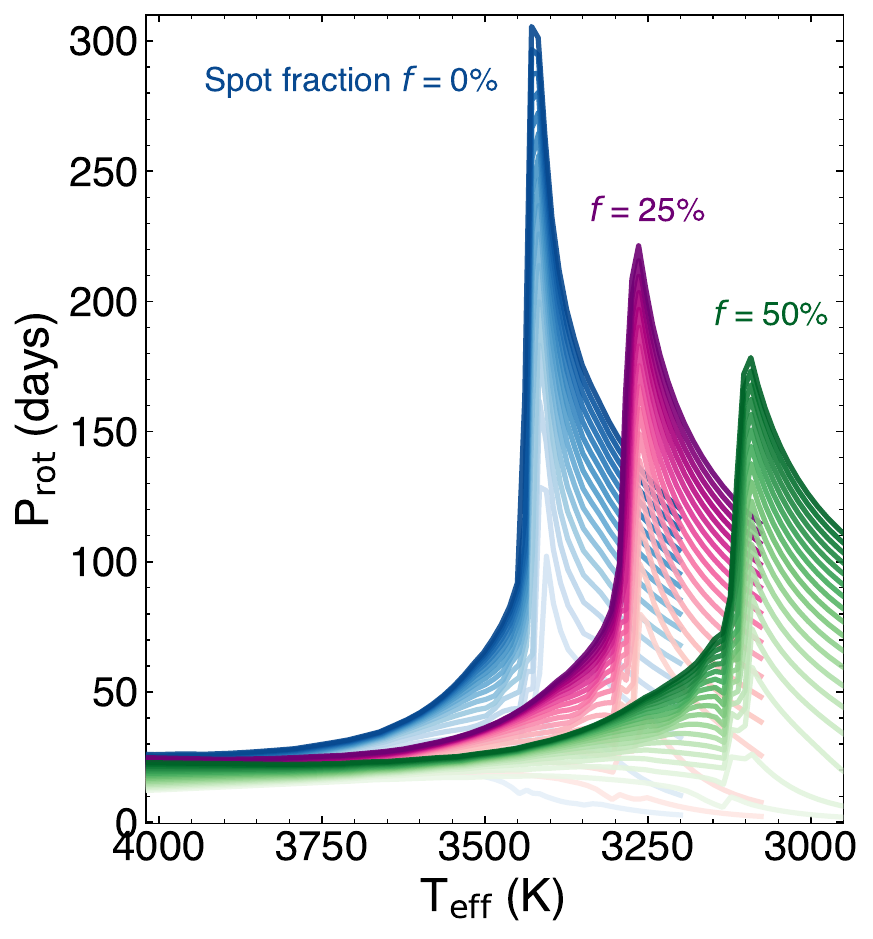}
    \caption{Solar-metallicity gyrochrones up to 10 Gyr constructed using three model grids with different spot covering fractions $f_{\mathrm{spot}}$. The sharp rise in rotation period occurs for the same 0.30-0.34 $M_{\odot}$ mass range across the three model grids, however the location of this feature varies across a $\sim$$350$\,K temperature window between the three model grids with different spot-covering fractions.}
    \label{spotted_models}
\end{figure}

While the location in mass (or temperature) of the sharp rise in $\tau_{cz}$ depends on properties like the metallicity and spot covering fraction, both produce only modest shifts in the precise location of the rise in $\tau_{cz}$, also shown in Fig.~\ref{lit_tau}. In the case of metallicity, stars are more convective at higher metallicity but fixed mass, shifting the rise in $\tau_{cz}$ and onset of full convection to slightly higher masses in metal-rich stars, although this vertical feature does not significantly move in temperature, as shown in Fig.~\ref{models}. 
Adding spots to the surface of the model --- which may be an important component in modeling young, low-mass stars \citep{cao2023} --- decreases the observed effective temperature, with only modest impacts on the structure of the deep interior \citep[see][]{Somers2015}, which shifts the onset of deep convection and large $\tau_{cz}$ values to lower effective temperatures but not significantly lower stellar masses. We find that the maximum stellar mass undergoing fully convective episodes decreases with increasing spot covering fraction: 0.35 $M_{\odot}$ for 0\% spot covering fraction; 0.34 $M_{\odot}$ for 25\% spot covering fraction; 0.32 $M_{\odot}$ for 50\% spot covering fraction. Furthermore, a higher spot covering fraction leads to a lower peak rotation period, which is a direct consequence of the onset of the $^3$He instability shifting to lower masses and therefore lower convective overturn timescales, as shown in Fig.~\ref{spotted_models}. While the precise location of the steep rise in $\tau_{cz}$ depends on the model physics, the existence of a steep rise does not.  

Finally, attempts to develop purely empirical calibrations of $\tau_{cz}$ also predict an increase in overturn timescale across the FCB. \citet{Wright2018} made the assumption that fully convective M dwarfs obeyed the same Ro-activity relation as partially convective stars, and then found the values of $\tau_{cz}$ as a function of color (mass) that minimized scatter in the Ro$-$X-ray luminosity relation. Although the resulting relation does not trace the model predictions exactly (Fig.~\ref{lit_tau}), it does indicate a reasonably steeply rising $\tau_{cz}$ across the FCB. 

\subsection{A Few Complications}
We find that 25\% of the MS stars shown in Fig.~\ref{compare} have $|\Delta P_{\mathrm{rot}}/\sigma_w| > 3$, i.e. their rotation rate as predicted by our spin-down model is at odds with the apparent system age given by the WD. This percentage goes down to 12\% if we exclude MS stars with an effective temperature within 200 K from the FCB. We argue that these discrepancies may have multiple potential sources. 
\subsubsection{Stellar Spots}
The model rotation periods of M dwarfs in our sample shown in Fig.~\ref{compare} were obtained using a model grid with solar metallicity and $f_{\mathrm{spot}}=0\%$. However, by adopting a model grid with a non-zero spot covering fraction, we can extend the region probed by our models to cooler temperatures, as shown in Fig.~\ref{spotted_models}. For instance, Fig.~\ref{compare} shows that some of the MS stars cooler than 3200 K that fall outside of the grey region bounded by the $f_{\mathrm{spot}}=0\%$ gyrochrones, are found within the region bounded by the $f_{\mathrm{spot}}=25\%$ and $f_{\mathrm{spot}}=50\%$ gyrochrones. Therefore, knowing the $f_{\mathrm{spot}}$ of these stars would be helpful to choose the most appropriate tracks to model their spin-down and improve the comparison between the predicted and observed rotation periods. 

\subsubsection{Metallicity}
Metallicity may also be responsible for some of the discrepancies observed in our data. In the absence of spectroscopic data, we have assumed a solar-metallicity for our sample. Modern braking laws, including the \cite{vansaders2013} prescription used in this work, suggest that metallicity can have a strong impact on the rotational evolution of low-mass stars \citep{claytor2020, amard2020}. The convective overturn timescale is a direct consequence of the stellar structure and therefore is affected by the chemical composition of the star. Stars with a high abundance of elements heavier than He have a higher opacity, which steepens radiative temperature gradients, leading to deeper convective envelopes \citep{vansaders2012,amard2019}, higher pressure scale height and, therefore, a longer $\tau_{\mathrm{cz}}$ and more efficient braking. We find that 57\% of the stars with $|\Delta P_{\mathrm{rot}}/\sigma_w|>3$ have a positive $\Delta P_{\mathrm{rot}}$ value, i.e. the MS stars observed rotation periods are longer than the model periods. This percentage does not significantly change when not accounting for stars within 200 K from the FCB (44\%).

\begin{figure}
    \centering
    \includegraphics[width=0.47\textwidth]{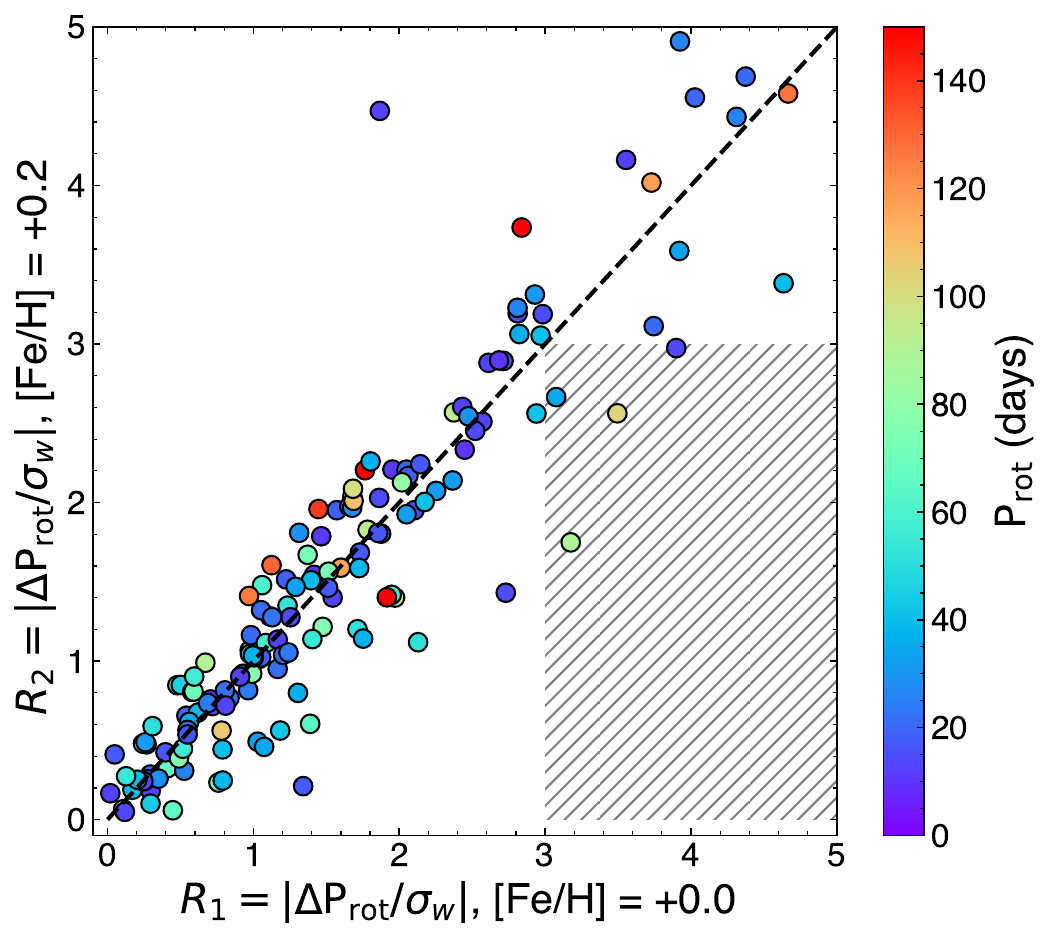}
    \caption{Comparison between $|\Delta\mathrm{P}_{\mathrm{rot}}/\sigma_w|$ computed for the MS stars in our sample using a [Fe/H] = +0.0 and a [Fe/H] = +0.2 model grids. Stars are color-coded by their observed rotation period. A 1:1 line is plotted as a black dashed line. The hatched region in the bottom right corner highlights stars for which $|\Delta\mathrm{P}_{\mathrm{rot}}/\sigma_w|$ decreases to $3\sigma_w$ or lower if a higher metallicity model grid is adopted instead of a solar metallicity one. Only a handful stars show improved values of $|\Delta\mathrm{P}_{\mathrm{rot}}/\sigma_w|$ when using an [Fe/H] = +0.2 model grid.}
    \label{met_comparison}
\end{figure}

Since higher metallicity leads to stronger braking, by adopting a metal-rich model grid we expect to recover longer model rotation period that may provide a better match to the measured rotation periods. Fig.~\ref{met_comparison} shows the difference between computing $|\Delta P_{\mathrm{rot}}/\sigma_w|$ using a solar-metallicity model grid, like the one used in Fig.~\ref{compare}, versus a higher-metallicity model grid. 
By adopting an [Fe/H] = +0.2 grid, we obtain an improvement in the predictions of rotation periods for only a handful of stars with $|\Delta P_{\mathrm{rot}}/\sigma_w|>3$ when computed with a solar-metallicity grid, as shown in Fig.~\ref{met_comparison}. Nevertheless, marginalizing over metallicity when constructing a model grid may improve the predictions of the rotation periods for systems with known metallicities. 

Of the full sample, only 28 MS stars have measured metallicities. We find no evident trend with metallicity in those stars where measurements are available.

\subsubsection{WD Age Resets in Triple Systems}\label{triples}
We have neglected the possibility of triple systems where the inner WD binary merges and resets the apparent system age. Modeling the evolution of single star and binary populations has shown that the age of a merger remnant can be underestimated by a factor of three to five if single star evolution is assumed for a WD \citep{temmink2020}. The same study found that WDs from binary mergers make up about 10-30\% of all observable single WDs and 30-50\% of massive ($>$0.9\,$M_{\odot}$) WDs. Similarly, \cite{heintz2022} estimated that $21-36$\% of WD+WD pairs likely started as a triple system. These values are consistent with the fraction of binaries in our sample ($\approx$30\%) for which the models are unable to predict the rotation periods of the MS companions. 

\begin{figure}
    \centering
    \includegraphics[width=0.47\textwidth]{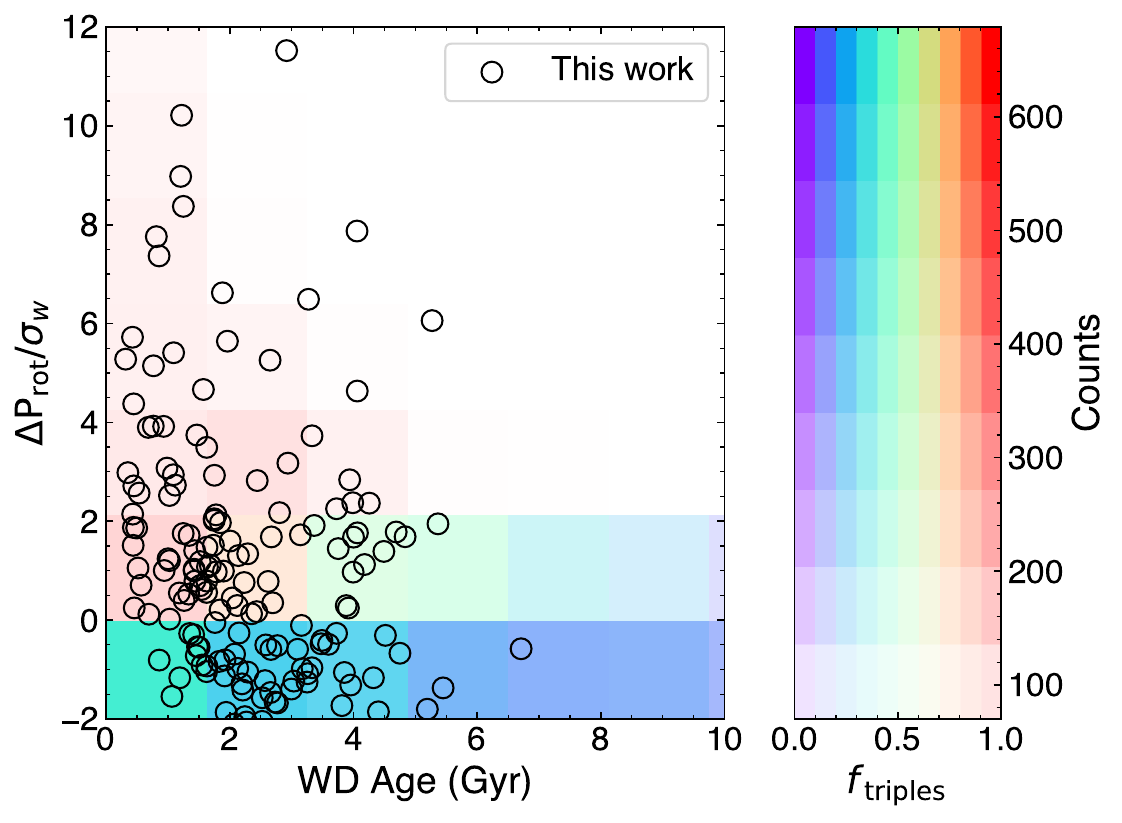}
    \caption{In the background, a 2D histogram shows the distribution of $\Delta P_{\mathrm{rot}}/\sigma_w$ relative to the ages of WDs in a sample drawn from a synthetic population. Each bin is color-coded based on the fraction of triples (i.e. the number of WDs in the bin that have had their age reset), while the transparency reflects the total count of WDs in the bin. Warmer colors signify a higher proportion of WDs merger products, and greater opacity indicates a larger overall WD count in the bin. The black circles represent the distribution of the observed sample. At young WDs ages, the observed sample exhibits an increased occurrence of systems with $\Delta P_{\mathrm{rot}}/\sigma_w>2$ in a region prone to triple contamination, as indicated by the synthetic population. }
    \label{sim}
\end{figure}
To test whether the discrepant systems in our sample may be reasonably accounted for by binary mergers, we create a synthetic population of 5000 stars with masses between 0.18 and 1.15 $M_{\odot}$ using the Chabrier functional form of the Initial Mass Function \citep{chabrier2003}. For each star, we set the age to a random number between 0 and its MS lifetime (i.e, $t_{MS}=t_{\odot}(M/M_{\odot})^{-2.5}$, $t_{\odot}\approx10$ Gyr, which is consistent with the models turn-off points). We find the point in our model grid that is closest to each age-mass combination and select the corresponding rotation period and effective temperature as the $P_{\mathrm{rot}}$ and $T_{\mathrm{eff}}$ of the stars in this synthetic population. Then, we randomly select 35\% of the stars and reset their ages to a number between 0 and the original age drawn from a uniform distribution. This percentage is determined using the observed and estimated fractions of WD mergers reported by \cite{temmink2020} and \cite{heintz2022}. We note that, for the purpose of this test, the age distributions of the synthetic population and our sample do not need to match since we are interested in testing whether we can reproduce the distribution of $\Delta P_{\mathrm{rot}}/\sigma_w$ rather than that of the WD ages. The reset ages are used to compute $\Delta P_{\mathrm{rot}}/\sigma_w$, as described in Equations \ref{pmodel} and \ref{likelihood}. 

We find that 16\% of the MS stars in our data sample have a positive $\Delta P_{\mathrm{rot}}/\sigma_w>3$, i.e the age that we would infer from the rotation period of the MS star is older than what we estimate from their WD companions. The synthetic population reveals that 5\% of the stars have a positive $\Delta P_{\mathrm{rot}}/\sigma_w>3$. The 2D histogram in the background of Fig.~\ref{sim} shows the distribution of $\Delta P_{\mathrm{rot}}/\sigma_w$ as a function of the WD age after reset in the synthetic population and highlights a tail of discrepant $\Delta P_{\mathrm{rot}}/\sigma_w$ values at $t<$ 5 Gyr in a region with a high fraction of triples per bin. Such a distribution is well matched by the distribution of $\Delta P_{\mathrm{rot}}/\sigma_w$ of our sample, as shown by the higher concentration of stars with $\Delta P_{\mathrm{rot}}/\sigma_w>3$ at $t<5$ Gyr and the decrease of the number of stars with $\Delta P_{\mathrm{rot}}/\sigma_w>3$ with age.

While we cannot identify with certainty which WDs in our sample may be the products of binary mergers, our findings suggest that a fraction of the systems showing $\Delta P_{\mathrm{rot}}/\sigma_w>3$ may have been triple systems that experienced merger events. Consequently, the ages we estimate for the WDs in such systems may underestimate the true system age, leading our models to predict shorter rotation periods for the companion MS stars than what is observed. We suspect, in particular, that the top panel of Fig.~\ref{compare} is subject to this bias, and that some significant portion of the long period outliers may be these former triple systems.

\subsection{Comparison to other datasets}
\subsubsection{Gyro-kinematic ages of Kepler stars}
We compare the WD ages with the empirical gyro-kinematic ages from \cite{lu2021}. Gyro-kinematic ages leverage on the idea that the velocity dispersion of a stellar population at a given age increases with time due to gravitational interactions between the star and gas clouds \citep{spitzer1951}. By making this assumption, \cite{lu2021} used the rotation periods of around $30{,}000$ Kepler stars to determine their coeval nature (i.e. they assigned the same age to stars showing similar rotation periods and temperatures) and applied age-velocity-dispersion relations to estimate average stellar ages for groups of coeval stars.

We restricted our analysis to gyro-kinematic ages of stars with $3350\,\mathrm{K}<T_{\mathrm{eff}}<5000\,\mathrm{K}$, where the gyro-kinematic ages should not be impacted by weakened braking \citep{vansaders2019}. To compare this sample to our WD + MS sample, for each MS star in our sample, we created a bin centered at its $P_{\mathrm{rot}}$ and $T_{\mathrm{eff}}$ and selected gyro-kinematic stars with a $P_{\mathrm{rot}}$ and a $T_{\mathrm{eff}}$ within 5 days and 100 K from the $P_{\mathrm{rot}}$ and $T_{\mathrm{eff}}$ of the MS star, respectively. If the bin did not contain a minimum number (10) of data points, we increased the size of the bin in the $P_{\mathrm{rot}}$ and $T_{\mathrm{eff}}$ directions by 10\%; we repeated this process up to three times and until the bin contained a sufficient number of data points to obtain a median age representative of the gyro-kinematic age of the bin. We computed the gyro-kinematic age associated to the MS star in our sample as the median of the gyro-kinematic ages of the stars within the bin. 

\begin{figure}
    \centering
    \includegraphics[width=0.47\textwidth]{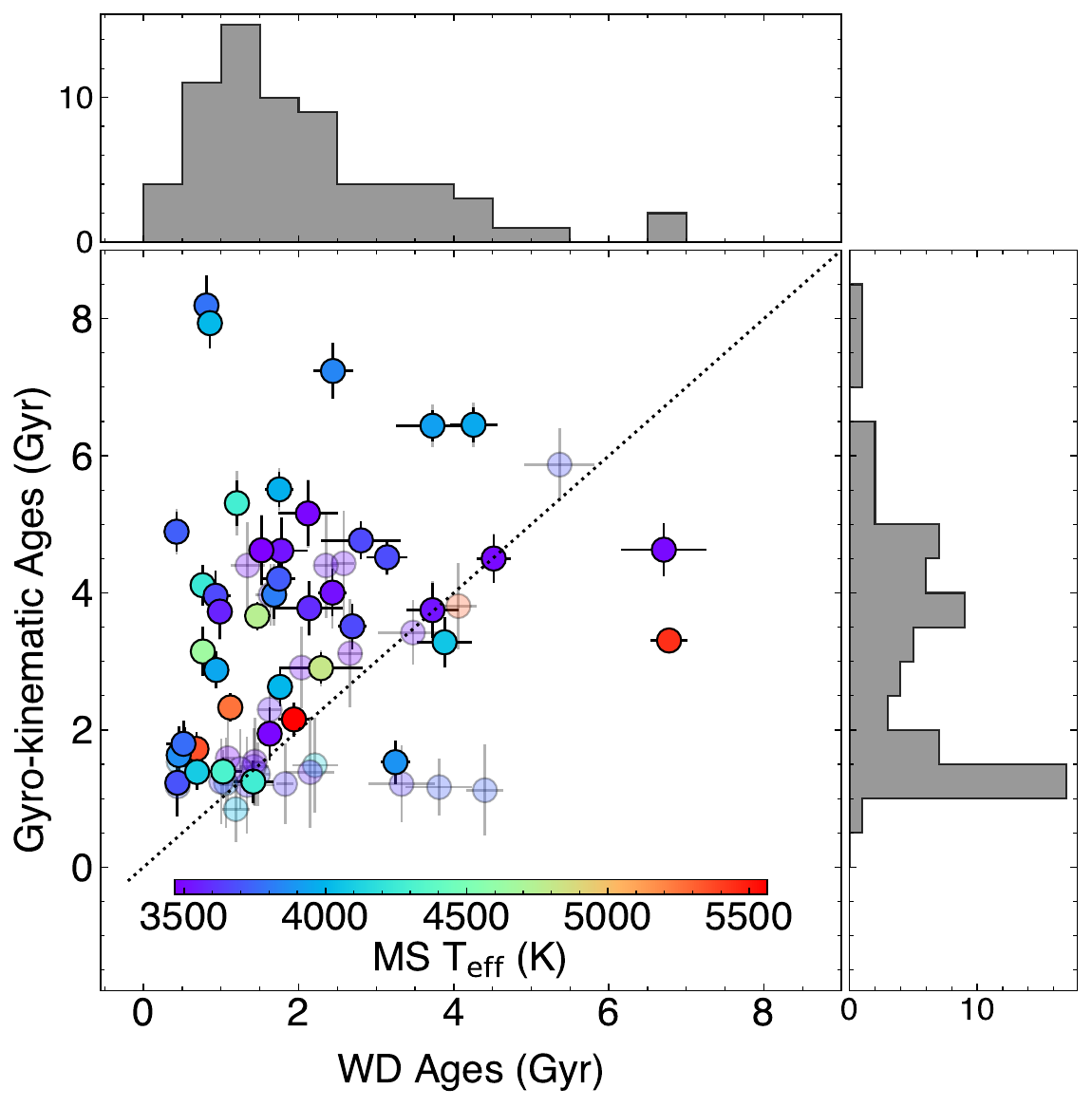}
    \caption{Comparison between the WD ages from this work and the median gyro-kinematic ages (Lu et al. 2021) for bins of Kepler stars around the MS stars in the sample. Data points are color-coded by the effective temperature of the MS stars, with more transparent points indicating bins with at least 10 Kepler stars and more opaque points indicating bins with 30 or more Kepler stars. Uncertainties represent the upper and lower bounds on the WD ages, inflated as prescribed by Heintz et al. (2022). Histograms show the distributions of WD and gyro-kinematic ages.}
    \label{gyro1}
\end{figure}

Fig.~\ref{gyro1} shows that there is a general disagreement between WD and gyro-kinematic ages. In particular, we identify two bands in the plot: a lower band, where the age predicted by the WD companion varies between 0.1 and 4 Gyr while the gyro-kinematic age is roughly constant at 1.5 Gyr, and an upper band, where the age inferred from the WD companions is younger than the gyro-kinematic age. Our hypothesis is that the elongation in the lower band may be due to core-envelope decoupling, which would cause these stars to have a similar rotation period but different ages. WDs more accurately track the true system age, while the gyro-kinematic age is confused by groups of stars with different ages having similar rotation periods. 
The discrepancies in the stars populating the upper band are likely caused by a combination of two factors: 1) some fraction of the WDs in our sample are WD merger products, therefore the age inferred from the WD age is an underestimate of the true age of the system; 2) at the FCB, the gyrochrones are compressed, therefore stars at the same rotation period along this boundary are not necessarily coeval.

\begin{figure*}
    \centering
    \includegraphics[width=\textwidth]{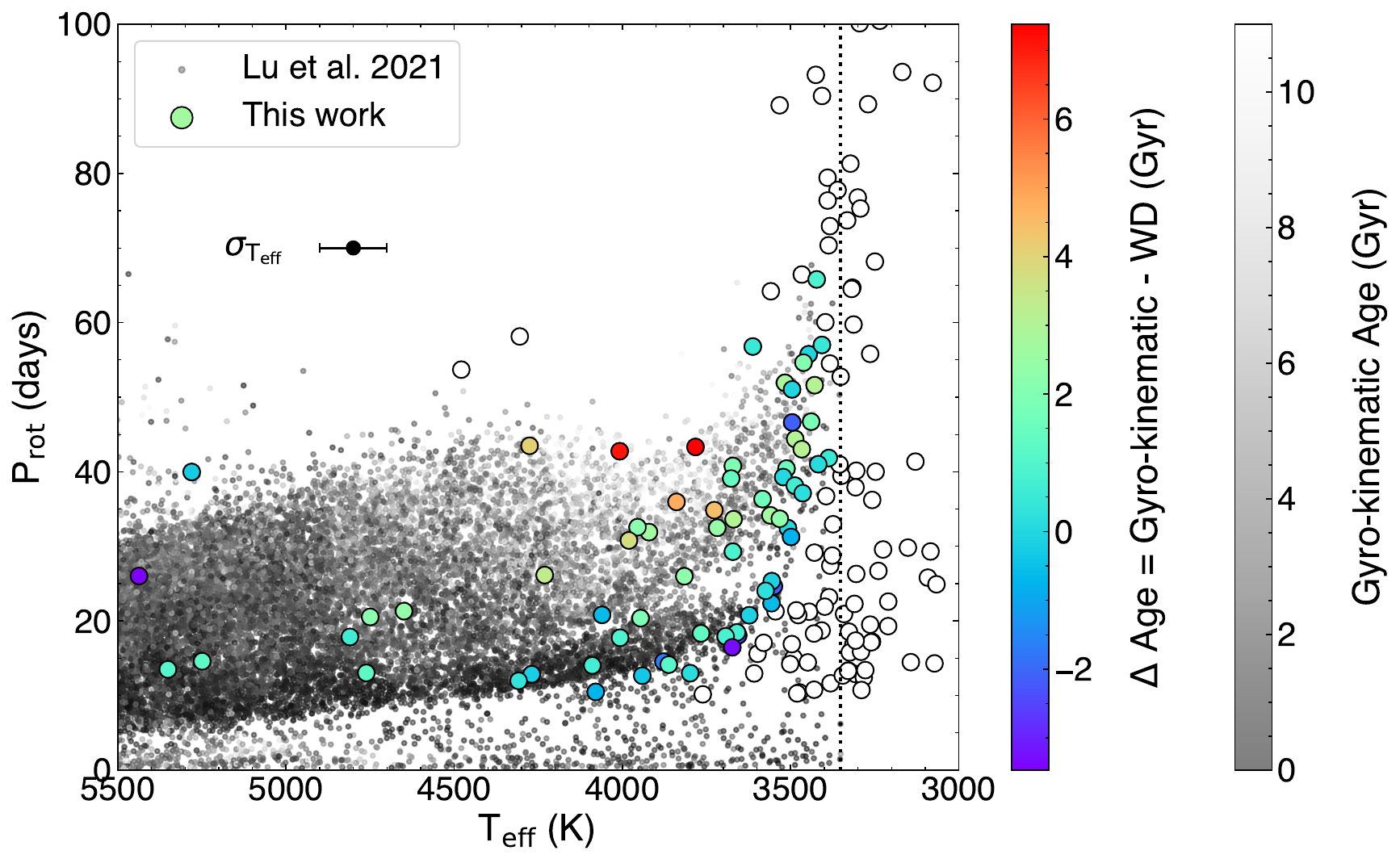}
    \caption{Kepler stars are color-coded by their gyro-kinematic ages \citep{lu2021} in greys. The MS stars in our sample are plotted as larger circles and color-coded by the difference between the median of the gyro-kinematic ages of the neighboring binned Kepler stars and the age inferred from the WD companions. The dashed vertical line shows the $T_{\mathrm{eff}}$ of the coolest star in the Kepler sample. MS stars for which age comparisons are not feasible due to an insufficient number of nearby Kepler stars are indicated with white circles.}. The typical uncertainty in $T_{\mathrm{eff}}$ is shown at the top. 
    \label{gyro2}
\end{figure*}

We find that 30\% of the stars located in this upper band show a $|\Delta P_{\mathrm{rot}}/\sigma_w| > 3$ and a positive $\Delta P_{\mathrm{rot}}$, which supports the WD merger hypothesis for these systems, as discussed in Section \ref{triples}. Furthermore, the MS stars populating the upper band of Fig.~\ref{gyro1} are distributed along the sharp rise in rotation period at the FCB and the disagreement between the WD and gyro-kinematic ages increases as we move toward longer periods, as shown in Fig.~\ref{gyro2}. Therefore, gyro-kinematic ages, which assume that stars at similar periods and temperatures have the same ages, are likely not reliable for stars at the FCB.

Other factors that affect the precision of WD total ages are their mass and the IFMR. Precise mass measurements and well-constrained IFMRs are required to obtain precise ages of low-mass WDs ($M<0.63\,M_{\odot}$) since their progenitor lifetimes represent a major part of their total age. However, the WD companions of the MS stars in the upper band of Fig.~\ref{gyro2} are all high-mass WDs ($M>0.67\,M_{\odot}$), which have short ZAMS progenitor lifetimes, thus their age precision is not significantly affected by their mass and choice of IFMR. Furthermore, because formal age uncertainties of higher-mass WDs are often underestimated, we have applied inflation factors \citep{heintz2022}. 
Lastly, the distribution of WD ages in our sample (grey histogram along the $x$-axis in Fig.~\ref{gyro1}) more closely resembles that of the Kepler-APOGEE Cool Dwarfs sample \citep{claytor2020} and Kepler field stars \citep{silvaaguirre2018}. These age distributions peak at around 1-2 Gyr and do not exhibit the double peak observed in the gyro-kinematic age distribution (grey histogram along the $y$-axis in Fig.~\ref{gyro1}).

The sharp increase of rotation periods at the FCB also challenges the hypothesis that the closing of the intermediate period gap detected in the Kepler distribution \citep{mcquillan2014} is caused by the disappearance of the radiative core, as proposed by \cite{lu2022}. We argue instead that invoking the mechanics of core-envelope decoupling may not be necessary to produce a closure of the gap within the temperature range associated with the FCB. On one hand, the steep incline of gyrochrones in the neighbourhood of this boundary is such that a large range of rotational periods would be consistent with any single gyrochrone within this range of temperatures. This spread of permissible rotational periods is also far larger than the rotational-period separation between gyrochrones near the fully-convective boundary, even for gyrochrones which, at higher temperatures, would be separated by the intermediate period gap. This being the case, points on the period-temperature diagram drawn randomly from two such gyrochrones near the FCB would appear to overlap, thereby apparently closing the gap. This is true irrespective of the actual underlying physical origin for the gap to begin with. In turn, this may imply that we cannot interpret the closing of the intermediate period gap at this location as strong evidence for core-envelope decoupling, although the feature is still fundamentally tied to the loss of a radiative core.

\subsubsection{Gyro-kinematic ages from \citet{lu2023}}
The shearing flows of the solar tachocline, the transition region between the convective zone and the underlying radiative core \citep{schou1998}, are considered to play a key role in the process of magnetic field generation. Fully convective stars do not have a tachocline and therefore are expected to have a different dynamo mechanism. Recent observations of X-ray emissions from fully convective stars reveal that these stars host a dynamo with a rotation-activity relationship that closely resembles that of solar-like stars \citep{Wright2018}, implying that the presence of the tachocline may not be a critical factor in the creation of the stellar magnetic field. However, a recent work by \citet{lu2023} suggested that the dynamos of partially and fully convective stars may be fundamentally different.

Using gyro-kinematic ages of a dataset that combines the Kepler stars from \cite{lu2021} and stars with ZTF rotation periods from \cite{lu2022} and \cite{lu2023b}, they found that fully convective stars exhibit a $1.51\times$ higher AM loss rate than partially convective stars. To account for this, they suggest that fully convective stars necessitate a dipole field strength approximately 1.26 times greater, or a $1.44\times$ increase in the rate of mass loss, or a blend of both factors. 

We use solar-metallicity tracks with $f_{\mathrm{spot}}=0\%$ to compute the ratio of AM loss rate of a 0.28 $M_{\odot}$ star and a 0.40 $M_{\odot}$ star at the same rotation period. We chose these masses to represent fully and partially convective stars, respectively, while also avoiding stars along the vertical feature in rotation period that we find at the FCB. For a range of rotation periods between 10 and 90 days, we find that the fully convective star always shows a higher AM loss rate than the partially convective star by at least a factor of 2, except when its rotation period is shorter than 16 days. Therefore, we suggest that invoking a modification to the stellar dynamo mechanism is not necessary to explain the stronger magnetic braking at the FCB. By scaling the torque with Rossby number, our models naturally reproduce the sharp rise in rotation period at the FCB. 

\subsection{Activity signatures}\label{sec:activity}

Because we are invoking changes in rotation period, Rossby number, and convective overturn timescales to explain the observed behavior, it is natural to ask whether there are observable activity signatures of such a physical transition.

Although the rotation periods increase across the fully-convective boundary, the convective overturn timescales also increase, meaning that we expect very modest values of the Rossby number. Stars near the “spike” in the gyrochrones achieve Rossby numbers less than solar ($\sim 2$) but greater than saturation ($\sim 0.1$) despite the extremes they represent in both rotation period and convective overturn timescale. Rotation-activity correlations have emerged from a variety of activity signatures such as X-rays  \citep{wright2011,Wright2018}, H$\alpha$ emission \citep{newton2017}, UV \citep{france2018}, and show that the activity level decreases with increasing rotation period and increases with decreasing Rossby number. If magnetic activity levels truly do track Rossby number, we expect these stars to be active, but neither unusually active nor unusually quiet compared to field stars of mixed ages at slightly hotter or slightly cooler temperatures.

\begin{figure}
    \centering
    \includegraphics[width=0.45\textwidth]{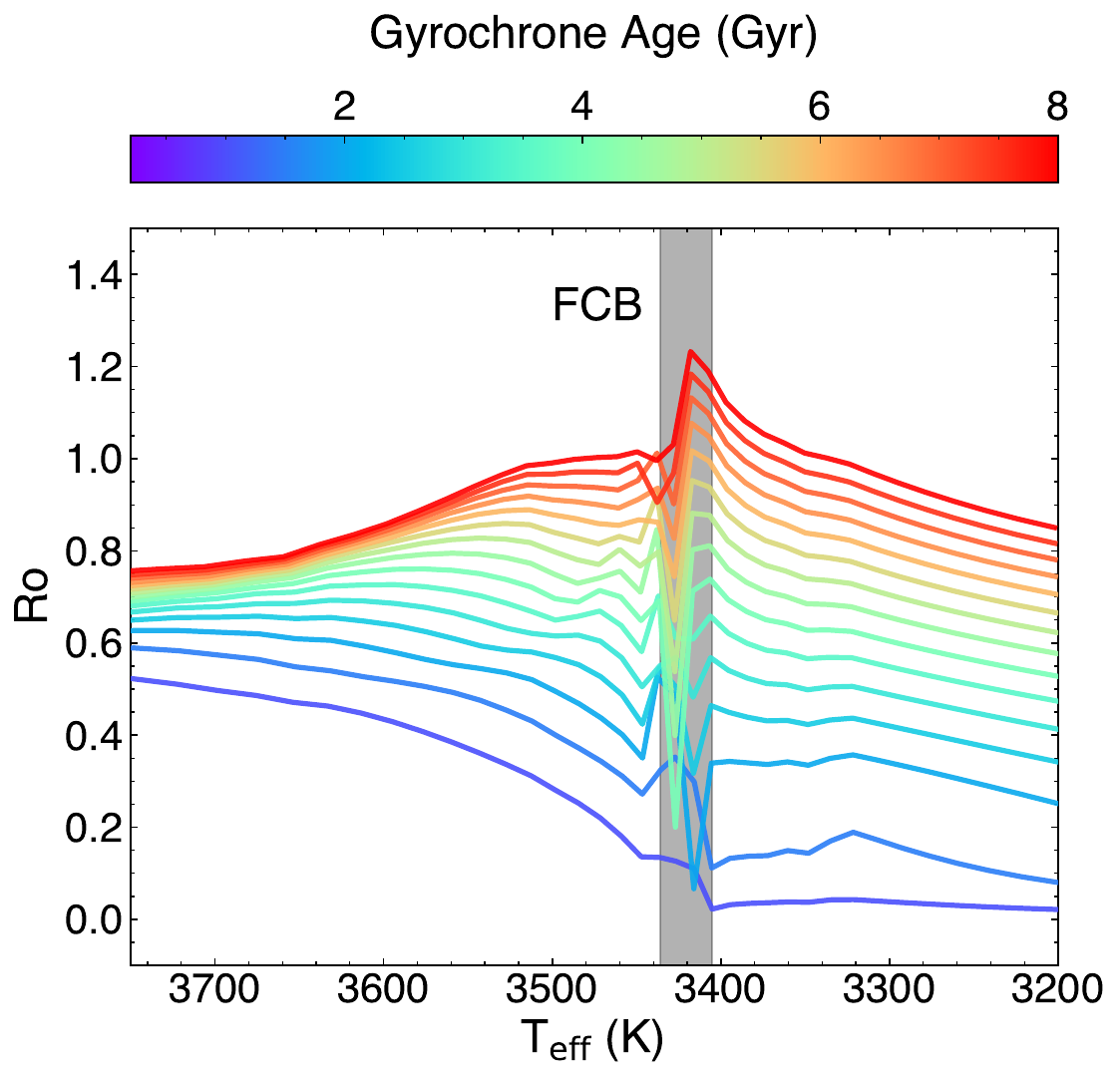}
    \caption{Rossby number as a function of effective temperature as shown by solar-metallicity, $f_{spot}=0\%$ gyrochrones color-coded by their age. At the FCB, highlighted in grey, the Rossby number sees a sudden decrease followed by a sharp increase over the span of $\approx 50$ K.}
    \label{rossbyvsteff}
\end{figure}

In Fig.\ref{rossbyvsteff} we show that stars “on the spike” do have Rossby numbers that are $\sim 0.2-0.7$ lower than stars immediately hotter or cooler than the feature, depending on the age. If we assume, for example, that magnetic activity scales as $ \textrm{Ro}^{-2}$, then this corresponds to a factor of $\sim 2-20$ enhancement in activity for stars in a narrow mass range at the FCB. Observational $L_{X}/L_{bol}-\textrm{Ro}$ relations display an order of magnitude of spread in the X-ray luminosities at fixed Rossby number, making the predicted activity signature at the FCB relatively subtle in comparison. Precision activity measurements in controlled environments like open clusters may represent the best hope of detecting an activity feature at the FCB.

There have been observational efforts to examine the activity of stars in the vicinity of the FCB. \citet{jao2023} claims that stars above the observed M dwarf luminosity gap (higher mass) are more active. In contrast, our models do not predict a lower Rossby number and higher activity rates just above the gap; instead, they predict that fully convective stars coolward (less massive) than the gap have lower Rossby numbers, and would thus appear more active (their long rotation periods are balanced by a larger $\tau_{cz}$). Because existing field samples are relatively small and challenging to control for binarity, it is not yet obvious if this apparent tension is robust.

\citet{boudreaux2024} similarly studied the activity of gap stars, finding that there was a larger scatter in both the observed rotation rates and activity levels on the cool (lower mass) side of the gap. We create a simple stellar population where ages are drawn from a Gaussian centered on 3 Gyr with a width of 2 Gyr, truncated at 0 Gyr and 14 Gyr or the main sequence turnoff age, whichever is younger for each mass in our model grid. In this toy model the dispersion in rotation periods does indeed increase across the FCB (by about a factor of 3 across the 100K near the spike), as does the dispersion in predicted activity levels (again a factor $\sim 3$ in $L_x/L_{bol}$, assuming a $\textrm{Ro}^{-2}$ scaling for activity proxies).

\section{Conclusions}\label{sec:conclusions}
In this work, we constructed a sample of 185 wide, coeval WD + MS binaries with a measured rotation period for the MS companions, which are mostly K and M dwarfs. For the white dwarfs, we derived effective temperatures, surface gravities, and masses by fitting photometric data from various all-sky surveys with atmosphere models—either hydrogen-dominated, pure helium, or mixed—depending on the spectral classification available for each white dwarf. Using these atmospheric parameters, we computed the total age of each WD using WD cooling models, a theoretically motivated and observationally calibrated IFMR, and stellar evolution model grids. Our sample is dominated by massive ($M>0.67\, M_{\odot}$) WDs for which the total age is primarily governed by cooling processes. This allowed us to achieve an average uncertainty of 10\% on the WD total age.  

To model the rotational evolution of the MS stars, we adopted an angular momentum loss prescription for
magnetized winds from \cite{vansaders2013} and modelled the internal angular momentum transport as in the standard two-zone model from \cite{deni2010}. We calibrated gyrochronology models to reproduce the rotational sequences of the open clusters Pleiades, Praesepe, NGC6811, Ruprecht 147, NGC6819 and M67 and the rotation period of the Sun at solar age. We used the calibrated gyrochrones to predict the rotation periods of the MS stars in the sample given their effective temperature and the age from their WD companions. 

We find that the rotation period steeply increases across a narrow temperature range for stars near the FCB and up to $\sim 8$ Gyr. This sharp rise in rotation period is evident in both the models and the data and suggests that stars rotating slowly at the FCB are not necessarily old.

We propose that the rise in rotation period at this boundary is driven by an increase in convective overturn timescale due to structural differences between partially and fully convective stars. As the convective envelope extends deeper into the star, encompassing a larger fraction of the overall stellar mass, it results in a rise in the pressure scale height and a reduction in convective velocity, leading to an increase in $\tau_{cz}$. Furthermore, we argue that the sharpness of such rise in $\tau_{cz}$ is induced by non-equilibrium $^3$He burning occurring for stars just short of the FCB \citep{vansaders2012}. Although the exact location of this vertical feature in $\tau_{cz}$, and consequently $P_{\mathrm{rot}}$, depends on properties like metallicity, spot covering fraction and $\tau_{cz}$ prescriptions, the existence of this feature does not. 

Due to the current uncertainties in temperature measurements, the rotation periods of stars situated along this distinct feature can be associated to a broad spectrum of gyrochrones, spanning a range of $\sim 6$ Gyr. Consequently, despite gyrochronology being regarded as a promising approach for determining the ages of low-mass stars, our findings suggest that age estimation via this method might pose greater challenges when applied to stars located at the FCB.

Future work is planned to obtain spot covering fractions for the MS stars in our sample to allow for a better comparison between the observed and the model rotation periods. Furthermore, as discussed in Sec.~\ref{sec:results}, metallicity has a non-negligible impact on stellar spin-down, therefore more robust predictions of rotation period will be achieved by taking into account the inherent variability and uncertainty associated with this parameter in our models. Having metallicity values for more stars in our sample would also allow for more accurate comparison between the observed and model rotation periods as well as better WD age estimates, since the IMFR is likely to be sensitive to the metallicity of the progenitor stars \citep{cummings2019}. Knowing the metallicity of some of the WDs in the sample will be useful to refine the IFMRs and obtain even more precise WD ages \citep{raddi2022}. Thus, these systems represent optimal targets for wide field spectroscopic surveys. 
For example, the Milky Way Mapper of the Sloan Digital Sky Survey V \citep{kollmeier2017} is collecting APOGEE infrared spectra for 6 million stars across the entire Milky Way and will provide metallicities for all these stars.

Having kinematic ages that remain unaffected by the rotation of main sequence stars would be advantageous. Such ages could serve as a supplementary assessment for the rapid increase in rotation period observed at the FCB. Furthermore, they have the potential to offer insights into the underlying causes of the discordance observed between the gyro-kinematic ages detailed in \cite{lu2021} and the ages deduced from our WD sample.

Finally, our sample represents a subset drawn from a pool of 5005 WD + MS binary systems with measured rotation periods. It is worth noting that there exists a total of $22{,}563$ such systems \citep{elbadry2021}. Access to a greater number of rotation periods would prove invaluable in expanding our sample size, probing even older age ranges, and enhancing our understanding of the rotational evolution of low-mass stars. While the count of TESS-derived rotation periods for cool, MS stars through machine learning techniques is on the rise \citep{claytor2023}, there is also promise in forthcoming space missions like the Nancy Grace Roman Space Telescope \citep{spergel2015}  which will enable many new rotation period measurements.

\bigskip
We thank the anonymous referees for their careful review and valuable feedback, which greatly improved the quality of this manuscript. We would like to thank Dhvanil Desai, Zachary Claytor, Lyra Cao and Nicholas Saunders for helpful discussions. We acknowledge support from the National Science Foundation under Grants No. AST-1908119 and AST-1908723. JMJO acknowledges support from NASA through the NASA Hubble Fellowship grant HST-HF2-51517.001, awarded by STScI, which is operated by the Association of Universities for Research in Astronomy, Incorporated, under NASA contract NAS5-26555.

\bibliographystyle{aasjournal-compact}
\bibliography{main}

\newpage
\appendix

\section{Supporting figures for Section 4.3} \label{sec:firstapp}
\begin{figure}
    \centering
    \includegraphics[width=0.5\textwidth]{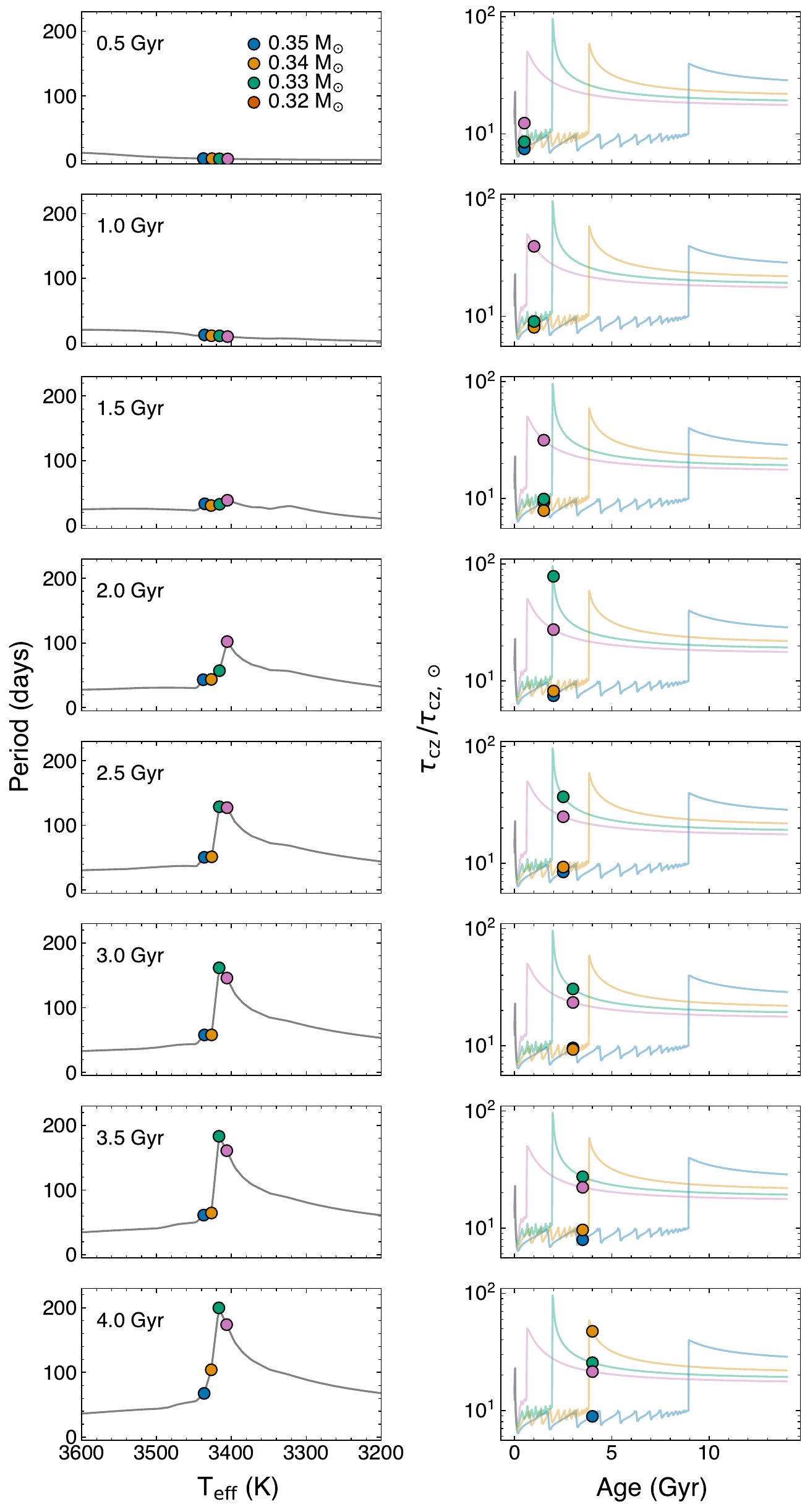}
    \caption{\textbf{Left:} Rotation period as a function of effective temperature for a specific age, computed using a solar-metallicity model grid with $f_{spot}=0\%$. Grey solid lines represent gyrochrones. Markers indicate stars undergoing non-equilibrium $^3$He burning. \textbf{Right:} the convective overturn timescale, normalized by the solar value, as a function of age for the same stars. Markers display $\tau_{cz}$ at the corresponding ages shown in the left plots. The saw-toothed curves represent fully convective episodes. The jump in $\tau_{cz}$ from the saw-toothed to the smooth curve marks the transition to a fully convective state.  Once the stars become fully convective, they reach the peak of the P$_{\mathrm{rot}}$ spike in the plot on the left.}
    \label{app_fig1}
\end{figure}

\begin{figure}
    \centering
    \includegraphics[width=0.6\textwidth]{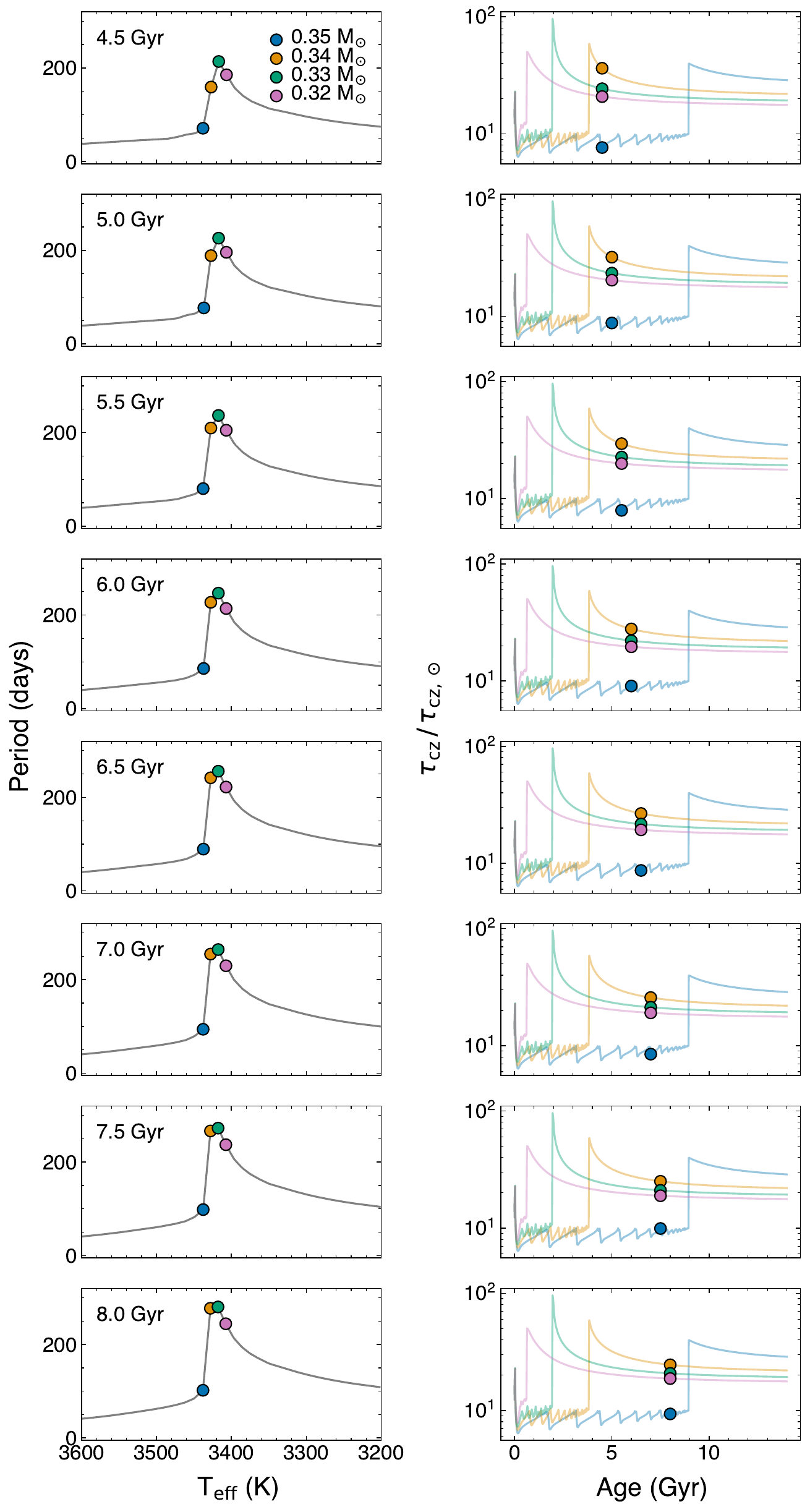}
    \caption{Same as in Fig.\ref{app_fig1} extended to older ages.}
    \label{app_fig2}
\end{figure}

Figures \ref{app_fig1} and \ref{app_fig2} show that the formation of the spike in P$_{\mathrm{rot}}$ at the fully convective boundary is caused by an increase in $\tau_{\mathrm{cz}}$ due to internal structural changes between partially and fully convective stars.  
\newpage

\newpage
\section{Data Sample}
Properties of the MS + WD sample used in this work are
reported in Table \ref{samp}. The table lists the Gaia DR3 IDs, effective
temperatures, observed surface rotation periods of the MS stars,
WD surface gravities, the computed WD total ages and WD
spectral types.
\begin{longtable}{>{\raggedright\arraybackslash}p{3cm}>{\raggedright\arraybackslash}p{3cm}>
{\raggedright\arraybackslash}p{1cm}>
{\raggedright\arraybackslash}p{1cm}>{\raggedright\arraybackslash}p{1.5cm}>{\raggedright\arraybackslash}p{1.5cm}>{\raggedright\arraybackslash}p{2cm}>{\raggedright\arraybackslash}p{2cm}}

\caption{\label{samp} Sample of 185 MS + WD wide binaries and their properties. In the last column, we report the spectral type of the WDs, when available, and their references in brackets.}\\
\hline\hline
MS Gaia DR3 ID & WD Gaia DR3 ID & MS $\mathrm{T_{eff}}$ (K) & MS $\mathrm{P_{rot}}$ (days) & WD $\mathrm{T_{eff}}$ (K) & WD log(g) & WD Total Age (Gyr) & WD spectral type\\
\hline
\endfirsthead
\caption{continued.}\\
\hline\hline
MS Gaia DR3 ID & WD Gaia DR3 ID & MS $\mathrm{T_{eff}}$ (K) & MS $\mathrm{P_{rot}}$ (days) & WD $\mathrm{T_{eff}}$ (K) & WD log (g) & WD Total Age (Gyr) & WD spectral type\\
\hline
\endhead
\hline
  18493721155296640 &   18493721155296768 &         3316 &                 64.7 &    $13230_{-200}^{+200}$ &  $8.09_{-0.04}^{+0.04}$ &                $1.09_{-0.12}^{+0.18}$ &       DA (1)\\
  28185744355533824 &   28185744355056128 &         3371 &                116.3 &    $ 9120_{-520}^{+60}$ &  $8.47_{-0.28}^{+0.26}$ &                $2.01_{-0.25}^{+0.41}$ &      --- \\
  52639226555782656 &   52639153540042496 &         3921 &                 31.9 &    $ 6590_{-250}^{+230}$ &  $8.28_{-0.08}^{+0.08}$ &                $3.73_{-0.20}^{+0.28}$ &      --- \\
  77783305135054720 &   77782544925602432 &         3249 &                 68.2 &    $ 6270_{- 90}^{+ 80}$ &  $8.16_{-0.04}^{+0.04}$ &                $3.49_{-0.20}^{+0.25}$ &      --- \\
  85649623637225856 &   85649413183595392 &         3129 &                 41.4 &    $ 6480_{-140}^{+130}$ &  $8.25_{-0.05}^{+0.05}$ &                $3.70_{-0.14}^{+0.19}$ &       DA (2) \\
 115344202888681216 &  115344095513403520 &         3558 &                 64.2 &    $ 6220_{-110}^{+110}$ &  $8.31_{-0.04}^{+0.04}$ &                $4.49_{-0.10}^{+0.11}$ &       DA (2) \\
 166587938734739456 &  166587938734739328 &         3225 &                 29.6 &    $ 9130_{-170}^{+170}$ &  $8.25_{-0.04}^{+0.04}$ &                $1.52_{-0.05}^{+0.05}$ &       DA (2) \\
 181306276262296192 &  181306168886213120 &         3761 &                 10.2 &    $ 7180_{-380}^{+350}$ &  $8.23_{-0.11}^{+0.11}$ &                $2.63_{-0.31}^{+0.36}$ &      --- \\
 192005963220544768 &  192005967510846848 &         3287 &                137.2 &    $11830_{-630}^{+520}$ &  $8.16_{-0.06}^{+0.06}$ &                $0.99_{-0.10}^{+0.21}$ &       DA (2) \\
 196589900204859392 &  196589861548739584 &         4060 &                 20.8 &    $ 6720_{-220}^{+200}$ &  $8.34_{-0.07}^{+0.07}$ &                $3.88_{-0.14}^{+0.20}$ &      --- \\
 267776887192982784 &  267777604451603328 &         3429 &                 29.2 &    $ 7220_{-170}^{+160}$ &  $8.50_{-0.05}^{+0.04}$ &                $3.95_{-0.05}^{+0.07}$ &      DA: (2) \\
 268123027195914368 &  268123332136902016 &         3128 &                200.0 &    $ 9180_{-270}^{+260}$ &  $8.36_{-0.05}^{+0.05}$ &                $1.57_{-0.06}^{+0.08}$ &      DQ: (2) \\
 271639471546035584 &  271639432884899712 &         3446 &                 55.8 &    $ 6770_{-260}^{+240}$ &  $8.28_{-0.08}^{+0.08}$ &                $3.47_{-0.19}^{+0.27}$ &      --- \\
 293717390146554624 &  293717454571113984 &         4058 &                169.2 &    $ 7810_{-320}^{+280}$ &  $8.21_{-0.09}^{+0.08}$ &                $2.06_{-0.14}^{+0.24}$ &      --- \\
 348345663302119936 &  348345659004467840 &         3223 &                144.9 &    $ 6640_{-120}^{+120}$ &  $8.22_{-0.04}^{+0.04}$ &                $3.27_{-0.16}^{+0.17}$ &       DA (2) \\
 374254624016996352 &  374254619721557504 &         4269 &                 12.8 &    $ 9860_{-240}^{+250}$ &  $8.50_{-0.04}^{+0.05}$ &                $1.41_{-0.12}^{+0.33}$ &       DC (2) \\
 377335386879208704 &  377335451303175808 &         3955 &                 32.6 &    $ 5970_{- 90}^{+ 90}$ &  $8.20_{-0.04}^{+0.04}$ &                $4.26_{-0.19}^{+0.23}$ &      --- \\
 390689402277884544 &  390689397980291072 &         3532 &                 89.1 &    $ 7140_{-150}^{+150}$ &  $8.28_{-0.04}^{+0.04}$ &                $2.94_{-0.13}^{+0.13}$ &      DA: (2) \\
 395377200164554368 &  395377513698597888 &         3142 &                 14.5 &    $ 6000_{- 90}^{+ 90}$ &  $8.33_{-0.04}^{+0.04}$ &                $5.09_{-0.07}^{+0.08}$ &       DA (2) \\
 395932320399086464 &  395932247377032320 &         4308 &                 12.0 &    $11250_{-680}^{+560}$ &  $8.20_{-0.08}^{+0.09}$ &                $1.03_{-0.10}^{+0.20}$ &      --- \\
 411948871922982528 &  411951822555835648 &         3598 &                 15.6 &    $ 8300_{-350}^{+330}$ &  $8.33_{-0.08}^{+0.08}$ &                $2.07_{-0.17}^{+0.13}$ &      DA: (2) \\
 420483143738322304 &  420483139441443200 &         3474 &                 10.4 &    $ 8790_{-520}^{+460}$ &  $8.23_{-0.12}^{+0.12}$ &                $1.62_{-0.12}^{+0.35}$ &      --- \\
 506944889856368256 &  506944889847883008 &         3671 &                 40.8 &    $ 6820_{-260}^{+230}$ &  $8.18_{-0.08}^{+0.08}$ &                $2.80_{-0.28}^{+0.45}$ &      --- \\
 544555471783803136 &  544555467486553984 &         4079 &                 10.5 &    $ 7670_{-280}^{+260}$ &  $8.24_{-0.08}^{+0.08}$ &                $2.21_{-0.17}^{+0.16}$ &      --- \\
 545940409758573824 &  545940164944375296 &         3512 &                 40.4 &    $ 7360_{-140}^{+140}$ &  $8.23_{-0.04}^{+0.04}$ &                $2.44_{-0.11}^{+0.11}$ &      DA: (2) \\
 546388323308593280 &  546388319012071040 &         3263 &                 19.5 &    $ 6260_{- 90}^{+ 90}$ &  $8.22_{-0.04}^{+0.04}$ &                $3.85_{-0.18}^{+0.20}$ &      --- \\
 549835601498785536 &  549788563014393728 &         3521 &                 39.3 &    $ 6270_{-120}^{+110}$ &  $8.20_{-0.05}^{+0.05}$ &                $3.73_{-0.20}^{+0.24}$ &       DA (2) \\
 564321564114510080 &  564509271364217344 &         3425 &                 93.2 &    $ 8480_{-250}^{+240}$ &  $8.30_{-0.06}^{+0.06}$ &                $1.85_{-0.09}^{+0.08}$ &       DA (2) \\
 569795207875481344 &  569795203578976128 &         3383 &                 72.9 &    $ 5470_{- 70}^{+ 70}$ &  $8.14_{-0.04}^{+0.04}$ &                $5.45_{-0.18}^{+0.31}$ &      --- \\
 583948877461123584 &  583949251122623232 &         3466 &                 66.4 &    $ 7480_{- 100}^{+ 100}$ &  $8.13_{-0.04}^{+0.04}$ &                $2.23_{-0.10}^{+0.25}$ &       DA (3) \\
 594439627139102720 &  594439519764767232 &         4761 &                 13.0 &    $20450_{-1000}^{+1030}$ &  $8.20_{-0.04}^{+0.04}$ &                $0.53_{-0.06}^{+0.12}$ &      DBA (4) \\
 642837139695905024 &  642837135401004672 &         3439 &                 46.7 &    $ 6970_{-80}^{+80}$ &  $8.13_{-0.04}^{+0.04}$ &                $2.58_{-0.10}^{+0.18}$ &       DA (5) \\
 657819944131129472 &  657820012850606080 &         3487 &                 38.1 &    $ 9040_{-140}^{+130}$ &  $8.29_{-0.04}^{+0.04}$ &                $1.63_{-0.08}^{+0.15}$ &       DB (4) \\
 678153006506212480 &  678153006506212608 &         3390 &                 79.4 &    $15060_{-710}^{+730}$ &  $8.50_{-0.05}^{+0.05}$ &                $0.60_{-0.04}^{+0.06}$ &      --- \\
 692919207148481920 &  692919202851887104 &         3311 &                130.4 &    $ 5820_{-70}^{+70}$ &  $8.11_{-0.04}^{+0.04}$ &                $4.18_{-0.31}^{+0.55}$ &       DA (2) \\
 701860809365640704 &  701860809367120640 &         3491 &                 15.0 &    $19890_{-530}^{+440}$ &  $8.48_{-0.08}^{+0.08}$ &                $0.35_{-0.05}^{+0.08}$ &      --- \\
 719440149164429824 &  719439423313644672 &         3381 &                 11.6 &    $ 7640_{-110}^{+110}$ &  $8.25_{-0.04}^{+0.04}$ &                $2.24_{-0.10}^{+0.30}$ &       DQ (5) \\
 739321307963291136 &  739321312258441216 &         3314 &                 16.8 &    $10230_{-420}^{+380}$ &  $8.41_{-0.07}^{+0.07}$ &                $1.35_{-0.08}^{+0.05}$ &      --- \\
 748247452595214336 &  748247456890075136 &         3092 &                 25.8 &    $ 6350_{-80}^{+80}$ &  $8.16_{-0.04}^{+0.04}$ &                $3.38_{-0.17}^{+0.21}$ &      --- \\
 765964300065414144 &  765965051684223360 &         3396 &                 60.0 &    $ 5980_{- 70}^{+ 70}$ &  $8.12_{-0.04}^{+0.04}$ &                $3.85_{-0.20}^{+0.30}$ &      DA: (2) \\
 793941132918315392 &  793917660919464960 &         3816 &                 26.0 &    $ 8950_{-250}^{+240}$ &  $8.33_{-0.06}^{+0.06}$ &                $1.68_{-0.06}^{+0.06}$ &      --- \\
 821363846267866368 &  821363807610959104 &         3351 &                 52.7 &    $ 6920_{-110}^{+110}$ &  $8.25_{-0.04}^{+0.04}$ &                $3.03_{-0.12}^{+0.15}$ &      --- \\
 823582763810855168 &  823582759515662592 &         3549 &                 24.6 &    $ 6520_{-150}^{+140}$ &  $8.20_{-0.06}^{+0.06}$ &                $3.33_{-0.26}^{+0.30}$ &      --- \\
 844085460212939392 &  844085490277157376 &         3726 &                 34.8 &    $22660_{-1020}^{+1960}$ &  $8.21_{-0.05}^{+0.05}$ &                $0.43_{-0.05}^{+0.06}$ &       DA (2) \\
 851088662087300992 &  851088662087301120 &         3612 &                 56.8 &    $ 5460_{-70}^{+ 70}$ &  $8.10_{-0.04}^{+0.04}$ &                $5.37_{-0.35}^{+0.56}$ &       DA (2) \\
 855190699452263040 &  855190695156439680 &         3210 &                 19.3 &    $ 5830_{- 70}^{+ 70}$ &  $8.15_{-0.04}^{+0.04}$ &                $4.28_{-0.18}^{+0.23}$ &       DA (2) \\
 856837802230613504 &  856837729215097984 &         4008 &                 42.7 &    $12890_{-660}^{+580}$ &  $8.18_{-0.05}^{+0.05}$ &                $0.86_{-0.07}^{+0.14}$ &      --- \\
 860411043221913984 &  860411038927355264 &         3623 &                 20.8 &    $ 9250_{-290}^{+270}$ &  $8.25_{-0.07}^{+0.07}$ &                $1.48_{-0.07}^{+0.10}$ &      --- \\
 861184515292492288 &  861184515291473536 &         3239 &                 26.7 &    $ 5300_{- 60}^{+ 60}$ &  $8.13_{-0.04}^{+0.04}$ &                $6.31_{-0.23}^{+0.47}$ &       DZ (2) \\
 861951493372439424 &  861951489076842880 &         3341 &                 20.9 &    $ 7040_{-120}^{+120}$ &  $8.13_{-0.05}^{+0.05}$ &                $2.52_{-0.16}^{+0.45}$ &      --- \\
 873994719110827136 &  873994719110827264 &         3326 &                 18.6 &    $ 7600_{- 90}^{+ 90}$ &  $8.37_{-0.04}^{+0.04}$ &                $2.27_{-0.11}^{+0.37}$ &       DC (6) \\
 889041879334169344 &  889041668879549696 &         3655 &                 18.1 &    $ 6660_{-220}^{+200}$ &  $8.31_{-0.08}^{+0.07}$ &                $3.81_{-0.18}^{+0.24}$ &      --- \\
 896972484905743872 &  898473932456801024 &         3304 &                 40.1 &    $ 9300_{-120}^{+130}$ &  $8.43_{-0.04}^{+0.04}$ &                $1.55_{-0.08}^{+0.16}$ &       DC (6) \\
 909162529804018688 &  909162525508848000 &         3258 &                 17.2 &    $10430_{-190}^{+190}$ &  $8.38_{-0.04}^{+0.04}$ &                $1.26_{-0.02}^{+0.02}$ &       DA (2) \\
 909192835093328896 &  909192830801295104 &         3324 &                 15.8 &    $ 6280_{- 70}^{+ 70}$ &  $8.10_{-0.04}^{+0.04}$ &                $3.36_{-0.26}^{+0.40}$ &      DA: (2) \\
 931487529290752000 &  931487524994983808 &         3486 &                 44.4 &    $ 8090_{-130}^{+120}$ &  $8.09_{-0.04}^{+0.04}$ &                $2.12_{-0.27}^{+0.49}$ &       DA (6) \\
 990883830321907456 &  990883830322597376 &         3409 &                120.0 &    $10560_{-250}^{+240}$ &  $8.52_{-0.04}^{+0.04}$ &                $1.25_{-0.03}^{+0.03}$ &      DC: (2) \\
 991811169597579520 &  991811165304213760 &         3331 &                 73.7 &    $ 8480_{-140}^{+140}$ &  $8.15_{-0.04}^{+0.04}$ &                $1.73_{-0.10}^{+0.17}$ &       DA (2) \\
1018436324699616000 & 1018436320403748480 &         3556 &                 22.4 &    $ 8810_{-140}^{+140}$ &  $8.35_{-0.04}^{+0.04}$ &                $1.83_{-0.05}^{+0.06}$ &      DAH (7) \\
1036217489305134848 & 1036217489305134976 &         3290 &                 15.9 &    $ 8760_{-130}^{+130}$ &  $8.16_{-0.04}^{+0.04}$ &                $1.63_{-0.08}^{+0.10}$ &       DA (2) \\
1055800238072103296 & 1055800165056700416 &         3234 &                100.5 &    $ 5970_{- 80}^{+ 70}$ &  $8.15_{-0.04}^{+0.04}$ &                $4.00_{-0.21}^{+0.27}$ &       DA (2) \\
1103650292622702976 & 1103650292623968640 &         3555 &                 25.3 &    $ 9690_{-520}^{+450}$ &  $8.22_{-0.11}^{+0.11}$ &                $1.33_{-0.11}^{+0.31}$ &       DA (2) \\
1175151533777264256 & 1175151538072196992 &         3422 &                 65.8 &    $ 7820_{-130}^{+120}$ &  $8.20_{-0.04}^{+0.04}$ &                $2.04_{-0.07}^{+0.10}$ &      DA: (2) \\
1202830540713288064 & 1202830540713287936 &         3386 &                 41.8 &    $10850_{-250}^{+240}$ &  $8.22_{-0.04}^{+0.04}$ &                $1.09_{-0.04}^{+0.05}$ &       DA (5) \\
1214994506568260608 & 1214994502274335360 &         3495 &                 46.6 &    $ 5820_{- 100}^{+ 90}$ &  $8.67_{-0.04}^{+0.04}$ &                $6.71_{-0.26}^{+0.24}$ &      --- \\
1222565537480885376 & 1222565533185185024 &         3353 &                177.6 &    $ 5900_{- 70}^{+ 70}$ &  $8.14_{-0.04}^{+0.04}$ &                $4.06_{-0.22}^{+0.35}$ &      DA: (2) \\
1233273646861764480 & 1233273646861782656 &         3360 &                 77.8 &    $ 9350_{-330}^{+320}$ &  $8.47_{-0.06}^{+0.07}$ &                $1.90_{-0.11}^{+0.13}$ &      --- \\
1291263917336163456 & 1291263913040087936 &         3407 &                 18.7 &    $ 7040_{- 80}^{+ 80}$ &  $8.13_{-0.04}^{+0.04}$ &                $2.52_{-0.10}^{+0.20}$ &       DA (3) \\
1299293685113944448 & 1299293680818146176 &         3300 &                 76.8 &    $ 6310_{- 90}^{+ 90}$ &  $8.19_{-0.04}^{+0.04}$ &                $3.60_{-0.18}^{+0.22}$ &       DA (2) \\
1319540848141886976 & 1319540843845667072 &         4479 &                 53.7 &    $ 9860_{-300}^{+300}$ &  $8.16_{-0.06}^{+0.07}$ &                $1.33_{-0.12}^{+0.29}$ &      --- \\
1322796261553157760 & 1322796261553165824 &         3798 &                 13.0 &    $11700_{-180}^{+180}$ &  $8.13_{-0.04}^{+0.04}$ &                $1.06_{-0.05}^{+0.11}$ &       DA (8) \\
1332138536976820096 & 1332138532682327552 &         3445 &                 21.2 &    $ 7360_{-160}^{+150}$ &  $8.16_{-0.05}^{+0.05}$ &                $2.29_{-0.13}^{+0.25}$ &      --- \\
1340289109302083584 & 1340289113593902080 &         3257 &                 36.2 &    $ 9110_{-210}^{+200}$ &  $8.33_{-0.04}^{+0.05}$ &                $1.62_{-0.04}^{+0.04}$ &      --- \\
1341558083155952512 & 1341557984372129536 &         3263 &                 55.8 &    $ 7340_{-120}^{+110}$ &  $8.29_{-0.04}^{+0.04}$ &                $2.77_{-0.12}^{+0.11}$ &       DA: (4) \\
1342071937339096192 & 1342071933043493120 &         3269 &                 89.3 &    $ 6320_{-90}^{+ 80}$ &  $8.24_{-0.04}^{+0.04}$ &                $3.92_{-0.12}^{+0.16}$ &      --- \\
1345042955196670208 & 1345041748308955648 &         3260 &                 17.1 &    $ 5920_{- 70}^{+ 70}$ &  $8.16_{-0.04}^{+0.04}$ &                $4.14_{-0.16}^{+0.18}$ &       DA (2) \\
1355692412505713152 & 1355692408211800704 &         3659 &                 18.5 &    $19610_{-890}^{+850}$ &  $8.30_{-0.05}^{+0.05}$ &                $0.44_{-0.03}^{+0.04}$ &       DA (2) \\
1365015515194773632 & 1365015545259769600 &         3168 &                 93.6 &    $ 6430_{-120}^{+120}$ &  $8.28_{-0.04}^{+0.04}$ &                $3.99_{-0.12}^{+0.14}$ &       DA (2) \\
1410448641324559744 & 1410448259071414528 &         3209 &                130.4 &    $ 8360_{-100}^{+110}$ &  $8.32_{-0.04}^{+0.04}$ &                $1.88_{-0.07}^{+0.11}$ &       DQ (3) \\
1412158789927363328 & 1412158789927364224 &         3480 &                 10.3 &    $16680_{-1050}^{+1030}$ &  $8.25_{-0.07}^{+0.07}$ &                $0.57_{-0.06}^{+0.09}$ &      --- \\
1444236977242948480 & 1444236972948340480 &         3583 &                 36.3 &    $ 7610_{-260}^{+230}$ &  $8.17_{-0.08}^{+0.08}$ &                $2.14_{-0.16}^{+0.51}$ &      --- \\
1444622768385439232 & 1444622768385439104 &         3322 &                 81.3 &    $ 8650_{-110}^{+120}$ &  $8.33_{-0.04}^{+0.04}$ &                $1.75_{-0.06}^{+0.10}$ &       DQ (5) \\
1526498829461895936 & 1526498863822158592 &         3345 &                 12.7 &    $ 7890_{- 100}^{+ 90}$ &  $8.32_{-0.04}^{+0.04}$ &                $2.37_{-0.07}^{+0.07}$ &       DA (3) \\
1541905598706074880 & 1541905560050370816 &         3496 &                 16.9 &    $18950_{-330}^{+250}$ &  $8.35_{-0.07}^{+0.08}$ &                $0.45_{-0.06}^{+0.05}$ &      --- \\
1588270938897733888 & 1588270934604149632 &         3980 &                 30.8 &    $ 8700_{-270}^{+260}$ &  $8.31_{-0.07}^{+0.07}$ &                $1.75_{-0.09}^{+0.08}$ &      --- \\
1604198258179547264 & 1604200899583092992 &         3406 &                 90.4 &    $ 5610_{- 70}^{+ 70}$ &  $8.11_{-0.04}^{+0.04}$ &                $4.75_{-0.24}^{+0.42}$ &       DA (2) \\
1659015063216647296 & 1659015063216647424 &         3318 &                 64.5 &    $ 8800_{-130}^{+120}$ &  $8.10_{-0.04}^{+0.04}$ &                $1.78_{-0.19}^{+0.42}$ &       DA (3) \\
1681875731024076160 & 1681875726730147968 &         3430 &                102.0 &    $ 8900_{-190}^{+180}$ &  $8.28_{-0.05}^{+0.05}$ &                $1.63_{-0.05}^{+0.06}$ &      --- \\
1748817160020646784 & 1748816983925915776 &         5280 &                 40.0 &    $ 5890_{- 70}^{+ 70}$ &  $8.09_{-0.04}^{+0.04}$ &                $4.06_{-0.20}^{+0.27}$ &       DA (8) \\
1766826194114207360 & 1766826194114406656 &         4649 &                 21.3 &    $13320_{-480}^{+470}$ &  $8.33_{-0.04}^{+0.04}$ &                $0.77_{-0.03}^{+0.03}$ &      DZA (10) \\
1787683727830825856 & 1787683723535019776 &         3477 &                 20.8 &    $ 6440_{- 90}^{+ 90}$ &  $8.15_{-0.04}^{+0.04}$ &                $3.17_{-0.16}^{+0.25}$ &       DA (2) \\
1803336714670003840 & 1803336749029801728 &         4231 &                 26.1 &    $15410_{-540}^{+500}$ &  $8.13_{-0.04}^{+0.04}$ &                $0.76_{-0.09}^{+0.18}$ &      --- \\
1812554680855696896 & 1812554676560349568 &         3478 &                292.4 &    $16227_{-770}^{+690}$ &  $8.36_{-0.05}^{+0.05}$ &                $0.56_{-0.03}^{+0.03}$ &       DA (2) \\
1819001289330307200 & 1819001289323819520 &         3507 &                 32.4 &    $ 9600_{-540}^{+490}$ &  $8.32_{-0.10}^{+0.10}$ &                $1.42_{-0.08}^{+0.11}$ &      --- \\
1822194099312577024 & 1822193996219126784 &         3312 &                 59.7 &    $ 6500_{- 100}^{+ 100}$ &  $8.18_{-0.04}^{+0.04}$ &                $3.26_{-0.15}^{+0.16}$ &       DA (2) \\
1830600484179388416 & 1830600484161582464 &         3693 &                 17.9 &    $17700_{-1090}^{+1750}$ &  $8.59_{-0.10}^{+0.10}$ &                $0.44_{-0.05}^{+0.07}$ &      --- \\
1874965739686844032 & 1874965499168713600 &         3878 &                 14.5 &    $ 7120_{-100}^{+ 100}$ &  $8.33_{-0.04}^{+0.04}$ &                $3.25_{-0.08}^{+0.10}$ &      DA: (2) \\
1880373172232064256 & 1880373172233144192 &         3483 &                 21.4 &    $19500_{-760}^{+710}$ &  $8.29_{-0.04}^{+0.04}$ &                $0.45_{-0.03}^{+0.04}$ &       DA (2) \\
1911106416309693952 & 1911106416309735552 &         3072 &                 14.3 &    $11750_{-370}^{+350}$ &  $8.46_{-0.04}^{+0.04}$ &                $1.02_{-0.05}^{+0.04}$ &       DA (2) \\
1937074166540526976 & 1937827194563435648 &         3516 &                 51.9 &    $ 9010_{-160}^{+160}$ &  $8.22_{-0.04}^{+0.04}$ &                $1.78_{-0.18}^{+0.49}$ &       DQ (2) \\
1969985710668480256 & 1969985710668477184 &         3285 &                143.5 &    $10280_{-560}^{+500}$ &  $8.64_{-0.08}^{+0.08}$ &                $1.96_{-0.05}^{+0.06}$ &      --- \\
1996725077535282944 & 1996725077535283200 &         3632 &                104.1 &    $ 5500_{- 70}^{+ 70}$ &  $8.08_{-0.04}^{+0.04}$ &                $5.27_{-0.27}^{+0.36}$ &       DA (8) \\
2010645375776770304 & 2010692306878145280 &         3448 &                 14.4 &    $24610_{-870}^{+860}$ &  $8.38_{-0.04}^{+0.04}$ &                $0.32_{-0.04}^{+0.02}$ &       DA (2) \\
2018812170226865664 & 2018818045723591680 &         3941 &                 12.7 &    $10660_{-700}^{+630}$ &  $8.37_{-0.09}^{+0.10}$ &                $1.19_{-0.09}^{+0.07}$ &       DA (2) \\
2020837264488052864 & 2020837157077399552 &         4749 &                 20.5 &    $ 9590_{-210}^{+240}$ &  $8.39_{-0.04}^{+0.04}$ &                $1.47_{-0.10}^{+0.17}$ &       DC (2) \\
2048599211506044160 & 2048599314570668544 &         3839 &                 36.0 &    $ 7320_{-160}^{+170}$ &  $8.39_{-0.05}^{+0.05}$ &                $2.44_{-0.13}^{+0.32}$ &      DC: (2) \\
2053240524958351360 & 2053240524960871680 &         3417 &                 41.0 &    $ 9400_{-180}^{+170}$ &  $8.16_{-0.04}^{+0.04}$ &                $1.43_{-0.08}^{+0.11}$ &       DA (2) \\
2053585565450552960 & 2053584770878226304 &         5351 &                 13.5 &    $14350_{-500}^{+500}$ &  $8.43_{-0.04}^{+0.04}$ &                $0.68_{-0.05}^{+0.03}$ &       DA (2) \\
2067604574919660416 & 2067604574919660544 &         3310 &                 22.3 &    $10350_{-590}^{+510}$ &  $8.25_{-0.10}^{+0.10}$ &                $1.18_{-0.08}^{+0.16}$ &      --- \\
2069622492291282176 & 2069622487994113408 &         3580 &                 17.1 &    $19430_{-620}^{+600}$ &  $8.36_{-0.04}^{+0.04}$ &                $0.43_{-0.02}^{+0.02}$ &      DA: (2) \\
2080422720138489344 & 2080422720133121152 &         3798 &                163.1 &    $ 7040_{-210}^{+190}$ &  $8.19_{-0.07}^{+0.07}$ &                $2.60_{-0.23}^{+0.28}$ &       DA (2) \\
2141224751076269824 & 2141224781139309312 &         3307 &                 12.9 &    $ 5480_{- 90}^{+ 90}$ &  $8.12_{-0.05}^{+0.05}$ &                $5.26_{-0.33}^{+1.17}$ &      --- \\
2142368311889141248 & 2142368346248880896 &         3430 &                 10.8 &    $ 7330_{-260}^{+240}$ &  $8.27_{-0.08}^{+0.07}$ &                $2.66_{-0.22}^{+0.22}$ &      --- \\
2187901837176956800 & 2187901832880159488 &         3498 &                 31.3 &    $ 8260_{-270}^{+250}$ &  $8.34_{-0.07}^{+0.07}$ &                $2.15_{-0.17}^{+0.13}$ &      --- \\
2187942828344053888 & 2187942789690550272 &         3390 &                 76.4 &    $ 9610_{-600}^{+530}$ &  $8.43_{-0.11}^{+0.11}$ &                $1.63_{-0.17}^{+0.12}$ &      --- \\
2205790165505947392 & 2205790169802867200 &         3349 &                 39.5 &    $ 6640_{-240}^{+220}$ &  $8.17_{-0.09}^{+0.09}$ &                $3.00_{-0.34}^{+0.64}$ &      --- \\
2238010155466716544 & 2238010185526881792 &         3382 &                 27.4 &    $ 6890_{-120}^{+120}$ &  $8.19_{-0.04}^{+0.04}$ &                $2.73_{-0.16}^{+0.15}$ &      --- \\
2255996069750790528 & 2255996069748606976 &         3544 &                 21.3 &    $ 6070_{-130}^{+120}$ &  $8.23_{-0.05}^{+0.05}$ &                $4.32_{-0.19}^{+0.25}$ &      --- \\
2275852459474001664 & 2275852455177465088 &         3376 &                 28.8 &    $ 7960_{-230}^{+220}$ &  $8.41_{-0.06}^{+0.06}$ &                $2.77_{-0.11}^{+0.13}$ &      DA: (2) \\
2299189284536000256 & 2299190006090503808 &         3501 &                 14.2 &    $17890_{-550}^{+520}$ &  $8.30_{-0.04}^{+0.04}$ &                $0.49_{-0.02}^{+0.02}$ &       DA (8) \\
2508478574102156544 & 2508478569806892928 &         3328 &                 13.4 &    $ 7470_{-170}^{+160}$ &  $8.32_{-0.05}^{+0.05}$ &                $2.79_{-0.13}^{+0.16}$ &      --- \\
2512689428758531456 & 2512689424463488256 &         3671 &                 29.3 &    $ 8490_{-260}^{+240}$ &  $8.51_{-0.06}^{+0.06}$ &                $2.69_{-0.06}^{+0.10}$ &      --- \\
2519281859960487424 & 2519281855665451008 &         3292 &                 75.3 &    $ 6890_{- 100}^{+ 100}$ &  $8.26_{-0.04}^{+0.04}$ &                $3.16_{-0.11}^{+0.14}$ &      --- \\
2539792321663876736 & 2539792317371533440 &         3329 &                 13.3 &    $ 9310_{-490}^{+460}$ &  $8.25_{-0.11}^{+0.11}$ &                $1.46_{-0.10}^{+0.20}$ &      --- \\
2576762025758278656 & 2576762266276447104 &         4089 &                 14.0 &    $14370_{-370}^{+370}$ &  $8.25_{-0.04}^{+0.04}$ &                $0.69_{-0.03}^{+0.03}$ &       DA (2) \\
2579247712310922240 & 2579247708016296448 &         3430 &                 18.3 &    $ 7060_{-260}^{+220}$ &  $8.19_{-0.09}^{+0.08}$ &                $2.57_{-0.26}^{+0.40}$ &      --- \\
2656915701868610944 & 2656915697573351680 &         3466 &                 43.0 &    $ 9320_{-210}^{+200}$ &  $8.32_{-0.04}^{+0.04}$ &                $1.52_{-0.04}^{+0.04}$ &      --- \\
2700089675200729216 & 2700089675200402048 &         3208 &                108.4 &    $ 5770_{-100}^{+100}$ &  $8.21_{-0.05}^{+0.05}$ &                $4.83_{-0.21}^{+0.25}$ &      --- \\
2724335723363858944 & 2724335723364297344 &         3282 &                 12.4 &    $ 9160_{-260}^{+230}$ &  $8.30_{-0.06}^{+0.05}$ &                $1.55_{-0.05}^{+0.07}$ &      --- \\
2739560214200407168 & 2739560218493488512 &         3669 &                 33.6 &    $12010_{-700}^{+600}$ &  $8.20_{-0.07}^{+0.07}$ &                $0.93_{-0.09}^{+0.15}$ &      --- \\
2781085405419253504 & 2781085401124115328 &         3454 &                159.7 &    $ 6490_{- 80}^{+ 80}$ &  $8.18_{-0.04}^{+0.04}$ &                $3.36_{-0.19}^{+0.45}$ &       DZ (6) \\
2816359731303136384 & 2816359727009816960 &         3151 &                 29.8 &    $ 5230_{- 60}^{+ 60}$ &  $8.21_{-0.04}^{+0.04}$ &                $7.46_{-0.10}^{+0.15}$ &      --- \\
2824510273561942272 & 2824510239202203392 &         3373 &                 33.0 &    $ 7590_{-150}^{+150}$ &  $8.21_{-0.04}^{+0.04}$ &                $2.20_{-0.08}^{+0.11}$ &      --- \\
2857544378863193472 & 2857547329505058176 &         3304 &                159.0 &    $ 6010_{- 70}^{+ 70}$ &  $8.16_{-0.04}^{+0.04}$ &                $3.94_{-0.15}^{+0.17}$ &      DA: (2) \\
2858833212649253504 & 2858833208354052096 &         3495 &                 51.0 &    $ 5810_{- 70}^{+ 70}$ &  $8.18_{-0.04}^{+0.04}$ &                $4.52_{-0.15}^{+0.18}$ &       DA (2) \\
3047076750159238016 & 3047076750150944512 &         3597 &                270.4 &    $ 5890_{-180}^{+170}$ &  $8.29_{-0.09}^{+0.09}$ &                $5.05_{-0.20}^{+0.33}$ &      --- \\
3096007732007670144 & 3096007762070816128 &         3862 &                 14.1 &    $23110_{-1870}^{+1070}$ &  $8.20_{-0.05}^{+0.05}$ &                $0.45_{-0.05}^{+0.11}$ &       DB (5) \\
3137340435682881920 & 3137340435679753728 &         3766 &                 18.3 &    $22880_{-3230}^{+3860}$ &  $9.12_{-0.11}^{+0.15}$ &                $0.52_{-0.13}^{+0.08}$ &      --- \\
3145894292546764544 & 3145894288249489152 &         3385 &                 23.2 &    $ 8060_{-260}^{+230}$ &  $8.20_{-0.07}^{+0.07}$ &                $1.92_{-0.10}^{+0.19}$ &      --- \\
3160143344767586176 & 3160143344767585536 &         3209 &                 22.6 &    $ 8520_{-300}^{+280}$ &  $8.33_{-0.07}^{+0.07}$ &                $1.92_{-0.13}^{+0.10}$ &      --- \\
3165392310198209664 & 3165392305902629504 &         3411 &                106.7 &    $ 8830_{-1560}^{+150}$ &  $8.56_{-0.04}^{+0.04}$ &                $2.62_{-0.03}^{+0.04}$ &      --- \\
3235085949939746304 & 3235085949939746176 &         3077 &                 92.1 &    $ 5650_{-100}^{+ 100}$ &  $8.14_{-0.05}^{+0.05}$ &                $4.69_{-0.27}^{+0.60}$ &      --- \\
3238194445407460864 & 3238194372390557312 &         4305 &                 58.1 &    $ 7230_{-290}^{+270}$ &  $8.30_{-0.08}^{+0.09}$ &                $2.92_{-0.22}^{+0.28}$ &      --- \\
3276414466021403008 & 3276414466021403264 &         3194 &                111.1 &    $ 7180_{-90}^{+ 90}$ &  $8.42_{-0.04}^{+0.04}$ &                $2.65_{-0.11}^{+0.18}$ &       DC (11) \\
3278241888706974336 & 3278241888707395584 &         3196 &                117.5 &    $ 6530_{-240}^{+230}$ &  $8.21_{-0.08}^{+0.08}$ &                $3.33_{-0.28}^{+0.39}$ &      --- \\
3283182338046329472 & 3283182265031074560 &         3306 &                 17.4 &    $ 9660_{-640}^{+580}$ &  $8.38_{-0.12}^{+0.12}$ &                $1.50_{-0.14}^{+0.10}$ &      --- \\
3317635947223274752 & 3317636703137518208 &         3317 &                125.4 &    $ 6340_{- 80}^{+ 80}$ &  $9.29_{-0.04}^{+0.04}$ &                $3.99_{-0.10}^{+0.08}$ &      --- \\
3361149463489086720 & 3361149527909420928 &         3354 &                 41.0 &    $ 9320_{-220}^{+210}$ &  $8.53_{-0.04}^{+0.04}$ &                $2.13_{-0.05}^{+0.06}$ &      --- \\
3369278874508583424 & 3369278767129521792 &         3673 &                 16.5 &    $ 6020_{- 80}^{+ 80}$ &  $8.23_{-0.04}^{+0.04}$ &                $4.40_{-0.14}^{+0.14}$ &       DA (2) \\
3369544303487186560 & 3369544234764481024 &         3278 &                 13.4 &    $ 5180_{- 100}^{+ 90}$ &  $8.12_{-0.06}^{+0.06}$ &                $7.07_{-0.33}^{+1.89}$ &       DA (2) \\
3387417517828617088 & 3387417513531897984 &         3428 &                 51.6 &    $ 9660_{-400}^{+370}$ &  $8.21_{-0.08}^{+0.08}$ &                $1.34_{-0.09}^{+0.17}$ &      --- \\
3389371380055187968 & 3389371384349749376 &         3084 &                 29.3 &    $ 6520_{-140}^{+130}$ &  $8.14_{-0.05}^{+0.05}$ &                $3.03_{-0.22}^{+0.50}$ &       DA (2) \\
3400087667066875648 & 3400087873225289728 &         3560 &                 34.2 &    $12680_{-370}^{+380}$ &  $8.12_{-0.04}^{+0.04}$ &                $0.98_{-0.09}^{+0.20}$ &       DA (2) \\
3413008990266533376 & 3413009093345760768 &         3387 &                 70.4 &    $ 6520_{-160}^{+140}$ &  $8.16_{-0.06}^{+0.06}$ &                $3.10_{-0.25}^{+0.41}$ &       DA (2) \\
3423222461152705536 & 3423222559935120000 &         4007 &                 17.7 &    $ 9580_{-160}^{+160}$ &  $8.74_{-0.04}^{+0.04}$ &                $1.76_{-0.06}^{+0.06}$ &      DC: (2) \\
3435597013551204736 & 3435597013551206784 &         3289 &                 10.7 &    $ 7180_{-290}^{+290}$ &  $8.25_{-0.08}^{+0.09}$ &                $2.73_{-0.25}^{+0.28}$ &      --- \\
3630646463602449664 & 3630648387747801088 &         3398 &                 21.9 &    $12230_{-150}^{+150}$ &  $8.57_{-0.04}^{+0.04}$ &                $1.02_{-0.01}^{+0.02}$ &      DA: (2) \\
3724362340763265536 & 3724362336470991488 &         3066 &                 24.9 &    $ 7930_{-110}^{+100}$ &  $8.26_{-0.04}^{+0.04}$ &                $2.08_{-0.09}^{+0.27}$ &       DQ (10) \\
3726944921778222080 & 3726944951843168768 &         3464 &                 37.1 &    $10480_{-140}^{+140}$ &  $8.13_{-0.04}^{+0.04}$ &                $1.25_{-0.07}^{+0.13}$ &       DA (2) \\
3727701042180797952 & 3727701007821059072 &         3247 &                 40.0 &    $ 8500_{-170}^{+160}$ &  $8.28_{-0.05}^{+0.05}$ &                $1.84_{-0.12}^{+0.32}$ &       DQ (4) \\
3733304978070296320 & 3733305179932909312 &         3098 &                173.7 &    $ 5810_{- 70}^{+ 70}$ &  $8.21_{-0.04}^{+0.04}$ &                $4.73_{-0.11}^{+0.11}$ &      DA: (6) \\
3805246023875111680 & 3805246019580616960 &         3946 &                 20.4 &    $12204_{-790}^{+680}$ &  $8.46_{-0.07}^{+0.08}$ &                $0.94_{-0.07}^{+0.07}$ &      --- \\
3831257784633494016 & 3831257823288140544 &         3306 &                 37.9 &    $ 9680_{-180}^{+160}$ &  $8.33_{-0.04}^{+0.04}$ &                $1.42_{-0.03}^{+0.03}$ &      DA: (5) \\
3839256319408219008 & 3839256319408293248 &         5565 &                 12.5 &    $ 7980_{-200}^{+200}$ &  $8.19_{-0.06}^{+0.06}$ &                $1.94_{-0.10}^{+0.18}$ &      --- \\
3870805121940366720 & 3870804537824812800 &         3471 &                129.3 &    $10160_{-450}^{+420}$ &  $8.22_{-0.08}^{+0.09}$ &                $1.22_{-0.09}^{+0.17}$ &      --- \\
3897698321657960832 & 3897698317362400384 &         3574 &                 24.0 &    $12010_{-450}^{+390}$ &  $8.52_{-0.05}^{+0.05}$ &                $1.00_{-0.03}^{+0.05}$ &      --- \\
3907622513610306432 & 3907622479250567808 &         3406 &                 57.0 &    $ 6950_{- 90}^{+ 80}$ &  $8.19_{-0.04}^{+0.04}$ &                $2.66_{-0.12}^{+0.10}$ &       DA (2) \\
3919287679146000640 & 3919287674850641792 &         5249 &                 14.6 &    $10710_{-350}^{+310}$ &  $8.29_{-0.05}^{+0.05}$ &                $1.12_{-0.03}^{+0.06}$ &      --- \\
3931756484602030976 & 3931756484602030848 &         3294 &                100.1 &    $ 7150_{- 90}^{+ 90}$ &  $8.21_{-0.04}^{+0.04}$ &                $2.67_{-0.20}^{+0.51}$ &       DQ (6) \\
3944167939359417088 & 3944167939360964480 &         3608 &                 13.0 &    $12480_{-980}^{+890}$ &  $8.33_{-0.08}^{+0.10}$ &                $0.86_{-0.07}^{+0.08}$ &      --- \\
3956995877097644416 & 3956995666643307008 &         4276 &                 43.5 &    $10700_{-480}^{+440}$ &  $8.39_{-0.07}^{+0.07}$ &                $1.20_{-0.07}^{+0.04}$ &      --- \\
4283721804641828480 & 4283721800317089920 &         3326 &                138.5 &    $ 6790_{-100}^{+ 100}$ &  $8.34_{-0.04}^{+0.04}$ &                $3.75_{-0.07}^{+0.09}$ &       DA (2) \\
4310991582806231296 & 4310991685885480064 &         3343 &                126.7 &    $ 8910_{-180}^{+170}$ &  $8.17_{-0.04}^{+0.04}$ &                $1.57_{-0.08}^{+0.13}$ &       DA (2) \\
4371782034475387776 & 4371782030179914624 &         3304 &                 26.3 &    $ 6100_{- 80}^{+ 70}$ &  $8.14_{-0.04}^{+0.04}$ &                $3.67_{-0.22}^{+0.32}$ &      DA: (2) \\
4433378806164302464 & 4433378806161595904 &         3394 &                 36.7 &    $ 5500_{- 70}^{+ 70}$ &  $8.10_{-0.04}^{+0.04}$ &                $5.19_{-0.37}^{+1.12}$ &       DA (2) \\
4447660805076826880 & 4447663760014329344 &         3383 &                 54.5 &    $ 9660_{-500}^{+450}$ &  $8.34_{-0.09}^{+0.10}$ &                $1.43_{-0.07}^{+0.09}$ &      --- \\
4465117102652542336 & 4465117098356736000 &         3676 &                 39.1 &    $ 6460_{- 100}^{+ 90}$ &  $8.15_{-0.04}^{+0.04}$ &                $3.14_{-0.18}^{+0.27}$ &      --- \\
4517946196857223808 & 4517946192529261824 &         4810 &                 17.9 &    $ 8450_{-590}^{+510}$ &  $8.40_{-0.13}^{+0.13}$ &                $2.29_{-0.23}^{+0.27}$ &      --- \\
4563562151123622656 & 4563562426001533696 &         3461 &                 54.6 &    $ 7340_{-120}^{+110}$ &  $8.20_{-0.04}^{+0.04}$ &                $2.35_{-0.09}^{+0.11}$ &      --- \\
4565593739374133504 & 4565593735079515904 &         3232 &                178.0 &    $ 6440_{-120}^{+110}$ &  $8.30_{-0.05}^{+0.05}$ &                $4.06_{-0.15}^{+0.16}$ &      --- \\
4565737053842898304 & 4565737049547212544 &         3718 &                 32.5 &    $ 8730_{-300}^{+280}$ &  $8.31_{-0.07}^{+0.08}$ &                $1.75_{-0.11}^{+0.09}$ &      --- \\
4611559819405628672 & 4611559815111559552 &         3532 &                 33.7 &    $ 8740_{-280}^{+250}$ &  $8.23_{-0.07}^{+0.07}$ &                $1.64_{-0.09}^{+0.14}$ &      --- \\
6222883046075139968 & 6222882251504782720 &         3271 &                274.2 &    $15330_{-300}^{+290}$ &  $8.38_{-0.04}^{+0.04}$ &                $0.61_{-0.02}^{+0.01}$ &       DA (2) \\
6236729088627705216 & 6236729329145889792 &         5437 &                 26.0 &    $ 5290_{- 80}^{+ 80}$ &  $8.18_{-0.05}^{+0.05}$ &                $6.78_{-0.15}^{+0.22}$ &       DA (2) \\
 720018767158643072 &  720018767158642944 &         3782 &                 43.3 &    $12830_{-780}^{+690}$ &  $8.27_{-0.06}^{+0.06}$ &                $0.81_{-0.05}^{+0.08}$ &   DC+dM: (5) \\
\end{longtable}
\begin{quote}
    \textbf{References:} WD spectral types have been obtained from: (1) \citet{2003ApJ...596..477Z}; (2) Gaia XP spectra \citep{vincent2024}; (3) \citet{2006ApJS..167...40E}; (4) \citet{2015MNRAS.446.4078K}; (5) \citet{2013ApJS..204....5K}; (6) \citet{2023MNRAS.519.4529C}; (7) \citet{2023MNRAS.520.6111H}; (8) \citet{1999ApJS..121....1M}; (9) \citet{2016MNRAS.455.3413K}; (10) \citet{2019ApJ...885...74C}; (11) \citet{2015ApJS..219...19L}. A colon after spectral types denote uncertain spectral classification.
\end{quote}


\newpage
\section{Rotational Isochrones}\label{appc}
Selected rotational isochrones calculated with our models are
reported in Table \ref{isochrones}. The table lists the surface rotation period,
in days, as a function of stellar mass and age; effective
temperatures are also given, as computed from a 2.0 Gyr
gyrochrone.
\begin{longtable}{ccccccccccccc}
\caption{\label{isochrones} Rotational isochrones constructed using a solar-metallicity, $f_{\mathrm{spots}}=0\%$ model grid.}\\
\hline
\hline
    $M(M_{\odot})$ & $T_{\mathrm{eff}}$ (K) & \multicolumn{11}{c}{Ages (Gyr)} \\
\hline
    & & 0.5 &    1.0 &    2.0 &     3.0 &     4.0 &     5.0 &     6.0 &      7.0 &     8.0 &     9.0 &    10.0  \\
\hline
\endfirsthead
\caption{continued.}\\
\hline
\hline
    $M(M_{\odot})$ & $T_{\mathrm{eff}}$ (K) & \multicolumn{11}{c}{Ages (Gyr)} \\
\hline
    & & 0.5 &    1.0 &    2.0 &     3.0 &     4.0 &     5.0 &     6.0 &      7.0 &     8.0 &     9.0 &    10.0  \\
\hline
\endhead
0.18 &  3200 &   0.75 &   2.78 &   32.46 &   53.27 &   68.01 &   80.12 &   90.67 &  100.14 &  108.84 &  116.93 &  124.53 \\
0.19 &  3221 &   0.84 &   3.19 &    35.8 &   56.42 &   71.35 &    83.7 &    94.5 &  104.24 &  113.19 &  121.52 &  129.35 \\
0.20 &  3241 &   0.94 &   3.66 &   39.07 &   59.66 &   74.83 &   87.47 &   98.57 &  108.59 &   117.8 &   126.4 &  134.48 \\
0.21 &  3259 &   1.04 &   4.18 &   42.32 &   63.03 &   78.51 &   91.48 &   102.9 &  113.22 &  122.73 &  131.59 &  139.91 \\
0.22 &  3276 &   1.16 &   4.77 &   45.63 &   66.58 &   82.44 &   95.78 &  107.55 &   118.2 &   128.0 &  137.13 &  145.71 \\
0.23 &  3292 &   1.29 &    5.4 &    49.0 &   70.34 &   86.65 &  100.42 &  112.56 &  123.55 &  133.66 &  143.07 &   151.9 \\
0.24 &  3307 &   1.42 &    6.1 &   52.56 &   74.43 &   91.26 &  105.49 &  118.04 &   129.4 &  139.83 &  149.52 &  158.61 \\
0.25 &  3321 &   1.56 &   6.84 &   56.34 &    78.9 &   96.34 &  111.09 &  124.08 &  135.82 &  146.58 &  156.58 &  165.93 \\
0.26 &  3335 &   1.69 &   7.08 &   57.58 &   81.81 &  100.31 &  115.83 &  129.44 &  141.68 &  152.88 &  163.25 &  172.95 \\
0.27 &  3348 &   1.72 &   6.57 &   57.95 &   84.48 &   104.3 &  120.75 &  135.08 &  147.91 &  159.61 &  170.41 &  180.49 \\
0.28 &  3360 &   1.85 &   7.14 &   62.79 &   90.81 &  111.63 &  128.84 &  143.77 &   157.1 &  169.23 &  180.42 &  190.85 \\
0.29 &  3372 &    2.0 &   7.45 &   66.35 &   96.98 &  119.27 &   137.5 &  153.21 &  167.18 &  179.85 &  191.52 &  202.37 \\
0.30 &  3383 &   2.12 &   8.03 &   73.43 &  106.78 &  130.58 &  149.84 &  166.34 &  180.95 &  194.17 &  206.32 &  217.61 \\
0.31 &  3394 &    2.3 &   8.85 &   84.02 &  121.16 &  146.81 &  167.25 &  184.62 &  199.92 &  213.73 &   226.4 &  238.15 \\
0.32 &  3405 &   2.49 &   9.74 &  102.15 &  145.87 &  173.89 &  195.64 &   213.9 &  229.89 &  244.27 &  257.42 &  269.61 \\
0.33 &  3416 &    2.7 &  10.92 &   57.15 &  161.66 &  199.83 &  225.93 &  246.77 &  264.56 &   280.3 &  294.56 &   307.7 \\
0.34 &  3426 &   2.94 &  11.15 &   43.67 &   57.95 &  104.08 &  188.55 &  227.14 &  254.86 &  277.38 &  296.76 &  314.01 \\
0.35 &  3437 &   3.05 &  12.33 &   43.15 &   57.87 &   67.56 &   76.75 &   85.68 &   94.08 &  102.01 &  116.45 &  194.68 \\
0.36 &  3446 &   3.19 &  10.99 &   30.03 &   43.71 &   55.65 &   64.93 &   72.68 &   79.42 &   84.58 &   89.38 &   94.24 \\
0.37 &  3458 &   3.61 &   13.4 &   30.65 &   43.47 &   53.35 &   61.13 &    65.2 &   70.98 &   76.22 &   80.69 &   84.59 \\
0.38 &  3470 &    4.1 &  15.24 &   30.85 &   42.64 &   50.44 &   54.57 &   60.82 &   66.18 &   70.58 &   74.34 &   77.44 \\
0.39 &  3484 &   4.69 &  16.46 &   30.94 &   40.87 &   45.81 &   51.97 &   57.81 &   62.34 &   66.05 &   68.99 &   71.47 \\
0.40 &  3497 &   5.38 &  17.33 &   30.89 &   38.59 &   43.83 &   50.25 &   55.22 &   59.04 &   62.01 &   64.26 &   66.11 \\
0.41 &  3512 &   6.24 &  18.31 &   30.89 &   37.09 &    43.2 &   48.97 &   53.17 &   56.29 &   58.64 &   60.43 &   61.73 \\
0.42 &  3526 &   7.27 &  18.88 &   30.31 &   36.05 &   42.16 &   47.17 &   50.61 &   53.05 &   54.81 &   56.11 &   57.02 \\
0.43 &  3542 &   8.53 &  19.38 &   29.55 &   35.44 &   41.04 &   45.25 &    48.0 &   49.86 &   51.14 &    52.0 &   52.71 \\
0.44 &  3558 &   9.96 &  19.81 &    28.8 &   34.85 &   39.74 &   43.21 &   45.33 &    46.7 &   47.64 &   48.24 &   48.77 \\
0.45 &  3575 &  11.05 &  20.15 &   28.28 &   34.09 &   38.33 &   41.11 &   42.71 &   43.72 &   44.41 &   44.86 &    45.3 \\
0.46 &  3592 &   11.9 &  20.41 &   27.93 &   33.21 &   36.83 &   38.95 &   40.13 &   40.88 &   41.43 &   41.87 &    42.2 \\
0.47 &  3611 &  12.57 &  20.53 &   27.57 &   32.25 &   35.21 &   36.89 &   37.77 &   38.36 &   38.82 &   39.22 &   39.55 \\
0.48 &  3631 &  13.14 &  20.49 &   27.17 &   31.17 &    33.6 &   34.83 &   35.52 &   36.03 &   36.47 &   36.88 &   37.22 \\
0.49 &  3651 &  13.42 &  20.15 &   26.38 &   29.79 &   31.68 &   32.64 &   33.21 &   33.67 &   34.04 &   34.47 &    34.9 \\
0.50 &  3673 &  13.96 &   20.1 &   25.93 &   28.75 &   30.24 &   31.06 &   31.58 &   32.04 &   32.49 &   32.89 &   33.36 \\
0.51 &  3697 &  14.32 &  19.89 &   25.21 &   27.54 &   28.77 &   29.45 &   29.95 &   30.43 &   30.91 &   31.34 &   31.85 \\
0.52 &  3721 &  14.66 &   19.8 &   24.55 &   26.53 &   27.52 &   28.15 &   28.67 &   29.19 &   29.71 &   30.19 &   30.74 \\
0.53 &  3747 &  14.84 &  19.62 &   23.78 &   25.36 &   26.24 &   26.85 &    27.4 &   27.96 &   28.52 &   29.06 &   29.65 \\
0.54 &  3774 &  14.88 &  19.28 &   22.75 &   24.08 &    24.9 &   25.49 &   26.07 &   26.68 &   27.29 &   27.92 &   28.51 \\
0.55 &  3803 &  14.96 &  19.05 &    22.0 &   23.19 &   23.97 &    24.6 &   25.24 &    25.9 &   26.57 &   27.25 &   27.92 \\
0.56 &  3834 &  14.93 &   18.7 &   21.22 &   22.29 &   23.04 &   23.75 &   24.45 &   25.11 &   25.86 &   26.62 &   27.39 \\
0.57 &  3866 &  14.78 &  18.25 &   20.34 &   21.34 &   22.13 &   22.89 &   23.65 &   24.43 &   25.23 &   26.02 &   26.88 \\
0.58 &  3900 &  14.66 &  17.75 &   19.56 &   20.53 &   21.37 &   22.19 &   23.01 &   23.87 &   24.75 &   25.62 &   26.57 \\
0.59 &  3935 &  14.66 &  17.44 &   19.04 &   20.03 &   20.91 &   21.82 &   22.74 &   23.69 &   24.67 &   25.65 &   26.72 \\
0.60 &  3972 &  14.42 &  16.74 &   18.22 &   19.23 &   20.17 &   21.14 &   22.14 &   23.18 &   24.26 &   25.34 &   26.53 \\
0.65 &  4164 &  12.97 &  14.08 &   15.45 &   16.79 &   18.23 &   19.75 &   21.36 &   23.02 &   24.83 &   26.79 &   28.76 \\
0.70 &  4395 &  10.93 &   11.9 &   13.76 &    15.8 &   18.04 &   20.49 &   23.14 &   25.95 &    28.9 &   31.92 &   35.08 \\
0.75 &  4638 &   9.21 &  10.43 &   13.12 &   16.19 &   19.58 &   23.17 &   26.91 &    30.7 &   34.44 &    38.2 &   41.89 \\
0.80 &  4885 &   7.97 &   9.62 &   13.44 &   17.73 &   22.15 &   26.51 &   30.77 &    34.9 &   38.91 &   42.82 &   46.65 \\
0.85 &  5124 &   7.28 &   9.51 &   14.57 &   19.72 &   24.64 &   29.23 &   33.63 &   37.84 &   42.03 &   46.16 &    50.4 \\
0.90 &  5347 &   6.89 &   9.76 &   15.74 &   21.25 &   26.27 &   30.93 &   35.46 &   39.93 &   44.47 &   49.21 &   54.33 \\
0.95 &  5554 &   6.76 &  10.15 &   16.55 &   22.08 &   27.09 &   31.87 &   36.64 &   41.56 &   46.89 &   53.11 &     *** \\
1.00 &  5742 &   6.68 &  10.34 &   16.72 &   22.13 &   27.12 &   32.03 &   37.22 &   43.17 &     *** &     *** &     *** \\
1.05 &  5912 &   6.62 &  10.23 &   16.23 &   21.33 &    26.2 &   31.37 &     *** &     *** &     *** &     *** &     *** \\
1.10 &  6067 &   6.19 &   9.42 &   14.71 &    19.3 &   24.08 &     *** &     *** &     *** &     *** &     *** &     *** \\
1.15 &  6206 &   5.38 &   7.97 &   12.27 &   16.33 &   21.77 &     *** &     *** &     *** &     *** &     *** &     *** \\
\end{longtable}
\begin{quote}
    \textbf{Notes.} Models corresponding to masses greater than 0.6\,$M_{\odot}$ are core-envelope decoupling models whose cores are rotating too fast, therefore the surface rotation periods reported here are an underestimate. Missing entries in a gyrochrone (indicated with `***') correspond to stars that have already left the main sequence by that age and that have surpassed the critical Rossby number (Ro$_{\mathrm{crit}}=2.08$, \citealt{vansaders2019}).
\end{quote}

\end{document}